\documentclass[twocolumn]{aastex6}
\usepackage{graphicx}
\usepackage[version=3]{mhchem}
\usepackage{soul} 
\usepackage{amssymb,amsfonts,amsmath,amstext,amsgen,amsopn,amsxtra,indentfirst,times}
\usepackage{comment} 

\begin{document}

\title{\texttt{VULCAN}: An Open-source, Validated Chemical Kinetics \texttt{Python} Code for Exoplanetary Atmospheres}

\author{Shang-Min Tsai\altaffilmark{1}, James R. Lyons\altaffilmark{2}, Luc Grosheintz\altaffilmark{1}, Paul B. Rimmer\altaffilmark{3}, Daniel Kitzmann\altaffilmark{1}, Kevin Heng\altaffilmark{1}}
\altaffiltext{1}{University of Bern, Physics Institute, Center for Space and Habitability, Sidlerstrasse 5, CH-3012, Bern, Switzerland.  \\ Emails: shang-min.tsai@space.unibe.ch, kevin.heng@csh.unibe.ch}
\altaffiltext{2}{Arizona State University, School of Earth and Space Exploration, Bateman Physical Sciences, Tempe, AZ 85287-1404, U.S.A.  \\ Email: jimlyons@asu.edu}
\altaffiltext{3}{University of St. Andrews, School of Physics and Astronomy, St. Andrews, KY16 9SS, United Kingdom}

\begin{abstract}
We present an open-source and validated chemical kinetics code for studying hot exoplanetary atmospheres, which we name \texttt{VULCAN}.  It is constructed for gaseous chemistry from 500 to 2500 K using a reduced C-H-O chemical network with about 300 reactions. It uses eddy diffusion to mimic atmospheric dynamics and excludes photochemistry.  We have provided a full description of the rate coefficients and thermodynamic data used.  We validate \texttt{VULCAN} by reproducing chemical equilibrium and by comparing its output versus the disequilibrium-chemistry calculations of Moses et al. and Rimmer \& Helling.  It reproduces the models of HD 189733b and HD 209458b by Moses et al., which employ a network with nearly 1600 reactions. We also use \texttt{VULCAN} to examine the theoretical trends produced when the temperature-pressure profile and carbon-to-oxygen ratio are varied.  Assisted by a sensitivity test designed to identify the key reactions responsible for producing a specific molecule, we revisit the quenching approximation and find that it is accurate for methane but breaks down for acetylene, because the disequilibrium abundance of acetylene is not directly determined by transport-induced quenching, but is rather indirectly controlled by the disequilibrium abundance of methane.  Therefore, we suggest that the quenching approximation should be used with caution and must always be checked against a chemical kinetics calculation.  A one-dimensional model atmosphere with 100 layers, computed using \texttt{VULCAN}, typically takes several minutes to complete. \texttt{VULCAN} is part of the Exoclimes Simulation Platform (ESP; \url{exoclime.net}) and publicly available at \url{https://github.com/exoclime/VULCAN}.
\end{abstract}


\section{Introduction}

Atmospheric chemistry is a nascent subdiscipline of exoplanet science that is rapidly gaining attention, because of its importance in deciphering the abundances of atoms and molecules in exoplanetary atmospheres.  Unlike for the Earth and Solar System bodies, the bulk of the focus is on currently observable exoplanetary atmospheres, which fall into the temperature range of 500 to 2500 K (for reviews, see \citealt{sd10,madhu14,hs15}).  There exists a diverse body of work on the atmospheric chemistry of exoplanets, and the published models fall into two basic groups: chemical equilibrium \citep{bs99,lodders02,madhu12,blecic16} and photochemical kinetics \citep{koppa12,line11,vm11,moses11,hu12,moses13a,moses13b,line13,agundez14,zahnle14,hu15,venot12,venot15,rimmer16}.  Some of this work traces its roots back to the study of brown dwarfs and low-mass stars \citep{bs99,lodders02}.  The two types of models take on very distinct approaches.  Chemical-equilibrium models seek to minimize the Gibbs free energy of the system and do not require a knowledge of its chemical pathways.  They are hence able to deal with a large number of species with different phases \citep{eq_book}.  Chemical-kinetics models employ a network to calculate the change of every reaction rate with time, and requires the solution of a large set of stiff differential equations.  Chemical-equilibrium models are a simple starting point, but we expect hot exoplanets to host disequilibrium chemistry \citep{ks10,moses11}.

In the current work, we have constructed, from scratch, a computer code to calculate the chemical kinetics of hot exoplanetary atmospheres using a flexible chemical network. In theory, one could construct a single, complete chemical network that is valid for all temperatures.  In practice, chemical networks are constructed with a limited subset of reactions specifically for low or high temperatures, with the former being relevant for Earth and the Solar System bodies.  A low-temperature network typically omits the endothermic reactions (e.g., \citealt{liang03}), because they are very slow and hence do not affect the outcome, but their inclusion would slow down the calculation unnecessarily. Furthermore, extrapolating reaction rates measured at low temperatures to higher temperatures may result in errors at the order-of-magnitude level. In constructing this code, which we name \texttt{VULCAN}\footnote{Named after the Roman god of alchemy.}, we have built a reduced chemical network consisting of about 300 reactions for the temperature range from 500--2500 K, which is compatible with the currently characterizable exoplanets.  We perform a twofold validation of our network: by reproducing chemical equilibrium and by reproducing the disequilibrium-chemistry models of HD 189733b and HD 209458b by \cite{moses11} and also models by \cite{rimmer16}. Initially, our results disagreed with those of \cite{rimmer16}, but upon further investigation we were able to show that this is due to the different chemical networks used.

Chemical kinetics codes are typically proprietary (e.g., \citealt{allen81,line11}), which raises questions of scientific reproducibility. \texttt{VULCAN} is constructed to be completely open-source under the GNU Free Documentation License in the hope that this will accelerate scientific progress.  Furthermore, the user may elect to use a chemical network that is different from what we provide.  \texttt{VULCAN} is also constructed as part of a long-term hierarchical approach, which started with the re-examination of the theoretical foundations of atmospheric chemistry in \cite{hlt16}, analytical carbon-hydrogen-oxygen (C-H-O) networks of equilibrium chemistry in \cite{hl16} and analytical carbon-hydrogen-oxygen-nitrogen (C-H-O-N) networks of equilibrium chemistry in \cite{ht16}.  In the current study, we focus on C-H-O networks of chemical kinetics that take into account disequilibrium chemistry due to atmospheric mixing, which we approximately describe by diffusion in the one-dimensional (1D) limit.  We consider only the gas phase and do not include photochemistry.  We investigate the validity of the quenching approximation and how it is affected by the carbon-to-oxygen ratio (denoted by C/O).

In \S\ref{sect:theory}, we state the governing equations and boundary conditions used.  In \S\ref{sect:method}, we describe our numerical methods.  In \S\ref{sect:rates}, we provide a detailed description of the chemical rate coefficients used in our default, reduced network, as they are a key ingredient of any chemical kinetics code.  In \S\ref{sect:benchmarking}, we subject \texttt{VULCAN} to several tests, thereby validating it.  In \S\ref{sect:results}, we use \texttt{VULCAN} to study theoretical trends, including varying C/O, and also to revisit the quenching approximation.  In \S\ref{sect:discussion}, we provide a concise summary of our current work, compare it to previous work and suggest opportunities for future work.  

Appendix \ref{append:euler} describes our Rosenbrock method, which we use for temporal integration.  Appendix \ref{append:sensitivity} describes a method we developed to identify the key chemical reactions involved in producing a specific atom or molecule.  Appendix \ref{append:generate} describes how one may implement a different set of chemical reactions in \texttt{VULCAN}.  Appendix \ref{appendix:rates} provides the full set of forward rate coefficients.  Appendix \ref{append:nasapoly} describes the thermodynamics data we used to reverse our forward reaction rates.

\section{Theory}
\label{sect:theory}

\subsection{Governing Equations}

For 1D systems, chemical kinetics codes essentially solve a set of mass continuity equations,
\begin{equation}
\frac{\partial n_i}{\partial t} = {\cal P}_i - {\cal L}_i - \frac{\partial \phi_i}{\partial z},
\label{eq:master}
\end{equation}
where $n_i$ is the number density of the $i$-th species and $t$ denotes the time.  ${\cal P}_i$ and ${\cal L}_i$ are the production and loss rates of the $i$-th species (cm$^{-3}$ s$^{-1}$).  The transport flux is given by
\begin{equation}
\phi_i = -K_{\rm zz} n_{\rm total} \frac{\partial X_i}{\partial z},
\label{eq:flux}
\end{equation}
where $z$ denotes the sole spatial coordinate in the vertical/radial direction.  The mixing ratio ($X_i$) is the number density of the $i$-th species normalized by the total number density, which we denote by $n_{\rm total}$.  The number density of each species is given by $n_i = X_i n_{\rm total}$.  Collectively, equations (\ref{eq:master}) and (\ref{eq:flux}) describe a diffusion equation with chemical source/production and sink/loss terms.  Appendix A of \cite{hu12} includes a derivation of equation (\ref{eq:flux}) from the general diffusion equation for heterogeneous atmospheres.

The so-called ``eddy diffusion" coefficient is denoted by $K_{\rm zz}$.  It assumes that convection and turbulence occur on scales much smaller than the pressure scale height, such that atmospheric motion resembles diffusion.  Such an approach has been successful for Earth and the Solar System bodies, e.g., in explaining the over-abundance of carbon monoxide in the upper troposphere of Jupiter \citep{pb77,vis10}.  However, previous studies of the atmospheric circulation of hot Jupiters, in three dimensions, have demonstrated that it takes the form of equator-to-pole circulation cells that extend over several orders of magnitude in vertical/radial pressure \citep{hfp11,php12,psl13,jk16}, which renders the eddy-diffusion approximation suspect\footnote{It has been argued that convection may be approximately described by ``mixing length theory", but the fact remains that such an approach involves a free parameter, which is the mixing length.  It cannot be derived from first principles and requires calibration against more sophisticated calculations.}.  Nevertheless, it has become entrenched, within the atmospheric chemistry community, to use such an approximation \citep{allen81,moses11,moses13a,moses13b,line11,vm11,koppa12,madhu12,agundez14,hu12,hu15,venot15,rimmer16} and we will do the same for the purpose of comparison to previous work in the literature.  We consider the use of eddy diffusion to be a ``necessary evil" for 1D calculations.

The general goal is to solve a set of equations given by (\ref{eq:master}), for $N_i$ number of species, and seek a steady-state solution where $\partial n_i/\partial t=0$, given the initial and boundary conditions.

\subsection{Initial Conditions}
\label{subsect:ic}

\begin{figure}
\begin{center}
\includegraphics[width=\columnwidth]{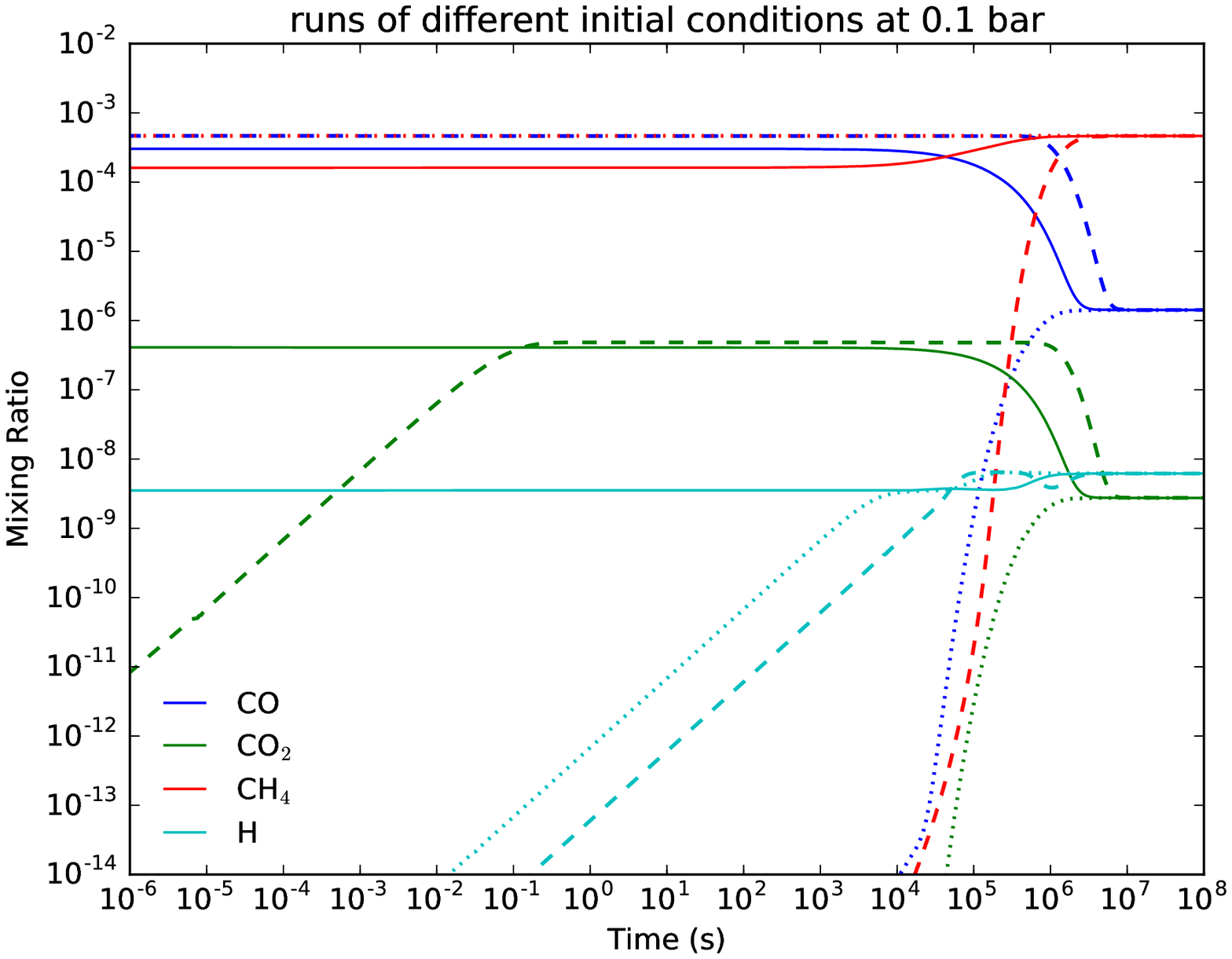}
\includegraphics[width=\columnwidth]{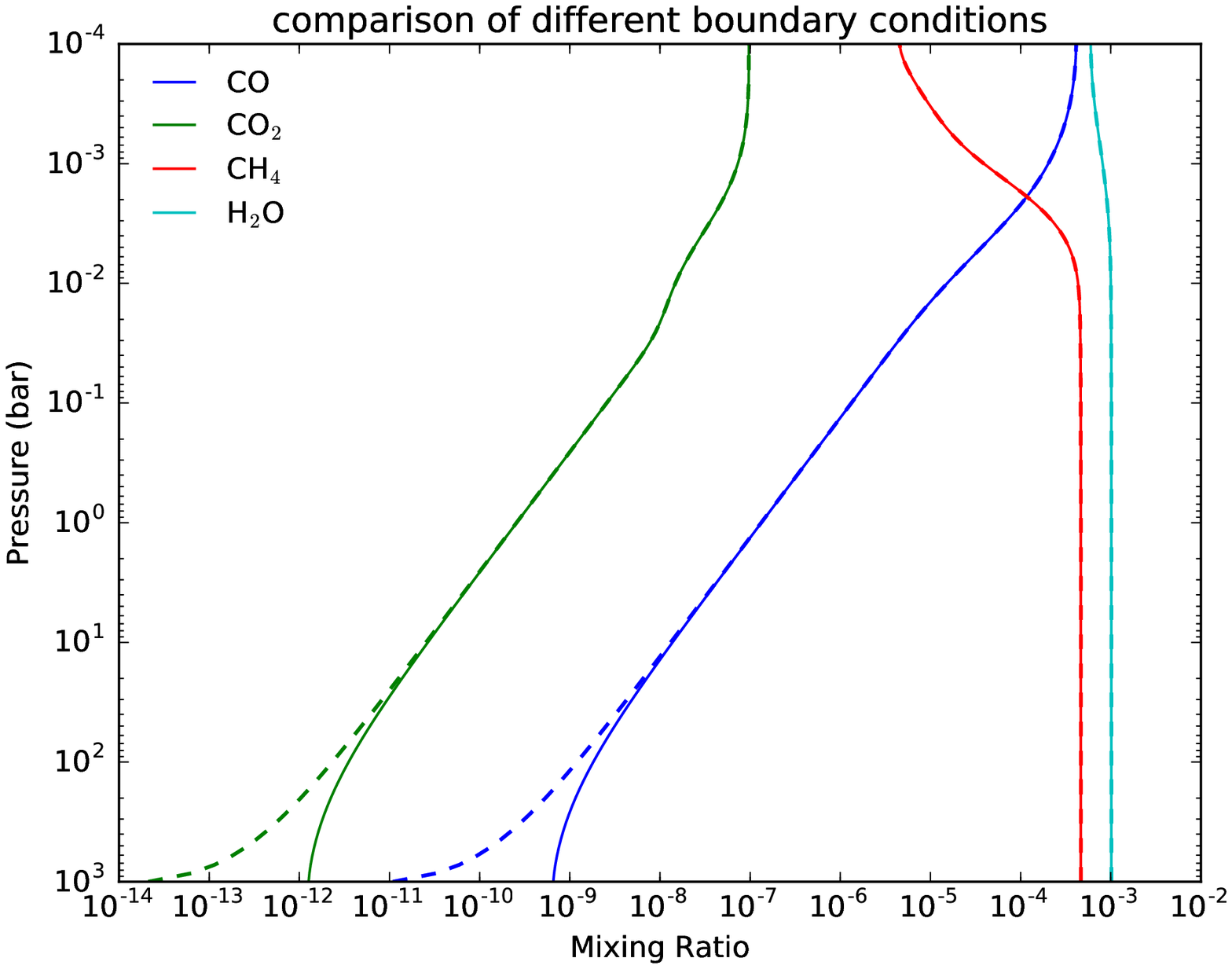}
\end{center}
\caption{Exploring the effects of different choices of initial and boundary conditions. The top panel shows the mixing ratios at $P$=$0.1$ bar from a set of calculations with different initial conditions: chemical equilibrium (solid curves), non-zero initial values for CO, \ce{H2O} and H$_2$ (dashed curves), and non-zero initial values for \ce{CH4}, \ce{H2O} and H$_2$ (dotted curves).  These calculations assume the temperature-pressure profile of Figure \ref{fig:tp2}-B, $K_{\rm zz}=10^{10}$ cm$^2$ s$^{-1}$ and zero-flux boundary conditions.  The bottom panel adopts the same temperature-pressure profile and compares zero-flux (solid curves) versus chemical-equilibrium (dashed curves) boundary conditions. Basically, the choice of boundary conditions affects the outcome, while the choice of initial conditions does not if zero-flux boundary conditions are chosen.}
\label{fig:bic}
\end{figure}

The initial conditions enter according to the chosen elemental abundances. In our C-H-O network, they are the ratios of carbon, oxygen and helium to that of hydrogen ($f_{\rm C}$, $f_{\rm O}$, $f_{\rm He}$). Helium is not sequestered by any molecule but only participates in chemical reactions as a third body.  We refer to the term ``metallicity" as the set of elemental abundances with atomic numbers that are larger than that of helium.  In our C-H-O network, these would be $f_{\rm C}$ and $f_{\rm O}$ only. We note that one may adopt different values of the metallicity, but still have the same value of
\begin{equation}
\mbox{C/O} \equiv \frac{f_{\rm C}}{f_{\rm O}}.
\end{equation}

For each elemental abundance (which we denote by $f_{\rm X}$), we write down the particle conservation equation \citep{hlt16},
\begin{equation}
\sum_i A_i n_i = f_{\rm X} n_{\rm H},
\end{equation}
where $n_{\rm H}$ is the total number of hydrogen atoms.  The quantity $A_i$ records the number of atoms of X that this $i$-th molecule contains.  For example, in conserving hydrogen, this quantity is 2 for H$_2$ and 4 for \ce{CH4}.

Operationally, since we have three elemental abundances for C, H and O, we have three particle conservation equations. Our method for specifying the initial conditions is to pick three molecules: at least one molecule needs to contain C, at least one molecule needs to contain H and at least one molecule needs to contain O.  For example, a set of choices would be \ce{CH4}, \ce{H2O} and H$_2$.  Given our input values of the elemental abundances, the set of three particle conservation equations can be uniquely solved to obtain the mixing ratios of \ce{CH4}, \ce{H2O} and H$_2$.  We use these mixing ratios as our initial conditions and set the mixing ratios of all of the other atoms and molecules, in the network, to zero.  When our calculations attain a numerical steady state, we find that the outcome is independent of our choice of these three molecules, as expected.

In the top panel of Figure \ref{fig:bic}, we show three sets of calculations that assume different initial conditions.  One of them adopts initial values for the mixing ratios of \ce{CH4}, \ce{H2O} and H$_2$, as we have just described.  Another instead adopts initial values for the mixing ratios of CO, \ce{H2O} and H$_2$.  The third set of calculations uses the analytical models of \cite{hl16} to specify the mixing ratios of all of the major molecules.  It is apparent that all three sets of curves, for the erent major molecules, converge to the same numerical solutions after they are evolved for $\sim 10^{7}$ s.  The time required to attain the numerical steady state depends on the shorter of the two timescales: chemical versus dynamical.

For the calculations we present in this study, we adopt the following values for protosolar abundance \citep{lodders09} : $f_{\rm C} \equiv n_{\rm C}/n_{\rm H}=2.7761 \times 10^{-4}$, $f_{\rm O} \equiv n_{\rm O}/n_{\rm H}=6.0618 \times 10^{-4}$ and $f_{\rm He} \equiv n_{\rm He}/n_{\rm H}=0.09691$, except for the comparison with \cite{moses11} for HD 189733b and HD 209458b, where we follow their assumption of multiplying $f_{\rm O}$ by a factor of 0.793, accounting for the effect of oxygen sequestration.  When the value of C/O is varied, we keep $f_{\rm O}$ fixed and vary the value of $f_{\rm C}$.

\subsection{Boundary Conditions}

Solving equation (\ref{eq:master}) also requires the specification of boundary conditions. Several combinations of choices are possible \citep{moses11,hu12}: constant deposition velocity, constant flux or constant mixing ratio at the lower and/or upper boundaries.  In \cite{hu12}, the boundary conditions were chosen to describe atmospheric escape (for the upper boundary) and surface emission or deposition (for the lower boundary), as these authors were modeling terrestrial exoplanets.  In the current study, we are interested in gas-giant exoplanets with no physical surface, which necessitates the choice of a zero-flux boundary condition at both boundaries.  Physically, we are imposing closed boundaries at the top and bottom of our model atmosphere, meaning that no exchange of mass is permitted across them.  For example, \cite{moses11} and \cite{venot12} both adopt zero-flux boundary conditions.  

Some previous works have made a different choice, which is to impose chemical equilibrium at the lower boundary (e.g., \citealt{line11,zahnle14}).  (It is not uncommon to encounter studies where the choice of lower boundary condition is unspecified.)  Physically, in making such a choice, one is arguing that the chemical timescale is short, compared to any dynamical timescale, deep within the atmosphere.  In the presence of vertical transport, molecules can freely flow across this open boundary until all other model levels reach a steady state with it.  While this seems like a plausible approach, we find in practice that some species do not reach chemical equilibrium even at high pressures.  

An example is shown in the bottom panel of Figure \ref{fig:bic}, where we assume the temperature-pressure profile of Figure \ref{fig:tp2}-B and $K_{\rm zz}=10^{10}$ cm$^2$ s$^{-1}$.  Due to the presence of a temperature inversion, the hotter upper atmosphere is in chemical equilibrium, while vigorous vertical transport prevents CO and \ce{CO2} from achieving chemical equilibrium near the bottom even when $P=100$--1000 bar and the temperature is close to 1000 K. The zero-flux boundary values of the molecular abundances deviate from their chemical-equilibrium values by over an order of magnitude. In fact, the chemical timescale associated with converting \ce{CO} to \ce{CH4} at 1000 K and 1000 bar is $\sim 10^6$ s, which is equivalent to the dynamical timescale associated with $K_{\rm zz}=10^9$ cm$^2$ s$^{-1}$ in this atmosphere. These results suggest that chemical-equilibrium boundary conditions lack generality.  Furthermore, we are able to demonstrate that our calculations with zero-flux boundary conditions naturally do not depend on the initial conditions assumed (see \S\ref{subsect:ic}).  For all of these reasons, we consider the zero-flux lower boundary condition to be a better choice than an equilibrium-chemistry one when modeling exoplanets without surfaces.

\section{Numerical Method}
\label{sect:method}

\subsection{Discretization of Equations}

To numerically solve equation (\ref{eq:master}), we discretize the spatial derivative (the diffusion term) using the finite difference method,
\begin{equation}
\frac{\partial n_{i,j}}{\partial t} = {\cal P}_{i,j} - {\cal L}_{i,j} - \frac{\phi_{i,j+1/2} - \phi_{i,j-1/2}}{\Delta z},
\label{eq:master2}
\end{equation}
where, as before, the index $i$ refers to the $i$-th species in our network.  The index $j$ refers to the $j$-th discretized layer of our model atmosphere.  The upper and lower boundaries of the $j$-th layer are marked by the $j+1/2$ and $j-1/2$ indices, respectively.  The spacing between the layers is given by $\Delta z$.  This staggered grid structure is a natural way to define the diffusion flux at the interface between two layers \citep{hu12}.  Equation (\ref{eq:flux}) can then be discretized as
\begin{equation}
\begin{split}
\phi_{i,j+1/2} =& - K_{{\rm zz},j+1/2} ~n_{{\rm total},j+1/2} ~\frac{X_{i,j+1} - X_{i,j}}{\Delta z}, \\
\phi_{i,j-1/2} =& - K_{{\rm zz},j-1/2}~n_{{\rm total},j-1/2} ~\frac{X_{i,j} - X_{i,j-1}}{\Delta z},
\end{split}
\label{eq:flux2}
\end{equation}
where we have approximated
\begin{equation}
n_{{\rm total},j \pm 1/2} = \frac{n_{{\rm total},j \pm 1} + n_{{\rm total},j}}{2}.
\end{equation}

With this approach, the set of partial differential equations in equation (\ref{eq:master}) is transformed into a set of ordinary differential equations with time as the independent variable.  Operationally, one begins with the initial conditions, which allows one to calculate all of the source and sink terms in equation (\ref{eq:master2}).  We then apply the Rosenbrock method (see Appendix \ref{append:euler}) to evolve $n_{i,j}$ forward in time.  At each timestep, we need to solve a block-tridiagonal systems of equations, the details of which we describe in Appendix \ref{append:euler}.  For a system of $N_i$ species and $N_j$ layers, a square matrix with a size of $N_i N_j$ by $N_i N_j$ has to be dealt with. As $n_{i,j}$ is evolved, the chemical production and loss terms, as well as the diffusion term, are updated. The process proceeds until all $n_{i,j}$ reaching a steady-state solution.  

The number of layers can be freely chosen for the required resolution and efficiency. We have tested the insensitivity of our results to specifying 20, 50, 100 and 200 layers.  We find that 20 layers is insufficient to accurately resolve the quench levels.  With 100 layers, we have about 6 layers for every pressure scale height (as our calculations span pressures of $10^{-4}$ to $10^3$ bar) and visibly smoother profiles for the mixing ratios. We first specify our grid in terms of pressure and then compute the corresponding grid in distance ($z$) by solving the hydrostatic balance equation.  At each grid point, we compute the mean molecular weight using
\begin{equation}
\mu = \sum_i \frac{m_i X_i}{m_{\rm u}},
\end{equation}
where $m_i$ is the mass of the individual molecules and $m_{\rm u}$ is the atomic mass unit.  We set the surface gravity to be $10^3$ cm s$^{-2}$.  For the comparison with \cite{moses11}, we use surface gravities of 2140 cm s$^{-2}$ and 936 cm s$^{-2}$ for HD 189733b and HD 209458b, respectively.

\subsection{Integration Routine for Stiff Equations}

Integrating equation (\ref{eq:master2}) is challenging, because the system of equations are ``stiff". This occurs because the chemical timescales vary over many orders of magnitude.  The stiffness of the system can be defined as the ratio 
\begin{equation}
r = \frac{t_s}{t_f} = \frac{\mbox{max}\left\vert \mbox{Re} \left(\xi \right) \right\vert}{\mbox{min}\left\vert \mbox{Re} \left(\xi \right) \right\vert}
\end{equation}
where $t_s$ and $t_f$ are the timescales of the slowest and fastest reactions, obtained by taking the ratio of the maximum to the minimum eigenvalues (denoted by $\xi$) of the Jacobian matrix, which is stated in equation (\ref{eq:jac}).  Our C-H-O network typically has a stiffness ratio $r \sim 10^{20} - 10^{30}$, which means that the fastest reaction is 20--30 orders of magnitude faster than the slowest ones.  The timestep used for explicit solvers is required to be smaller than the fastest reaction in the chemical network ($t_f$) to maintain stability, which creates a computational bottleneck. Implicit solvers are suitable for solving stiff equations, because the computational timestep is only limited by the desired accuracy \citep{nr}.  Among the implicit methods, the backward Euler is one of the simplest, lowest-order (first-order) and most stable.  It is widely used by atmospheric chemists (e.g., \citealt{hu12}).  Other chemical kinetics codes use canned solvers of higher order, e.g., \texttt{DLSODES} \citep{venot12,krome,rimmer16}.  \texttt{VULCAN} is implemented with the Rosenbrock method, which is of higher order (second-order or forth-order) than the backward Euler method (Appendix \ref{append:euler}).

The initial timestep is typically chosen as $\Delta t=10^{-8}$ s to better maintain positivity (otherwise negative values of the number density may result).  After each step, the following conditions have to be satisfied: 1. positive solutions obtain; 2. the estimated truncation error (see Appendix \ref{append:euler}) has to be smaller than the desired tolerance; 3. mass conservation has to be obeyed. If all of these conditions are satisfied, the solution is stored and evolved to the next step. If not, it is rejected and the stepsize is reduced, and the solver routine is called until the conditions are satisfied. 

The integration stepsize is adjusted according to the estimated truncation error.  It is crucial for solvers in kinetics to have an adaptive stepsize control since the chemical timescales span such a wide range. As the error becomes smaller in a typical run, the stepsize increases in the later stages.  The solver stops once the convergence condition is satisfied, for which we define the variation over a  period of time: After the $k$-th timestep and $\tau$ integration time, we compute the relative variation
\begin{equation}
\Delta \hat{n} \equiv \frac{\left \vert n_{i,j,k} - n_{i,j,k'} \right \vert}{n_{i,j,k}},  \quad  \Delta t \equiv t_k - t_{k'}
\end{equation}
where $k'$ refers to the timestep at $f \tau$, i.e., the variation of the solution from $f \tau$ to $\tau$ is examined.  For example, if $f$ is set as 0.5, it means the last half of the integration is examined for a steady state.  We declare our calculation to have reached a numerical steady state if
\begin{equation}
\Delta \hat{n} < \delta \mbox{ and } \frac{\Delta \hat{n}}{\Delta t} < \epsilon,
\end{equation}
In practice, we find that $\delta = 0.01$ and $\epsilon = 10^{-4}$ s$^{-1}$ is useful for ensuring that our calculations complete within a reasonable time ($\sim$minutes). \texttt{VULCAN} is written purely in \texttt{Python}, but still remains efficient\footnote{Our other codes from the ESP suite are written in \texttt{CUDA} in order to exploit GPU acceleration.}. For the model atmosphere of HD 189733b with 100 vertical levels, it takes about 3 minutes for \texttt{VULCAN} to attain steady state with a single 3.2 GHz Intel Core i5 processor running on an iMac.

In the current version of \texttt{VULCAN}, our temperature-pressure profiles are held fixed.  As such, after a successful integration we reset the total number density (if it has changed) in order to preserve the value of the pressure in each layer.

\section{Chemical Rate Coefficients}
\label{sect:rates}

Rate coefficients are the essential ingredients of a chemical kinetics model.  We adopt the generalized Arrhenius equation \citep{combustion,hlt16},
\begin{equation} 
k = A ~T^{b} ~\exp{\left( - \frac{E}{T} \right)},
\label{eq:arr}
\end{equation}
where $k$ is the rate coefficient in units of cm$^3$ s$^{-1}$ for bimolecular reactions and cm$^6$ s$^{-1}$ for termolecular reactions, such that the reaction rate has units of reactions per unit volume and time (cm$^{-3}$ s$^{-1}$). The values of $A$, $b$ and $E$ (the activation energy normalized by the specific gas constant) for the forward reactions are listed in Table \ref{tab:rates}. The size of the chemical network is determined by the number of reactions included.  We have kept our network as small and lean as possible without compromising the accuracy of our calculations (We will validate this claim in \S\ref{sect:benchmarking}.)  For our current C/H/O network, we include 29 molecules and radicals with up to two carbon atoms: \ce{H}, \ce{H2}, \ce{O}, \ce{OH}, \ce{H2O}, \ce{CH}, \ce{C}, \ce{^3CH2} (triplet), \ce{CH3}, \ce{CH4}, \ce{C2}, \ce{C2H2}, \ce{C2H3}, \ce{C2H}, \ce{C2H4}, \ce{C2H5}, \ce{C2H6}, \ce{CO}, \ce{CO2}, \ce{CH2OH}, \ce{H2CO}, \ce{HCO}, \ce{CH3O}, \ce{CH3OH}, \ce{CH3CO}, \ce{O2}, \ce{H2CCO}, \ce{HCCO} and \ce{He}.  In total, we have about 300 forward and reverse reactions involving these species.  A reduced chemical network greatly improves the computational efficiency and makes it easier to monitor and identify individual reactions within the network, as well as weed out suspect input data.  It is constructed with the aid of C$_0$--C$_2$ sets of chemical mechanisms (Appendix C in \citealt{combustion}), but with most of the rate coefficients obtained from the NIST database\footnote{\texttt{http://kinetics.nist.gov/kinetics/}} (validated from 500 to 2500 K).  Some of the irrelevant reactions are further removed by performing the sensitivity analysis (described in Appendix \ref{append:sensitivity}).  We typically select the exothermic reaction as the forward reaction, wherever possible, since they are usually more well characterized---with some exceptions (often thermo-dissociation reactions) where the endothermic reaction rates are better measured.  It is common to have inconsistent data for the same reaction in the database.  When there is more than one rate coefficient available, they are selected by the following criteria \citep{zahnle09}: 1. validation within a better temperature range; 2. date of the publication; 3. review versus experimental versus theoretical publications (in that order).  We note that although we benefit from the detailed investigation of the reactions from the combustion industry, some of the rate coefficients are not well studied or have been measured in a biased manner, meaning some of the combustion networks are optimized for a specific scheme and prefer a more effective pathway (see \citealt{venot12,moses14} for a detailed discussion).

As an example,  in the scheme for \ce{CH4} $\leftrightarrow$ \ce{CO}, which is one of the most important conversion schemes in the hot Jupiter regime, there are various pathways leading to the interconversion.  The slowest reaction in the fastest pathway effectively determine the efficiency of the conversion scheme, which is the control or the rate-limiting step/reaction. The uncertainties in these reaction rates have a major influence on the quenching behavior and the disequilibrium mixing ratios.  In particular, the reactions 
\begin{equation} \label{R1}
\ce{CH3 + OH + M -> CH3OH + M}
\end{equation}
and 
\begin{equation}
\ce{CH3OH + H -> CH3 + H2O} \label{R2}
\end{equation}
are often the control steps in \ce{CH4} $\rightarrow$ \ce{CO} and \ce{CO} $\rightarrow$ \ce{CH4}, respectively. Physically, they are responsible for the step with highest energy barrier, breaking and forming the bond between carbon and oxygen. Unfortunately, these reactions rates are not well studied and the measured rates reported in the literature are inconsistent \citep{bau94,hum94,hidaka89}.  (See the discussion in \citealt{vm11,moses11}.) In our network, we adopt the rate coefficients for reaction (\ref{R1}) from \cite{jas07} by considering its more recent publication date and more thorough treatment.  For reaction (\ref{R2}), we adopt the rate coefficient from the ab initio transition-state theory calculated in \cite{moses11}, which has an activation barrier much larger than other similar reactions and also those from the combustion literature \citep{hidaka89}.  Specifically, \cite{moses11} derived a rate coefficient of $4.91 \times 10^{-19} (T/1\mbox{ K})^{2.485} \exp{(-10380\mbox{ K}/T)}$ cm$^3$ s$^{-1}$.  We note that \cite{rimmer16} performed a similar quantum chemical calculation of the rate coefficient and obtained $9.41 \times 10^{-9} \exp{(-12400\mbox{ K}/T)}$ cm$^3$ s$^{-1}$, where the pre-exponential factor is much larger and hence yields a faster reaction rate.  The reason for this discrepancy remains unknown. Adopting different rate coefficients alters the \ce{CH4} $\rightarrow$ \ce{CO} pathways and eventually changes its chemical timescales.

In some cases, the rate coefficient of the reverse reaction is experimentally measured.  We refer to this as the ``backward rate coefficient" to distinguish it from the thermodynamically reversed rate coefficient.  One may then use either the backward or reversed rate coefficients (see \citealt{venot12} for a discussion).  In our initial calculations, we used both the forward and backward rate coefficients that were obtained from experiments, whenever possible, and found that it does not always reproduce chemical equilibrium in our benchmarking tests (see the discussion in Section 2.1.3 of \citealt{venot12}).  We then used the reversed rate coefficients for all of our calculations to ensure that chemical equilibrium can be consistently achieved. The forward rate coefficient is reversed using the procedure outlined in Appendix \ref{append:nasapoly} (see also \citealt{vm11}, \citealt{venot12} and \citealt{hlt16}\footnote{We note that the standard Gibbs free energy stored in the NASA polynomials are defined differently than the standard Gibbs free energy of formation. They have a different reference level that corresponds to zero energy. Such a difference vanishes when calculating the difference of the Gibbs free energy of a balanced reaction according to Hess's Law \citep{atkins}.}), which requires the use of thermodynamic data.  For completeness, Table \ref{tab:nasa9} lists the NASA polynomials used to compute the enthalpy and entropy, which are taken from {\scriptsize \texttt{http://garfield.chem.elte.hu/Burcat/burcat.html}}.

Having obtained the reversed rate coefficients, we do not fit them with the functional form of the generalized Arrhenius equation, because we find that the fitting procedure can fail or produce noticeable errors (especially when the values vary rapidly with temperature) and requires manual verification \citep{nagy11}. Instead, we reverse the rate coefficients, at given values of the temperature and pressure, on the fly and before the calculation starts.

\section{Benchmarking}
\label{sect:benchmarking}

In writing \texttt{VULCAN} from scratch, we have learned that benchmarking is an indispensible first step in demonstrating that a chemical kinetics code is accurate, especially when constructing a reduced chemical network.  We have used the \texttt{TEA} Gibbs free energy minimization code of \cite{blecic16} to compute equilibrium-chemistry results from 800 to 2500 K.  For disequilibrium chemistry, we compare the output from \texttt{VULCAN} with that from the \texttt{ARGO} code of \cite{rimmer16}.  Furthermore, we reproduce the models of HD 189733b and HD 209458b by \cite{moses11}.

\subsection{Comparison with Calculations of Equilibrium Chemistry}

\begin{figure}
\begin{center}
\includegraphics[width=\columnwidth]{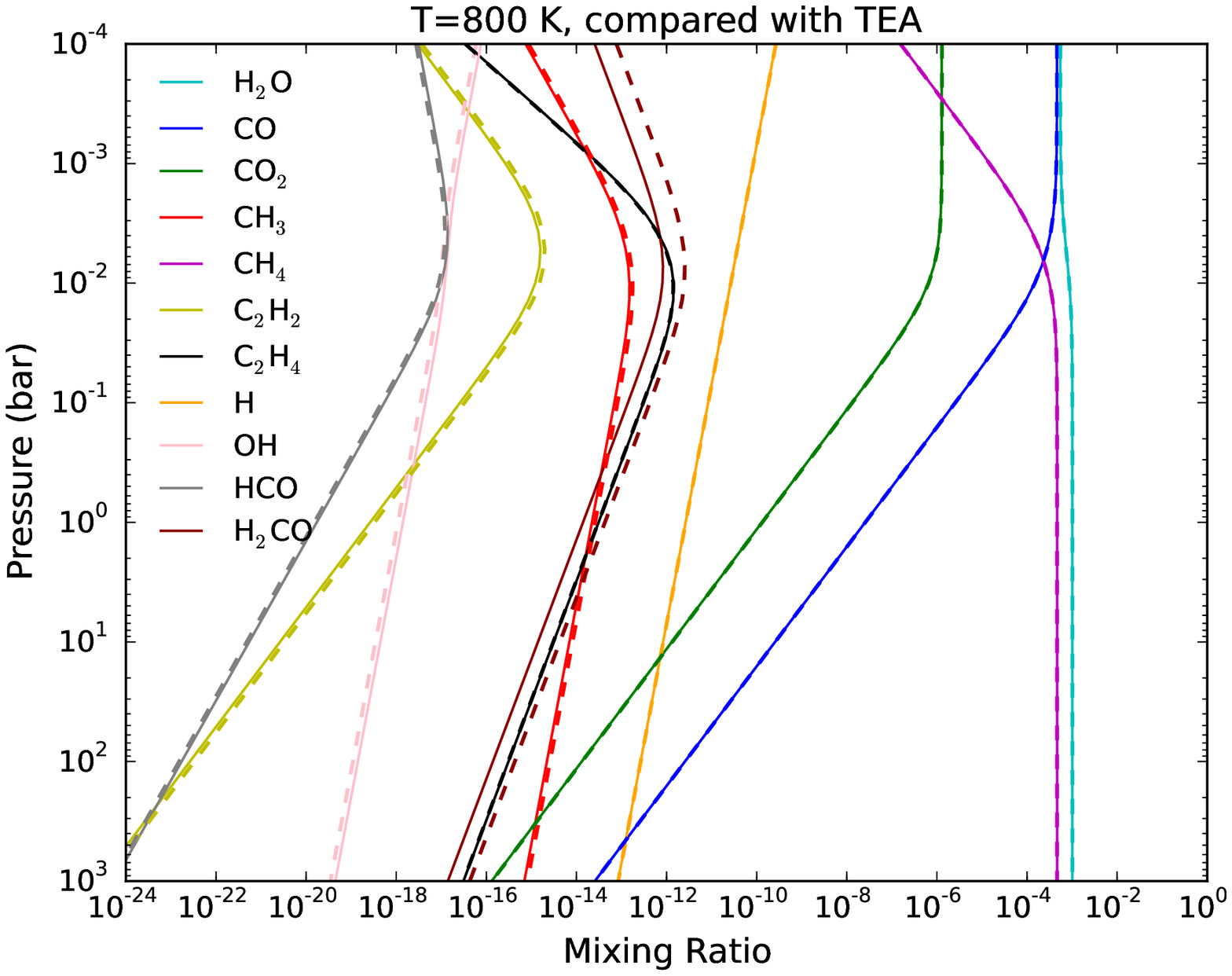}
\includegraphics[width=\columnwidth]{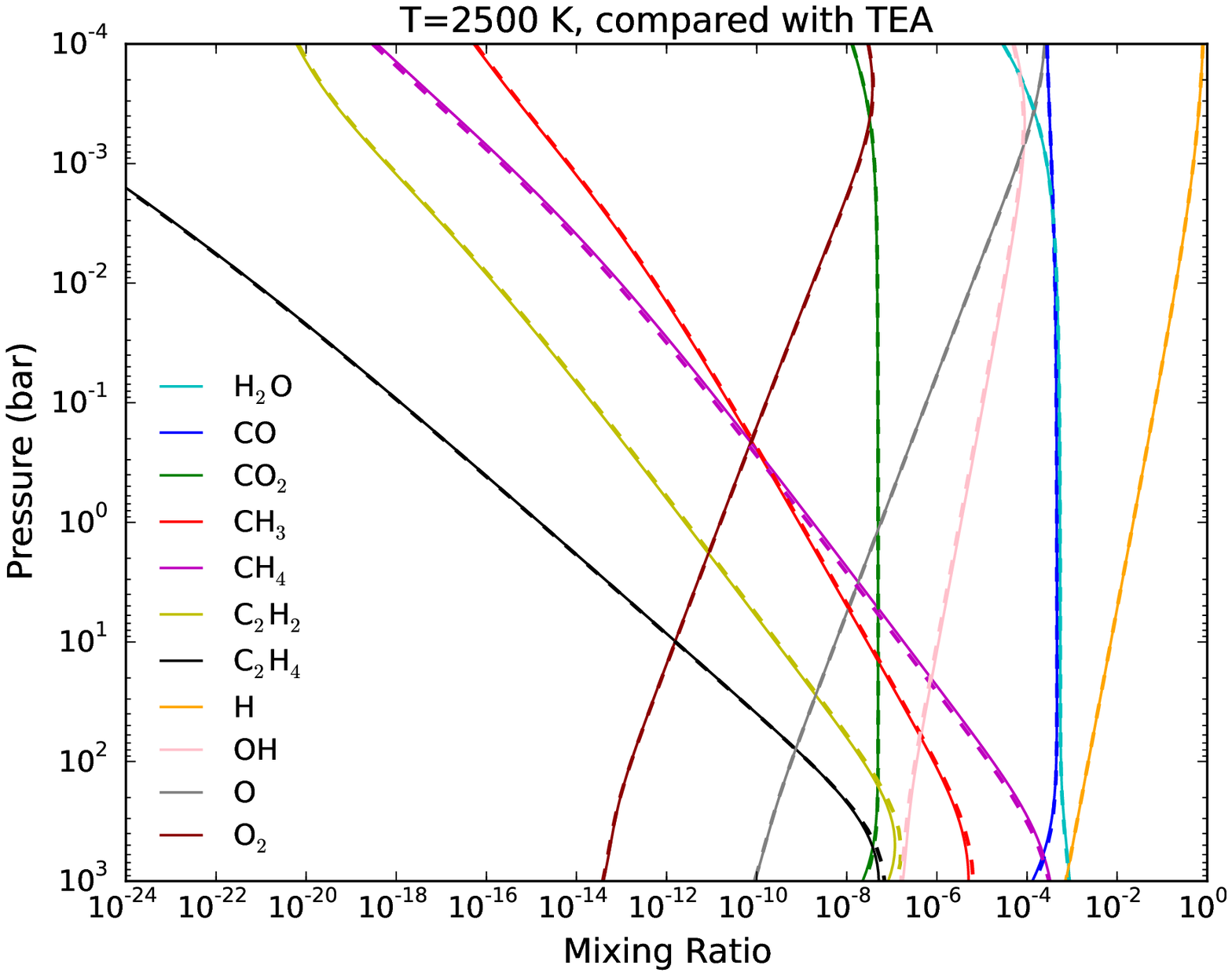}
\end{center}
\caption{Calculations of equilibrium chemistry at $T=800$ K (top panel) and 2500 K (bottom panel).  The solid and dashed curves represent calculations from \texttt{TEA} and \texttt{VULCAN}, respectively.}
\label{fig:tea}
\end{figure}

We perform a first validation of \texttt{VULCAN} by comparing its equilibrium-chemistry results---steady-state solutions of (\ref{eq:master}) without the diffusion term---to that from the \texttt{TEA} code, which we show in Figure \ref{fig:tea}.  Generally, the agreement is excellent except for \ce{H2CO} at $T=800$ K. We determined that the difference in \ce{H2CO} at 800 K between \texttt{VULCAN} and \texttt{TEA} is due to the different choices for the thermodynamic data: \texttt{VULCAN} uses the NASA polynomials, while \texttt{TEA} uses the JANAF tables\footnote{\texttt{http://kinetics.nist.gov/janaf/}}.  Table \ref{tab:nasa_janaf} illustrates the differences in the dimensionless standard Gibbs free energy of formation (normalized by the universal gas constant and temperature) between the two databases.  Since the standard Gibbs free energy provided by the NASA polynomials have a different reference level (where the zero energy is defined) than the standard Gibbs free energy of formation, we have to convert the former to the latter.  This is done by considering the standard Gibbs free energy change for the reactions forming the substance from the elements in their most stable form. For example, \ce{2C(graphite) + O2} $\rightarrow$ \ce{2 CO} is used to calculate the standard Gibbs free energy change of formation for \ce{CO}. For the purpose of discussion, more consistent data for \ce{CH4} is also listed. The differences between JANAF and the NASA polynomials are quite significant for \ce{H2CO} and \ce{C2H}, about 10--20 $\%$ for \ce{H2CO} and about 20--25 $\%$ for \ce{C2H}, while those for \ce{CH4} agree to within 2$\%$.  This discrepancy in \ce{H2CO} directly leads to the different equilibrium abundances in Figure \ref{fig:tea}. Work to update the thermodynamic databases, especially for these two species, is required to reduce the discrepancies.   

It is worth noting that the integration time required to attain a numerical steady state is highly sensitive to the temperature and pressure, ranging from $\sim 10^4$ s to $\sim 10^{20}$ s for 2500 K and 800 K, respectively.  An efficient, stable solver that is capable of integrating for such long times is desirable.  Nevertheless, reproducing chemical equilibrium is a necessary but insufficient first step.  It is useful for weeding out errors associated with implementation of the code, but does not by itself validate a chemical kinetics code \citep{venot12}.

\begin{table} [!h]
\begin{center}
\caption{Differences of normalized Gibbs free energy of formation: JANAF versus NASA polynomials}
\label{tab:nasa_janaf}
\begin{tabular}{c|cc|cc|cc}
\hline
\hline
T (K) & \multicolumn{6}{|c}{$\Delta_f G^0 / R T$} \\
 & \multicolumn{2}{|c}{\ce{C2H}} & \multicolumn{2}{c}{\ce{H2CO}} & \multicolumn{2}{c}{\ce{CH4}} \\
& JANAF & \multicolumn{1}{c}{NASA} & \multicolumn{1}{c}{JANAF} & \multicolumn{1}{c}{NASA} & \multicolumn{1}{c}{JANAF} & \multicolumn{1}{c}{NASA} \\ 
\hline
500 & -25.239 & -23.456 & 98.975 & 119.675 & -7.876 & -7.826 \\
700 & -16.954 & -15.674 & 66.179 & 80.682 & -2.172 & -2.145 \\
900 & -12.256 & -11.252 & 47.996 & 59.046 & 1.151 & 1.162 \\
1100 & -9.223 & -8.399 & 36.459 & 45.292 & 3.332 & 3.321 \\
1300 & -7.102 & -6.400 & 28.498 & 35.790 & 4.869 & 4.838 \\
1500 & -5.536 & -4.924 & 22.679 & 28.834 & 6.007 & 5.952 \\
1700 & -4.332 & -3.788 & 18.245 & 23.524 & 6.881 & 6.801 \\
1900 & -3.377 & -2.888 & 14.757 & 19.338 & 7.572 & 7.463 \\
2100 & -2.601 & -2.155 & 11.943 & 15.955 & 8.131 & 7.991 \\
2300 & -1.957 & -1.548 & 9.627 & 13.166 & 8.592 & 8.419 \\
\hline
\end{tabular}
\end{center}
\end{table}

\subsection{Comparison with the Disequilibrium Calculations of Rimmer \& Helling}
\label{subsect:rimmer}

\begin{figure*}[!h]
\begin{center}
\includegraphics[width=\columnwidth]{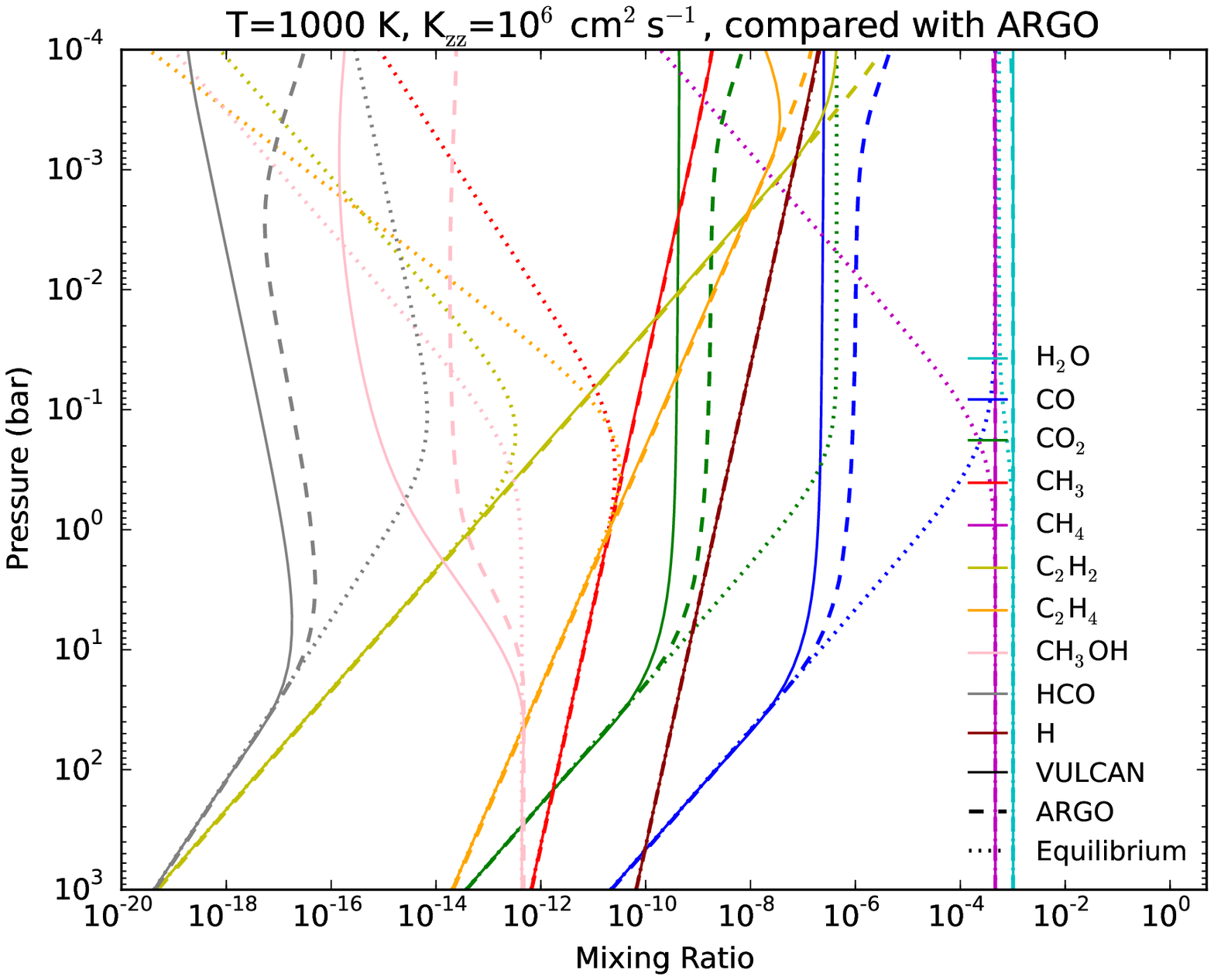}
\includegraphics[width=\columnwidth]{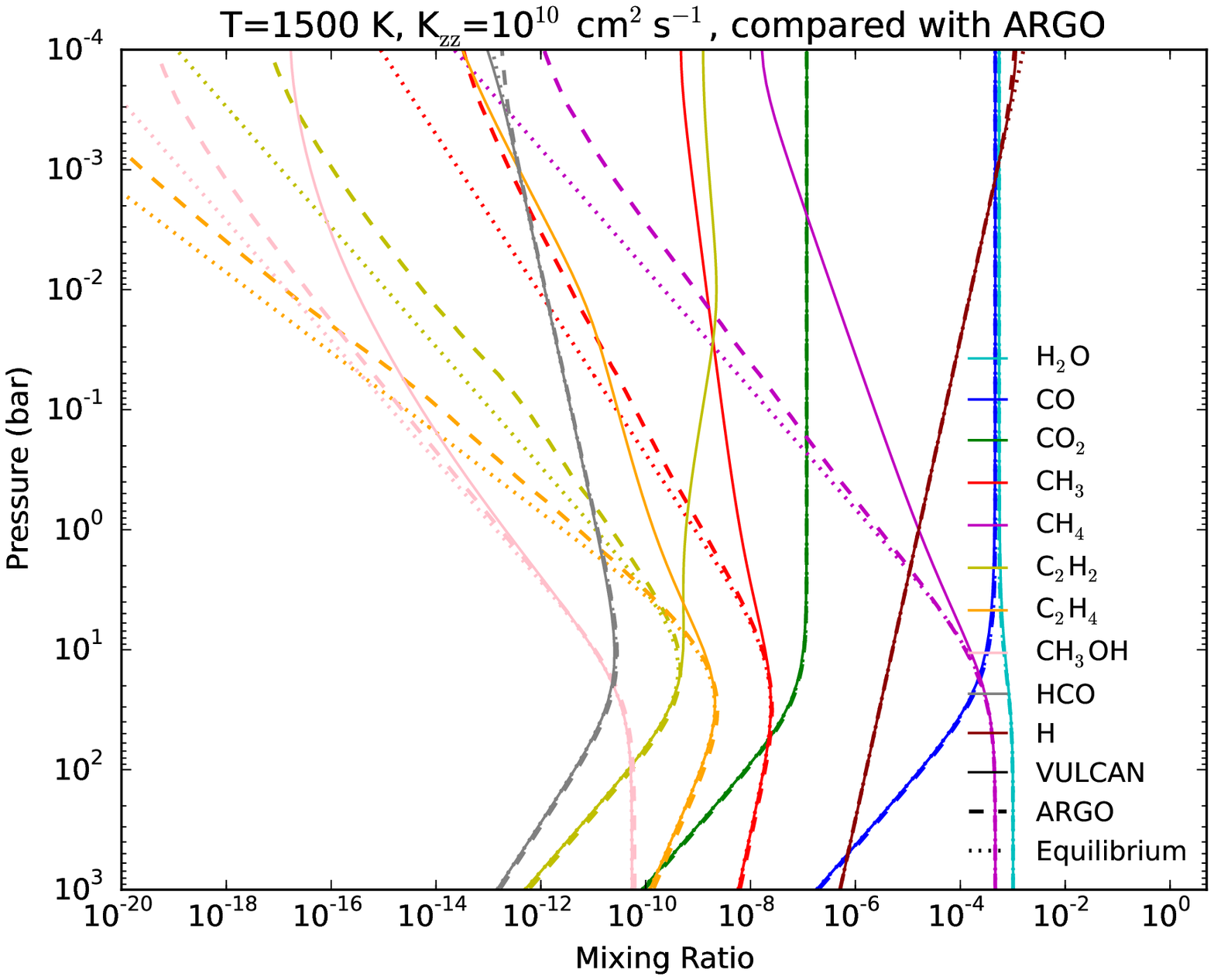}
\includegraphics[width=\columnwidth]{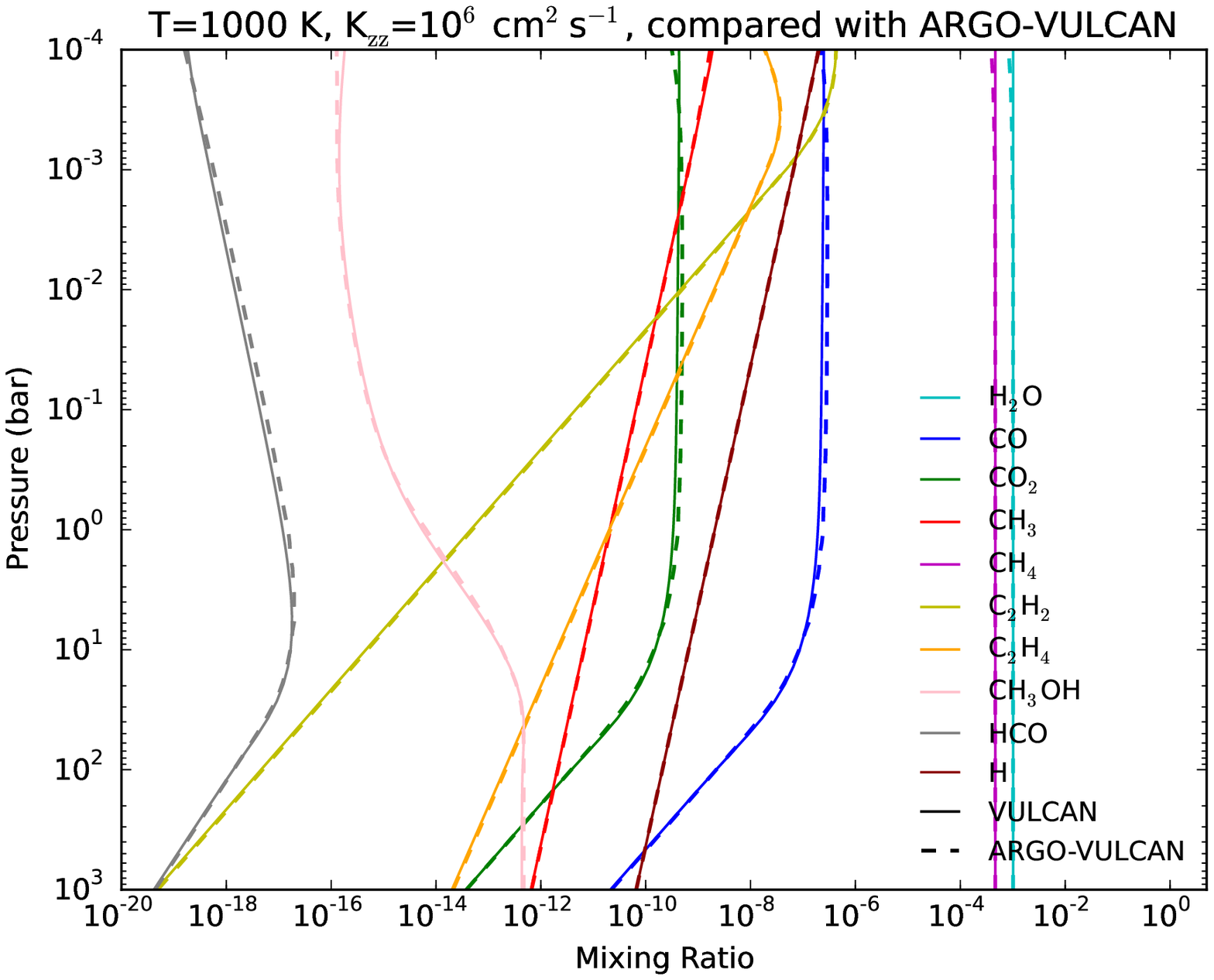}
\includegraphics[width=\columnwidth]{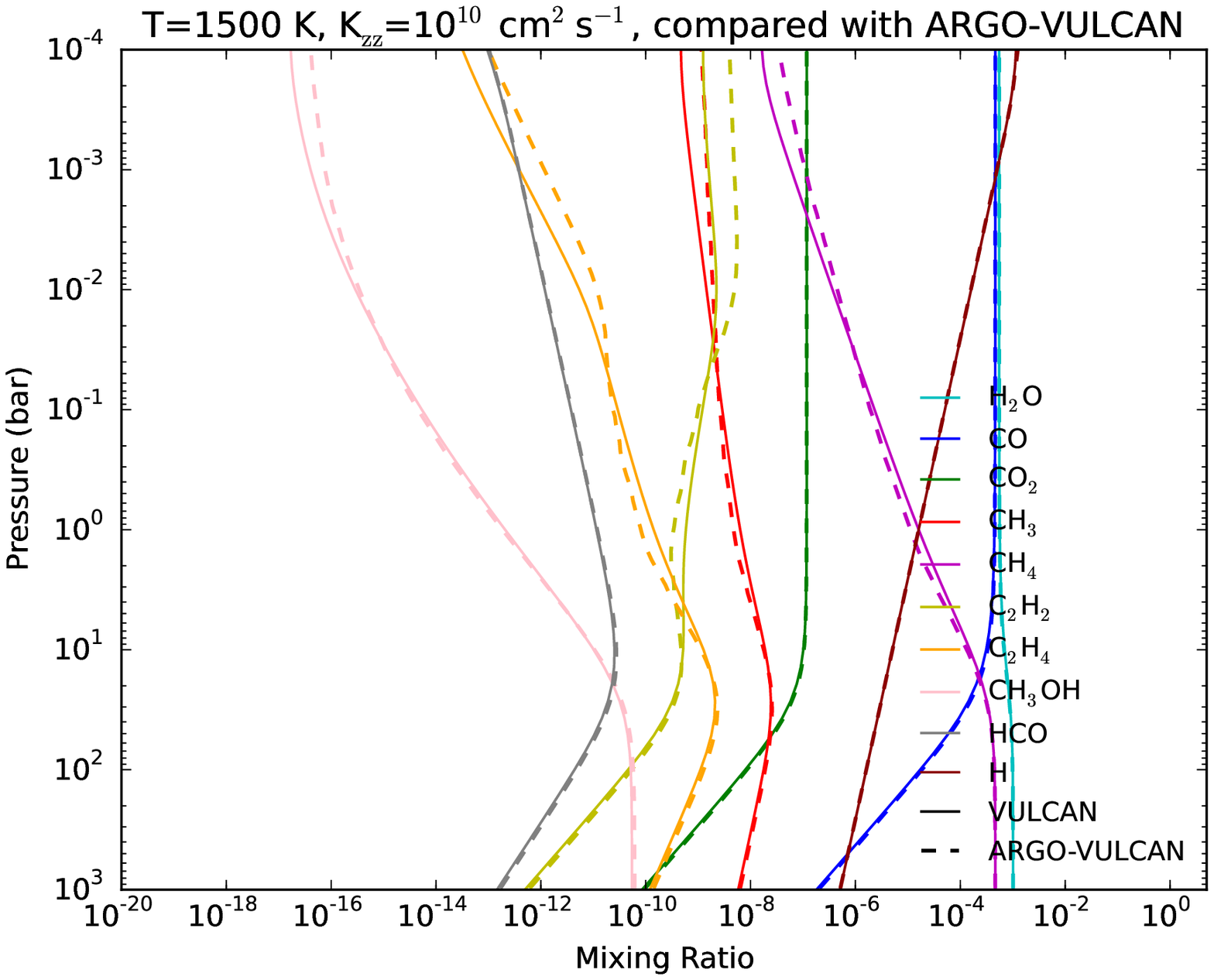}
\includegraphics[width=\columnwidth]{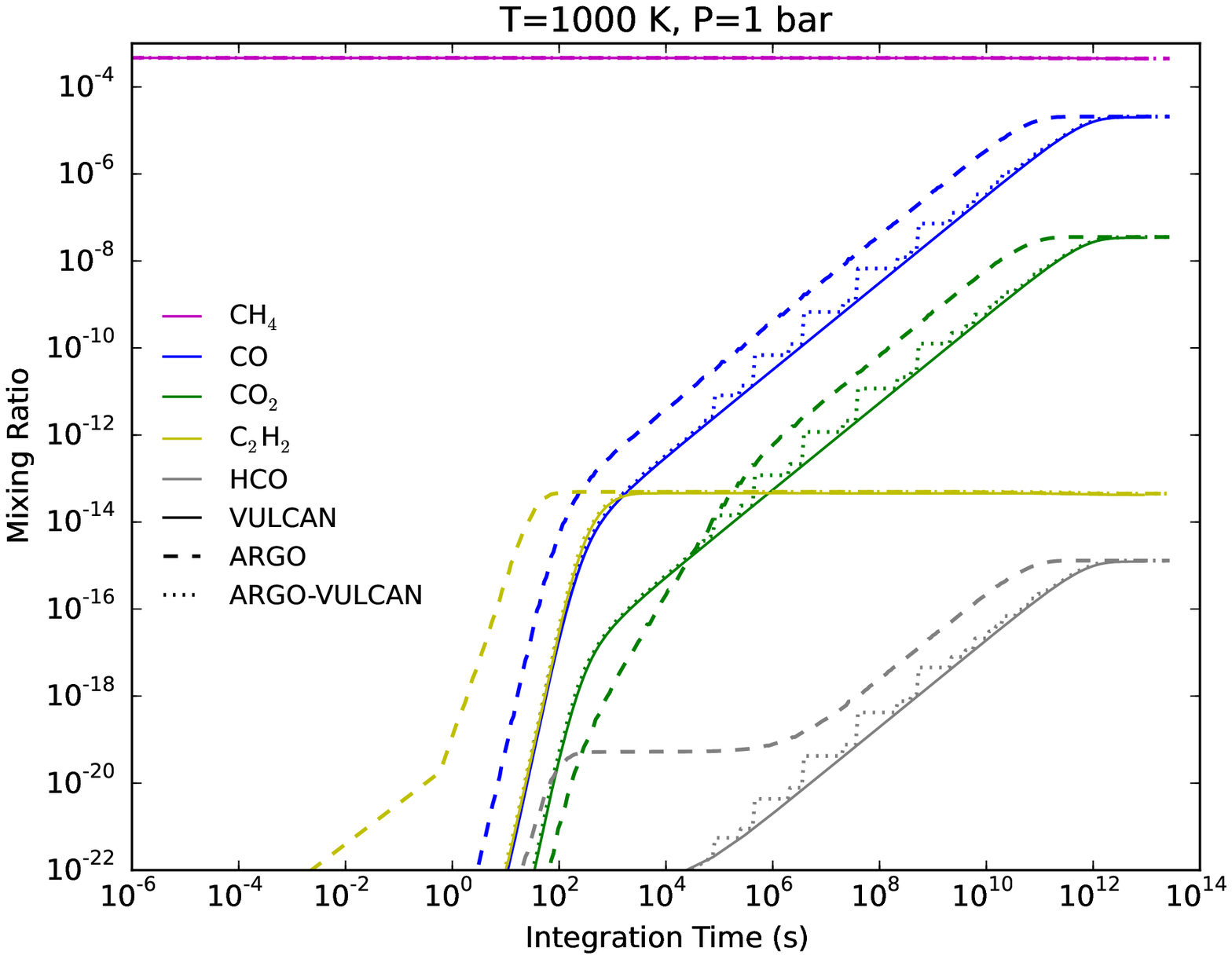}
\includegraphics[width=\columnwidth]{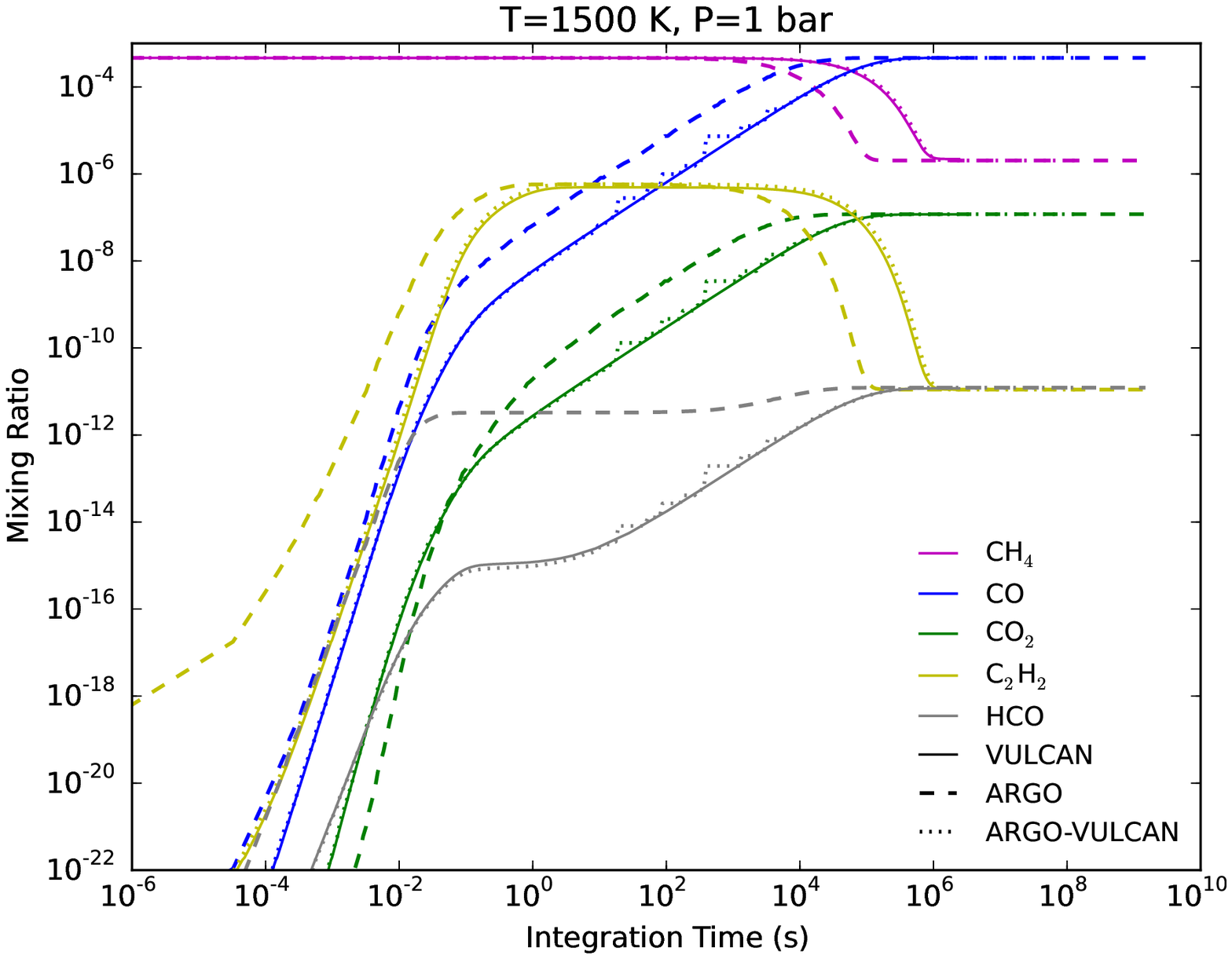}
\end{center}
\caption{Comparing \texttt{VULCAN} and \texttt{ARGO} calculations of disequilibrium chemistry at $T=1000$ K (left column) and 1500 K (right column).  The constant eddy diffusion coefficient is taken to be $K_{\rm zz}=10^6$ cm$^2$ s$^{-1}$ and $10^{10}$ cm$^2$ s$^{-1}$, respectively.  The top panels compare calculations of chemical equilibrium from \texttt{VULCAN} and \texttt{ARGO}.  The middle panels additionally show chemical-equilibrium calculations performed using \texttt{ARGO}, but using the chemical network of \texttt{VULCAN}, which we label \texttt{ARGO-VULCAN}.  The bottom panels show the evolution of mixing ratios at 1 bar, using all three models but without diffusion ($K_{\rm zz}=0$), with initial conditions determined using \ce{H2}, \ce{CH4} and \ce{H2O} (see text).}
\label{fig:argo}
\end{figure*}

A necessary second step for validating a chemical kinetics code is to compare its output with that of other chemical kinetics codes.  As already explained by \cite{venot12}, different chemical networks with backward reactions reversed from the same equilibrium constants will evolve toward the same equilibrium state, but they may take somewhat different paths en route.  These differences in the path taken will lead to discrepant results when chemical disequilibrium, due to atmospheric mixing, is included in the system.  Effectively, each setup has a different overall chemical timescale.

We compare the output from \texttt{VULCAN} against that from \texttt{ARGO} written by \cite{rimmer16}.  \texttt{ARGO} takes a different approach.  Instead of solving a set of 1D mass continuity equations, it uses a zero-dimensional (0D) Lagrangian ``box" approach and tiles these boxes to construct a 1D model atmosphere.  Each box takes a set of initial conditions and computes the output, which is then used as the initial condition for the next box.  \texttt{ARGO} utilizes a large network of about 1100 thermochemical reactions\footnote{If we include their reactions for photochemistry and ion chemistry, then the number of reactions is about 3000.}. 

The top panels of Figure \ref{fig:argo} shows calculations from \texttt{VULCAN} and \texttt{ARGO} at 1000 K and 1500 K.  Generally, the two sets of models agree well at 1000 K, but \texttt{ARGO} produces results much closer to chemical equilibrium at 1500K, even with stronger diffusion. Since \texttt{VULCAN} and \texttt{ARGO} have adopted different chemical networks and different numerical approaches, we run a separate set of models to track down the source of the discrepancies: we run \texttt{ARGO}, but using the reduced chemical network of \texttt{VULCAN} (labeled \texttt{ARGO-VULCAN} in the middle panels of Figure \ref{fig:argo}), which shows that the discrepancies are minor and may be entirely attributed to the differences in the numerical approaches.  Hence, we conclude that the discrepancies between the outputs of \texttt{VULCAN} and \texttt{ARGO} are mainly due to the different chemical networks employed, with the implication that the 0D box model in \texttt{ARGO} can mimic the 1D diffusion process.

The bottom panels of Figure \ref{fig:argo} show the chemical evolution of 0D calculations (without diffusion) from \texttt{VULCAN}, \texttt{ARGO} and \texttt{ARGO-VULCAN} (basically, \texttt{ARGO-VULCAN} and \texttt{VULCAN} only differed in their numerical solvers).  It demonstrates that the same chemical-equilibrium abundances are obtained, albeit over different integration times. Overall, \texttt{ARGO} exhibits shorter chemical timescales than \texttt{VULCAN} and \texttt{ARGO-VULCAN} by an order of magnitude. It confirms that the chemical network of \texttt{ARGO} produces results closer to chemical equilibrium, as discussed above. It reinforces the point that, for benchmarking chemical kinetics calculations with diffusion, reproducing chemical equilibrium is a necessary but insufficient condition.

In \S\ref{sect:rates}, we already discussed the control steps of the \ce{CH4} $\leftrightarrow$ \ce{CO} interconversion scheme.  For reaction (\ref{R1}), \texttt{ARGO} makes a different choice and uses \cite{oser92}, which is valid at only 300--480 K and may not be suitable for higher temperatures.  For reaction (\ref{R2}), \texttt{ARGO} uses a rate coefficient based on their own calculations (their reaction 646).  Overall, the more efficient \ce{CH4} $\leftrightarrow$ \ce{CO} scheme keeps the kinetics results of \texttt{ARGO} closer to chemical equilibrium and less affected by atmospheric diffusion.

\subsection{Comparison with the Disequilibrium Calculations of Moses et al. for HD 189733b and HD 209458b}

\begin{figure*}
\begin{center}
\includegraphics[width=\columnwidth]{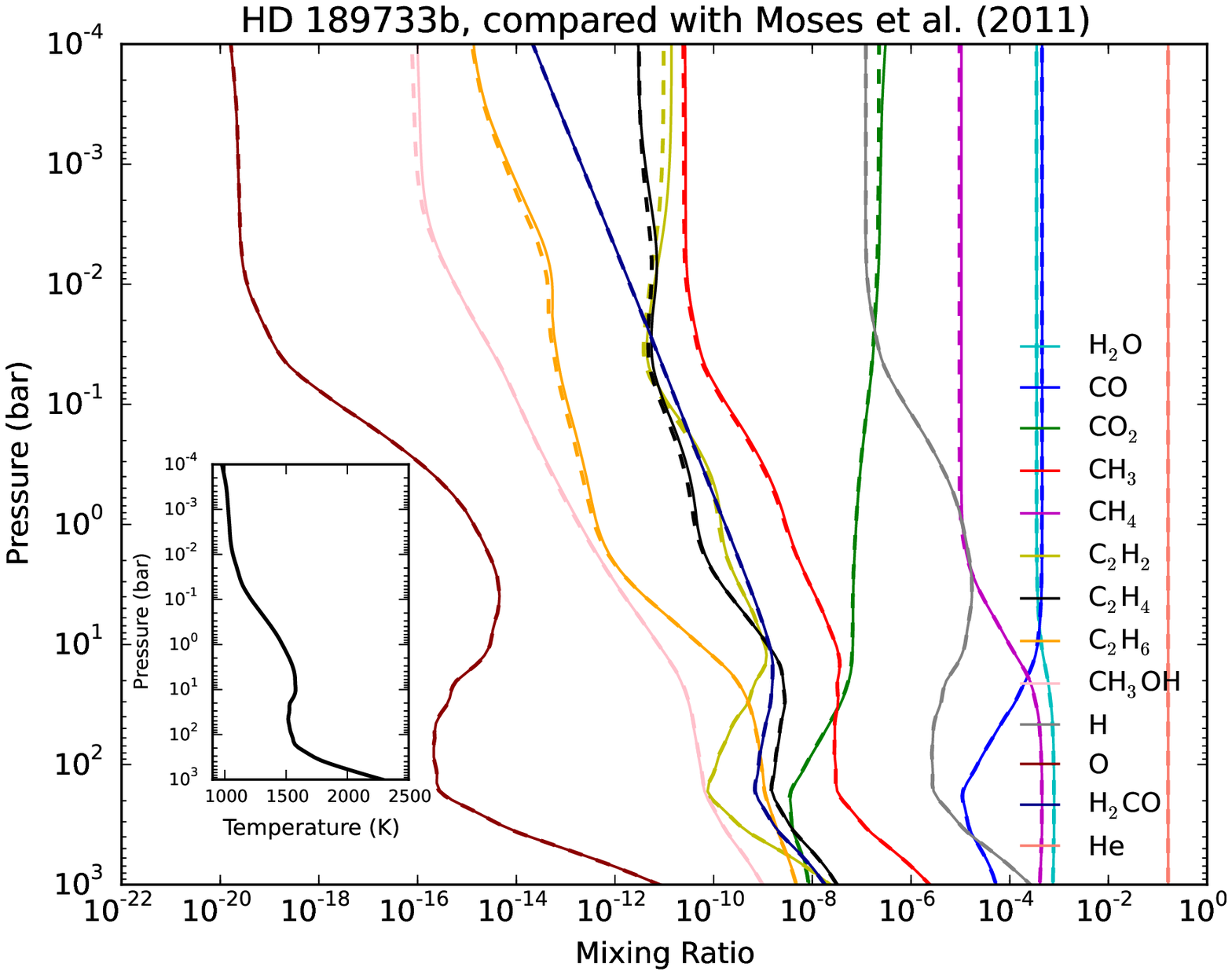}
\includegraphics[width=\columnwidth]{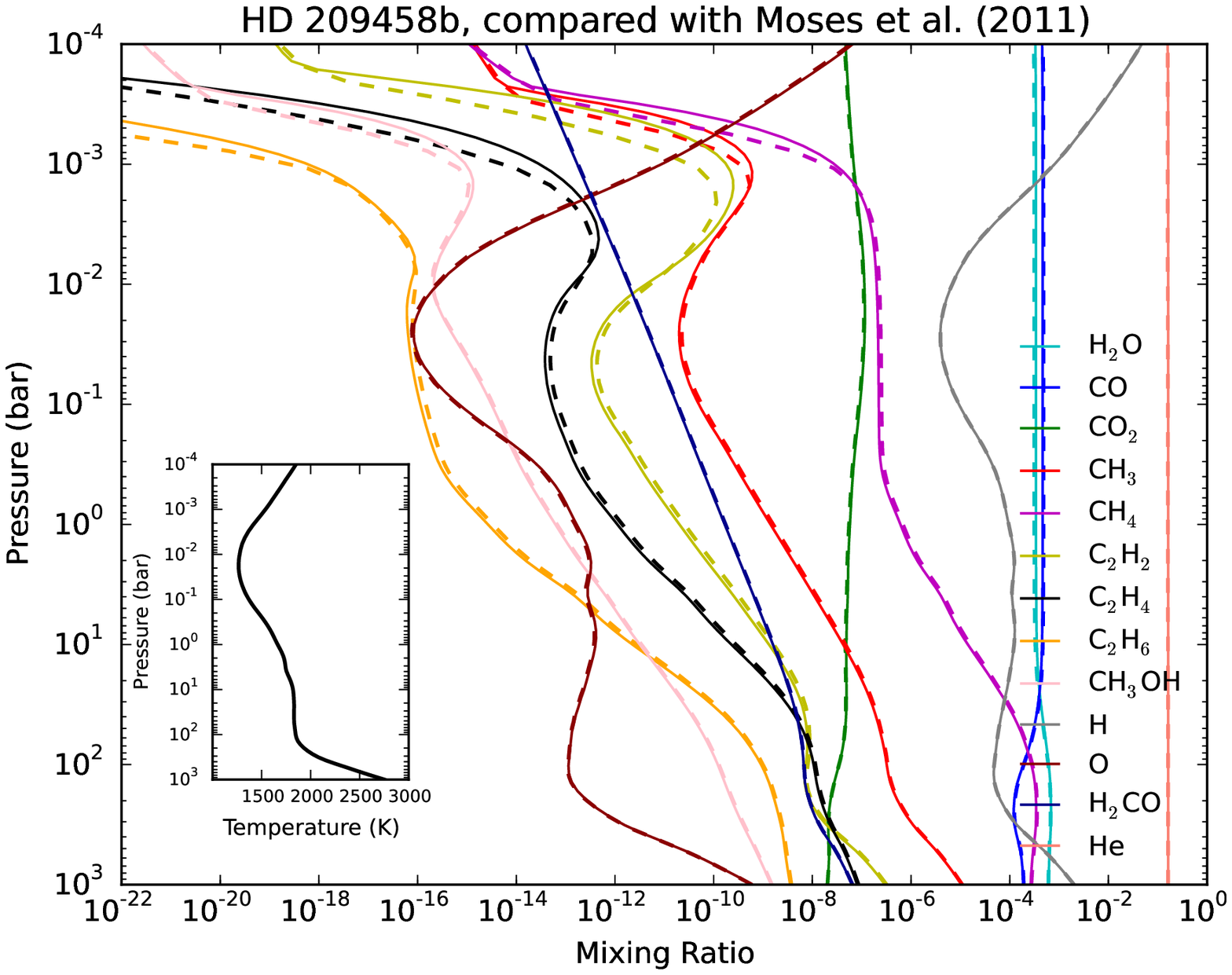}
\end{center}
\caption{Calculations of disequilibrium chemistry for models of the hot Jupiters HD 189733b (left panel) and HD 209458b (right panel), as originally constructed by \cite{moses11}.  We compare the results from \texttt{VULCAN} (solid curves) with those of \cite{moses11} (dashed curves).  For completeness, we show the dayside-average temperature-pressure profiles used (inserts).}
\label{fig:moses}
\end{figure*}

Our final benchmarking exercise is to produce calculations of disequilibrium chemistry with more realistic temperature-pressure profiles.  We pick the published models of \cite{moses11}, who considered the hot Jupiters HD 189733b and HD 209458b.  Note that, like \cite{moses11}, our calculations are not self-consistent in the sense that the temperature-pressure profiles are held fixed at their input values.  In reality, any change in the molecular abundances will lead to changes in the opacities, which will in turn alter the temperature-pressure profile \citep{drummond16}.  Fully self-consistent calculations take this interplay into account.

Figure \ref{fig:moses} demonstrates that our \texttt{VULCAN} results closely reproduce those of \cite{moses11}, despite our use of a reduced chemical network.  \cite{moses11} employ a network with nearly 1600 reactions (including nitrogen species), compared to about 300 for our current version of \texttt{VULCAN}.  Since we have omitted photochemistry in this current study, we have compared our results to the ``no photon" models  with constant $K_{\rm zz}=10^9$ cm$^2$ s$^{-1}$ of HD 189733b and HD 209458b by \cite{moses11}.  Discrepancies, if any, may thus be entirely attributed to differences in our chemical networks and numerical implementation.  For HD209458b, the abundances of most species match well, except that \ce{CH4} is quenched slightly differently and the discrepancy propagates to other species that are influenced by the abundance of \ce{CH4}, including \ce{CH3}, \ce{C2H2}, \ce{C2H4}, \ce{C2H6} and \ce{CH3OH}.  Specifically, CH$_3$OH (methanol) is a key intermediate molecule that determines the conversion rate of \ce{CH4} to \ce{CO}.  By contrast, species such as \ce{CO}, \ce{CO2} and \ce{H2O} are largely unaffected by this difference in quenching behavior.  The fact that \texttt{VULCAN} closely matches \cite{moses11} for the C-H-O species also informs us that nitrogen couples weakly with the other molecules in the hot Jupiter regime (but see \citealt{venot12} for the uncertainty of reactions for nitrogen compounds).

\cite{rimmer16} have included many more reactions than we have (about 1100 versus about 300), because of their desire to treat lightning chemistry and prebiotic photochemistry, which we have neglected in our current implementation of \texttt{VULCAN}.  Despite this simplification, we have reproduced the HD 189733b and HD 209458b models of \cite{moses11} better than \cite{rimmer16} have (see their Figure 9), because of similarities in our choices of rate coefficients for key reactions compared to \cite{moses11}, which we have already discussed.  It demonstrates that thermochemistry in the hot Jupiter regime may be represented by a reduced set of chemical reactions, which will have important implications for coupling chemical kinetics to three-dimensional general circulation models.

\subsection{Benchmarking Conclusions}

The results of this section lead us to conclude that we have successfully benchmarked \texttt{VULCAN}.  We are able to reproduce chemical equilibrium.  We are also able to reproduce the model atmospheres of HD 189733b and HD 209458b as originally constructed by \cite{moses11}.

\section{Results}
\label{sect:results}

\subsection{Exploring Trends with Analytical Temperature-Pressure Profiles}

\begin{table}[!h]
\begin{center}
\caption{Input Parameters for $T$-$P$ profiles in Figures \ref{fig:tp1}--\ref{fig:tp4}}
\label{tab:tp}
\begin{tabular}{|c|c|c|c|c|c|}
\hline
Parameter & T$_{\rm irr}$ & $\kappa_{\rm L}$ & $\kappa_{\rm S}$ & $\beta_{\rm S}$ (A$_B$) & $\beta_{\rm L}$ \\
\hline
Units & K & cm$^2$ g$^{-1}$ & cm$^2$ g$^{-1}$ & --- & --- \\
\hline
Figure \ref{fig:tp1}-A & 1200 & 0.01& 0.001& 1& 1\\
Figure \ref{fig:tp1}-B & 2200 & 0.01& 0.001& 1& 1\\
Figure \ref{fig:tp2}-A & 1500 & 0.01& 0.001& 1& 1\\
Figure \ref{fig:tp2}-B & 1500 & 0.01& 0.1& 1& 1\\
Figure \ref{fig:tp3}-A & 1500 & 0.01& 0.001& 1 (0)& 1\\
Figure \ref{fig:tp3}-B & 1500 & 0.01& 0.001& 0.333 (0.5)& 1\\
Figure \ref{fig:tp4}-A & 1500 & 0.01& 0.001& 1& 1\\
Figure \ref{fig:tp4}-B & 1500 & 0.01& 0.001& 1& 0.5\\
\hline
\end{tabular}\\
\end{center}
\end{table}

\begin{figure}
\begin{center}
\includegraphics[width=\columnwidth]{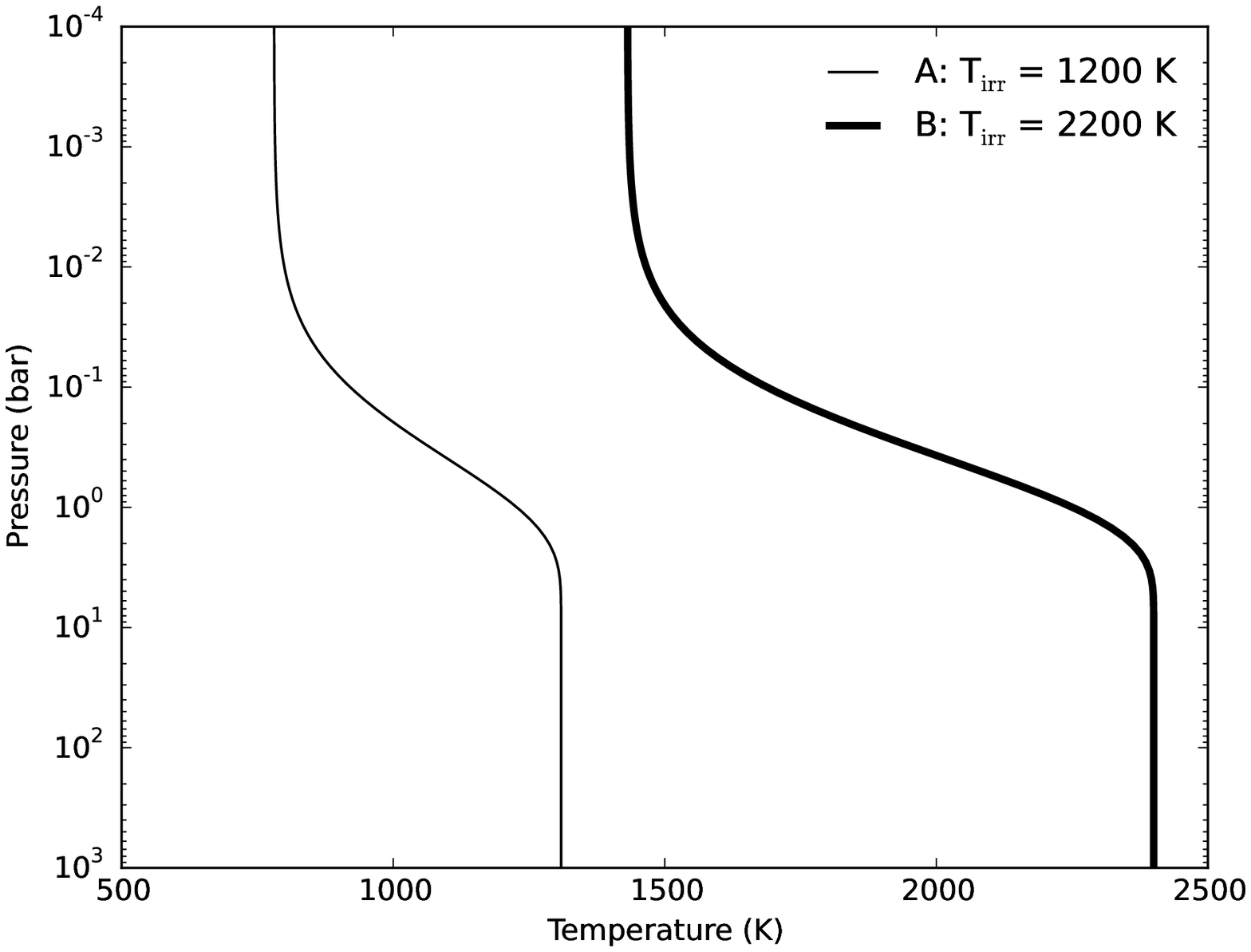}
\includegraphics[width=\columnwidth]{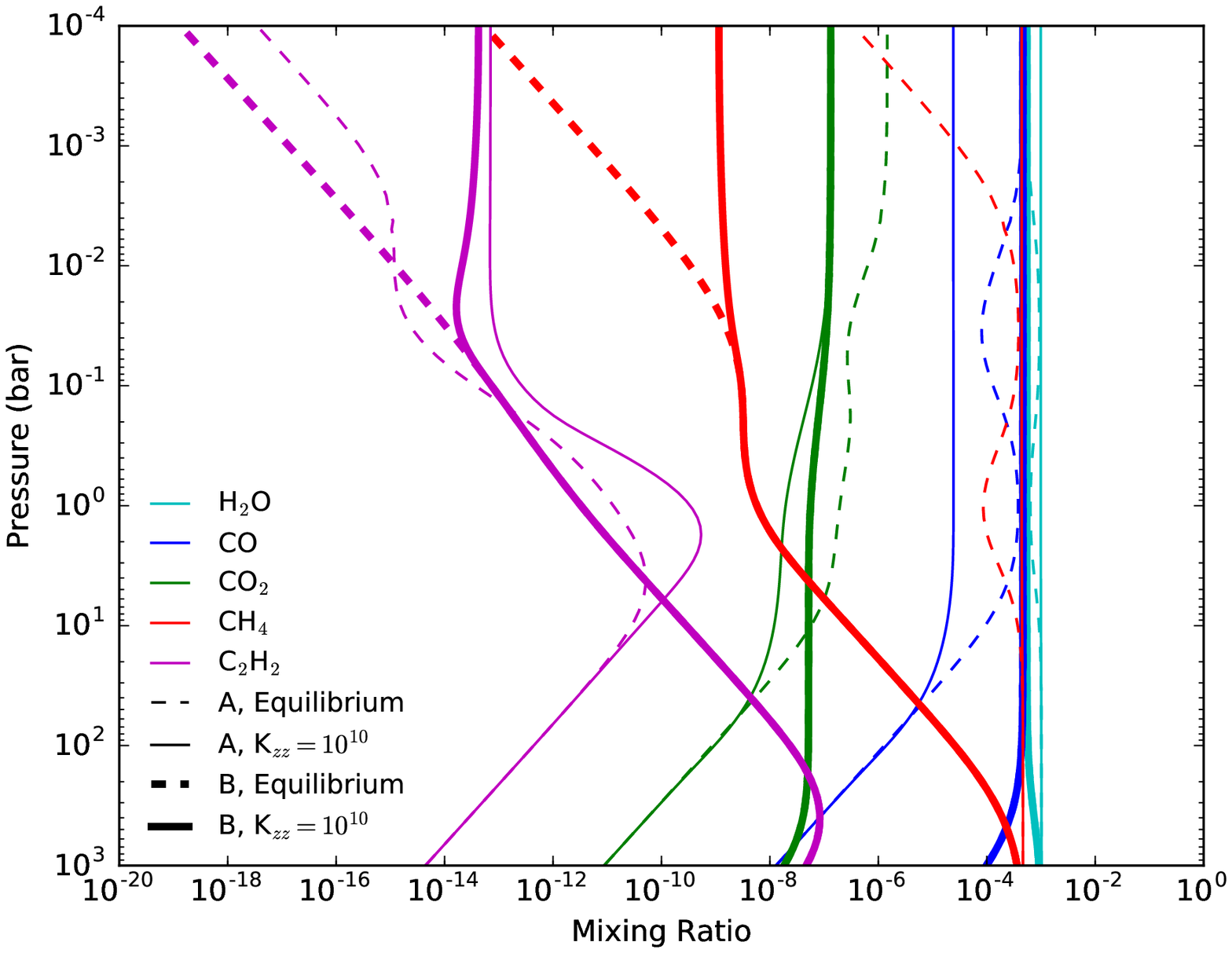}
\end{center}
\caption{Exploring the effects of varying the strength of stellar irradiation.  The top panel shows the temperature-pressure profiles used, while the bottom panel shows the corresponding mixing ratios as computed by \texttt{VULCAN}.  For the bottom panel, the solid curves assume $K_{\rm zz}=10^{10}$ cm$^2$ s$^{-1}$ and the dashed curves assume chemical equilibrium.}
\label{fig:tp1}
\end{figure}

\begin{figure}
\begin{center}
\includegraphics[width=\columnwidth]{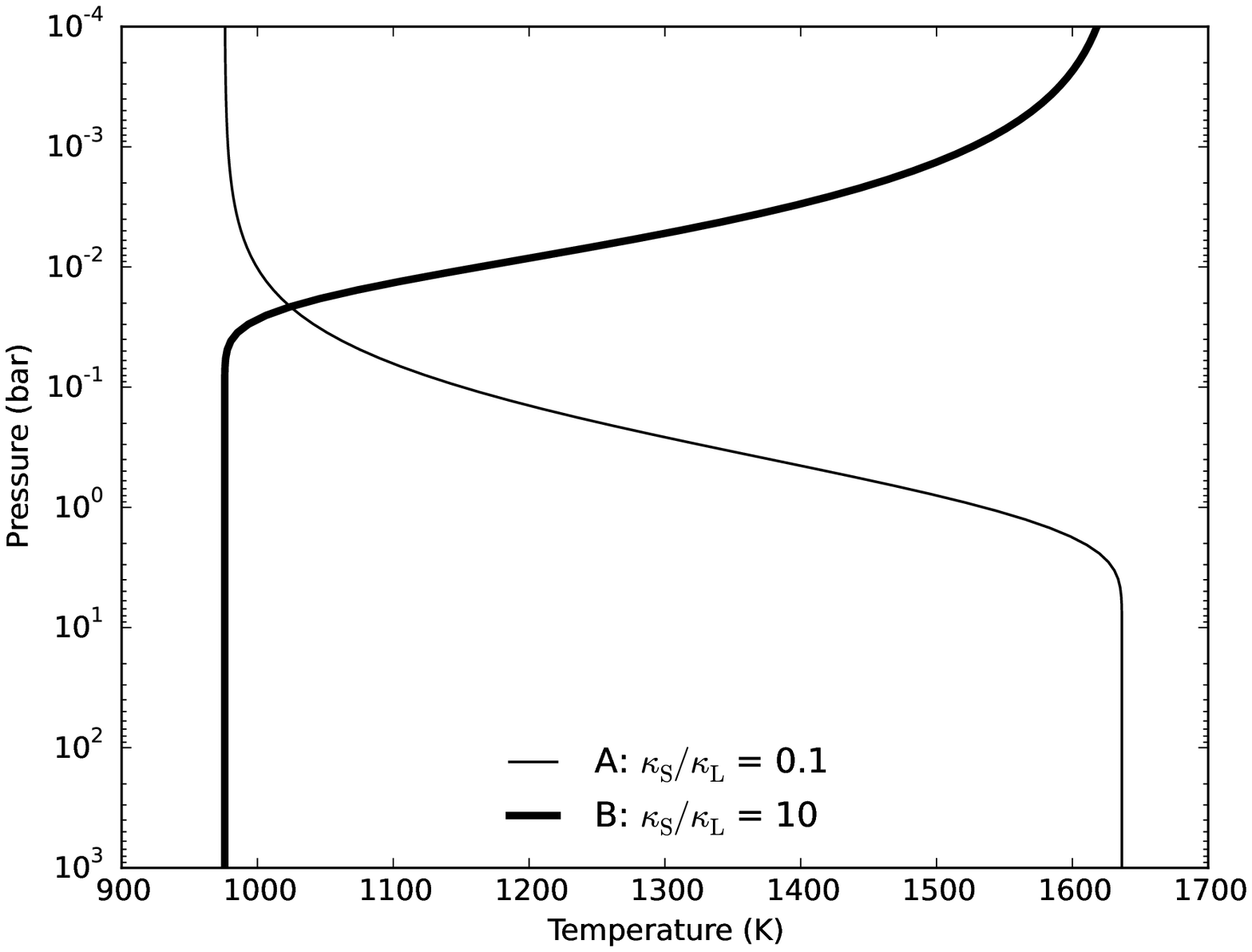}
\includegraphics[width=\columnwidth]{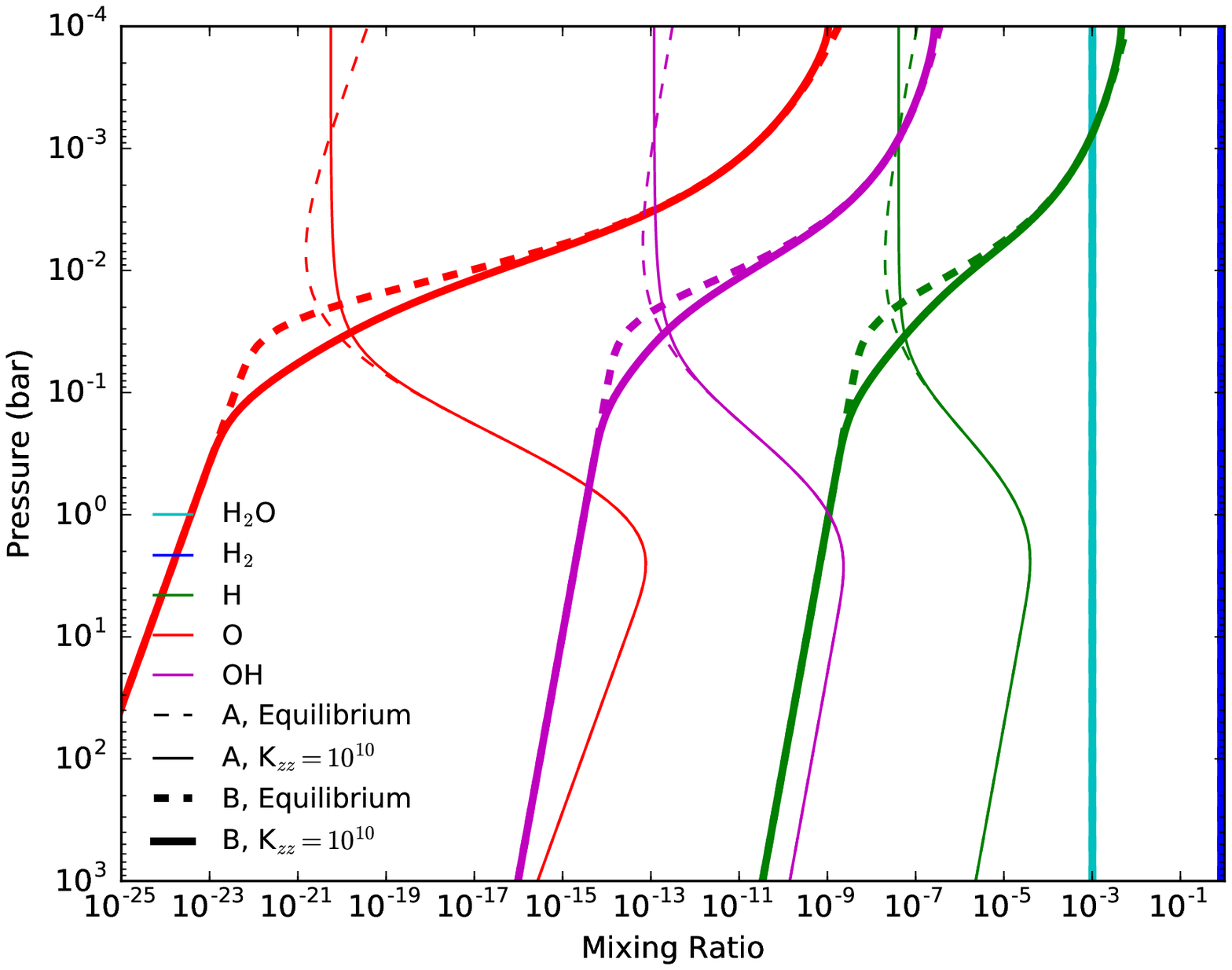}
\includegraphics[width=\columnwidth]{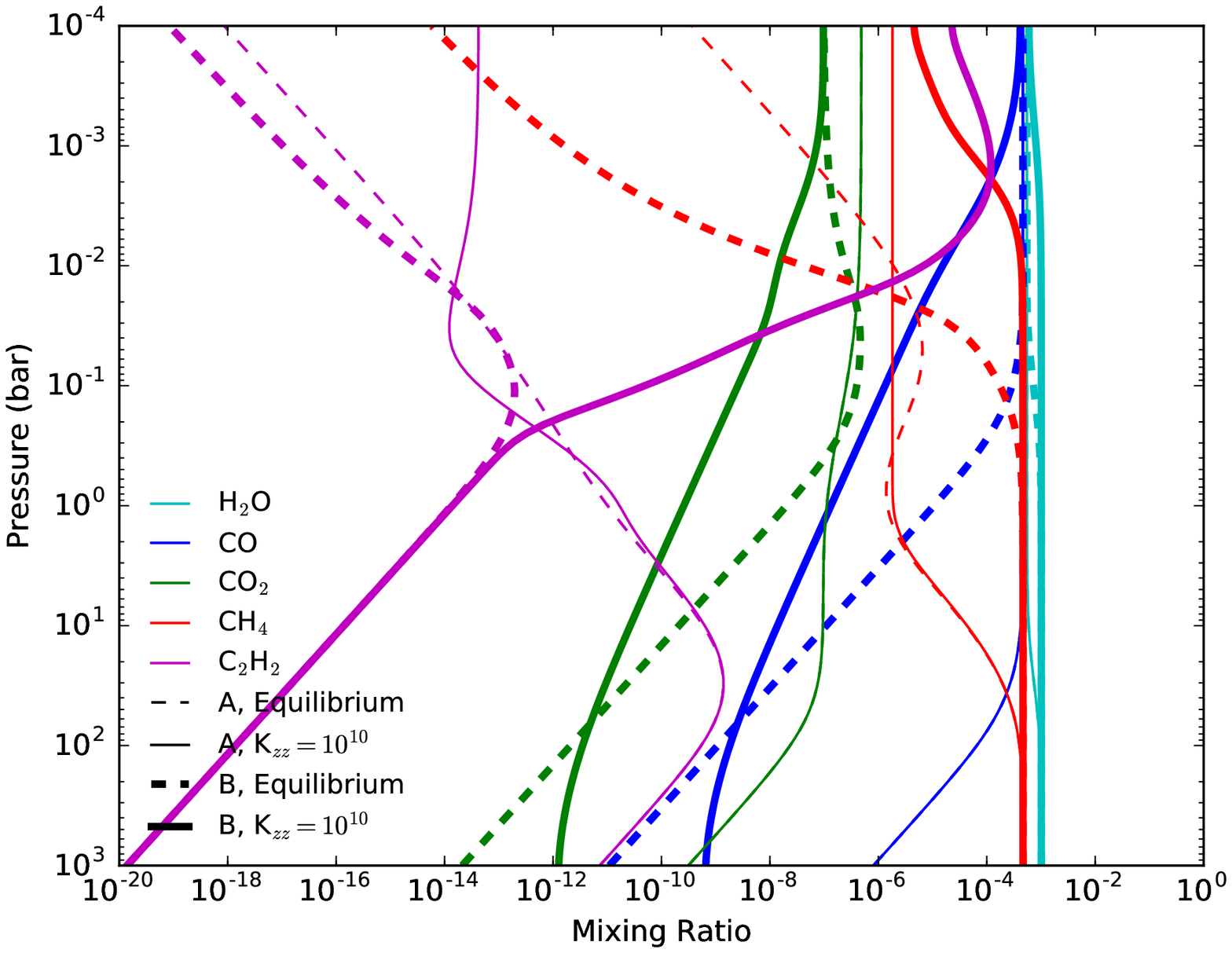}
\end{center}
\caption{Exploring the effects of a temperature inversion.  The top panel shows the temperature-pressure profiles used.  The middle and bottom panels shows the corresponding mixing ratios, as computed by \texttt{VULCAN}, for H-O and C-H-O networks, respectively.  Dashed curves assume chemical equilibrium.  The solid curves assume $K_{\rm zz}=10^{10}$ cm$^2$ s$^{-1}$ for the middle and bottom panels.}
\label{fig:tp2}
\end{figure}

We explore trends associated with model atmospheres computed using analytical temperature-pressure profiles.  The formulae for these profiles are taken from \cite{hml14}, which are based on generalizations of the work of \cite{guillot10} and \cite{hhps12}.  Our choice of parameter values for the irradiation temperature ($T_{\rm irr}$), longwave opacity ($\kappa_{\rm L}$), shortwave opacity ($\kappa_{\rm S}$), longwave scattering parameter ($\beta_{\rm L}$) and shortwave scattering parameter ($\beta_{\rm S}$) are fully listed in Table \ref{tab:tp}.  As already mentioned, these calculations are not self-consistent, because the temperature-pressure profiles are held fixed at their input values.  The advantage of using these analytical temperature-pressure profiles is that they are easily reproducible.

\subsubsection{The Effects of Stellar Irradiation}

Figure \ref{fig:tp1} shows the temperature-pressure profiles when the irradiation temperature takes on values of $T_{\rm irr} = 1200$ and 2200 K, while holding the other parameter values fixed.  The profiles are similar in shape, but their normalizations are shifted wholesale.  The first trend we expect is that the dominant carbon carrier shifts from being \ce{CH4} to CO as the model atmosphere becomes hotter \citep{madhu12,hlt16}.  Furthermore, \ce{CO} and \ce{CH4} are favored at low and high pressures, respectively, because of Le Ch\^{a}telier's principle.  These competing behaviors are evident in the equilibrium-chemistry abundance profile for \ce{CH4} (red dashed curves). The second trend we see is that \ce{CO2} is always subdominant compared to CO and \ce{H2O} for a solar-like metallicity.

\subsubsection{The Effects of a Temperature Inversion}
\label{subsect:inversion}

A much discussed issue in the literature is the absence or presence of temperature inversions in the atmospheres of hot Jupiters.  Intuitively, we expect the presence of a temperature inversion to negate the effects of disequilibrium chemistry produced by atmospheric mixing.  In Figure \ref{fig:tp2}, we vary the ``greenhouse parameter" ($\kappa_{\rm S}/\kappa_{\rm L}$) from 0.1 to 10 and examine its effect on the atomic and molecular abundances.  It is called the greenhouse parameter, because it controls the relative strength of optical/visible to infrared absorbers in the atmosphere.

We build our understanding in a hierarchical fashion.  The middle panel of Figure \ref{fig:tp2} shows the mixing ratios for a network with only hydrogen and oxygen, which has only about a dozen chemical reactions.  Water and molecular hydrogen are dominant species that are unaffected by atmospheric mixing.  For the model atmosphere without a temperature inversion, we see that H, O and OH are quenched at pressures about 0.1 bar.  When a temperature inversion is present, we see that the mixing ratios of H, O and OH eventually attain their chemical-equilibrium values at high altitudes.  Somewhat surprisingly, we were able to identify the reaction,
\begin{equation}
2\mbox{H} + \mbox{M} \rightarrow \mbox{H}_2 + \mbox{M},
\end{equation}
as the key reaction controlling the quench levels of O and OH using the technique outlined in Appendix \ref{append:sensitivity}.  Although this reaction only involves and controls \ce{H}, due to the simplicity of this chemical network, \ce{OH} and \ce{O} are predominantly controlled by the abundance of H.  It is an illustrative example of how the abundances of some species can be controlled by those of others, which has implications for the quenching approximation that is commonly invoked in the literature (and which we will revisit in \S\ref{subsect:quench}).

A similar qualitative behavior is seen for CO, \ce{CO2} and \ce{H2O}, when one examines a C-H-O network.  The mixing ratios of \ce{CH4} and \ce{C2H2} do not attain their chemical-equilibrium values at high altitudes, due to the longer timescale of converting \ce{CH4}.  We will elucidate the reasons for this seemingly strange behavior in \S\ref{subsect:quench}.

\subsubsection{The Effects of Scattering in the Visible (Varying the Bond Albedo)}

\begin{figure}
\begin{center}
\includegraphics[width=\columnwidth]{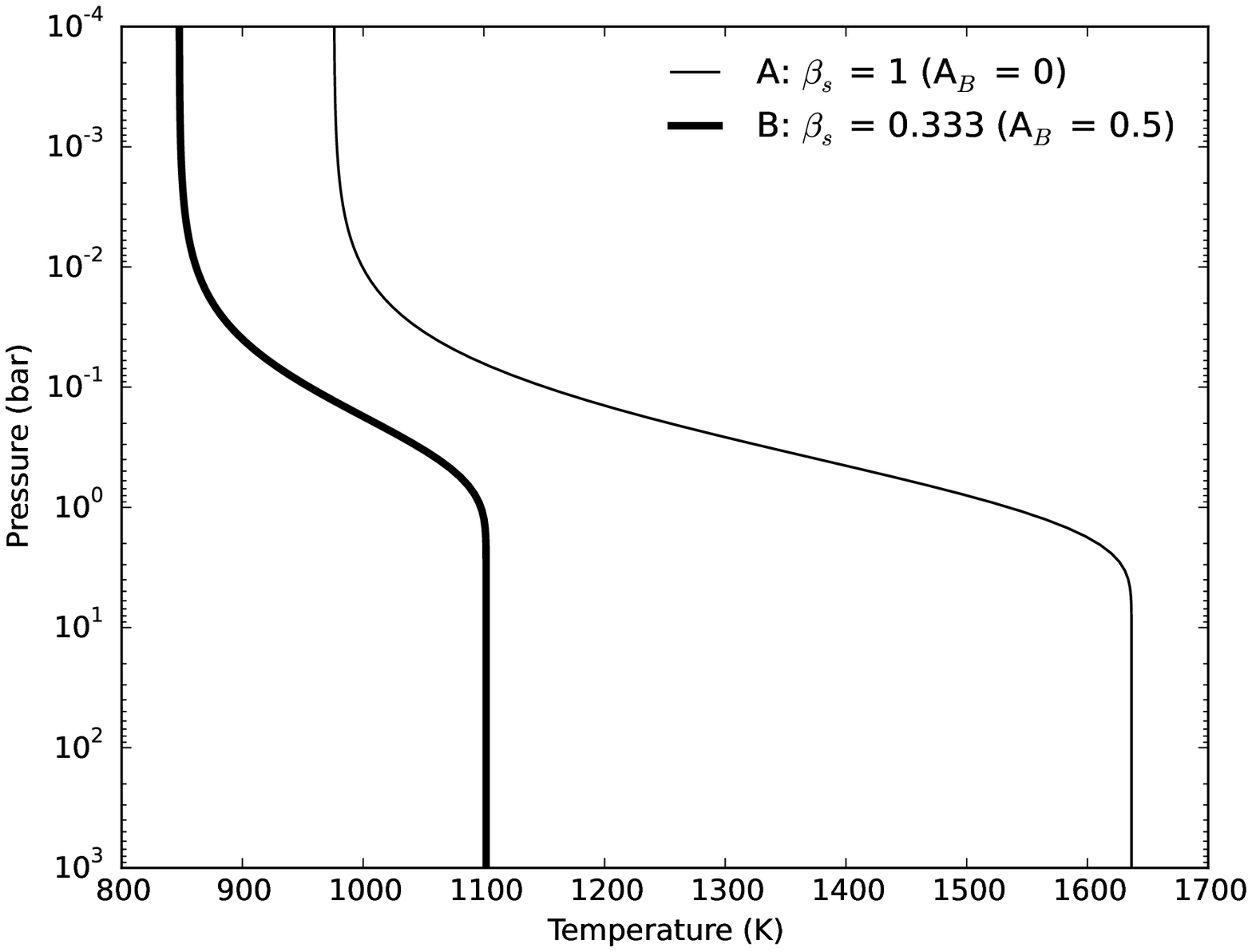}
\includegraphics[width=\columnwidth]{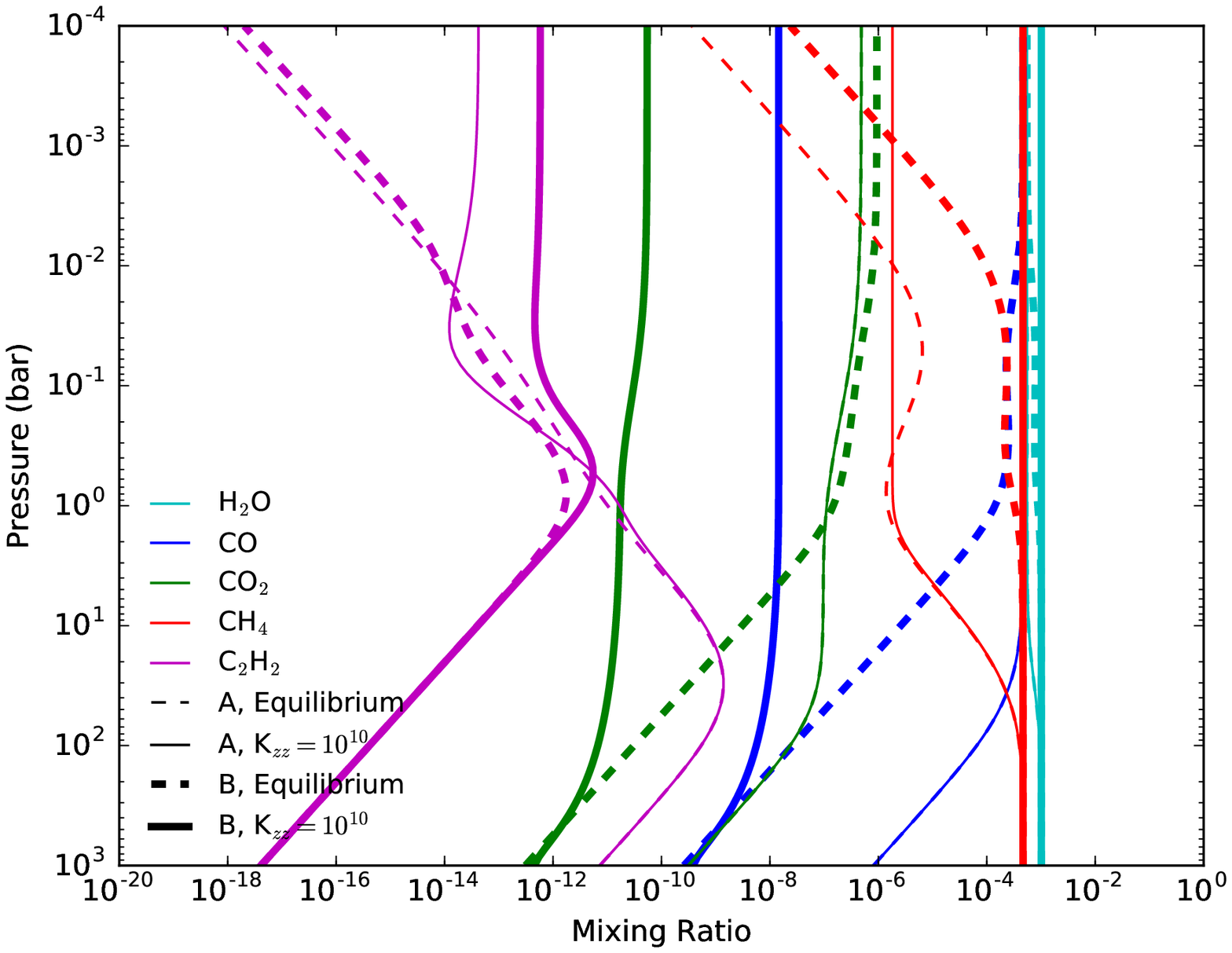}
\end{center}
\caption{Same as Figure \ref{fig:tp1}, but exploring the effects of varying the Bond albedo.}
\label{fig:tp3}
\end{figure}

Since the geometric albedos of exoplanetary atmospheres may be readily measured by detecting their visible/optical secondary eclipses, it is relevant to understand the effects of varying the Bond albedo.  As derived by \cite{hml14}, the Bond albedo ($A_{\rm B}$) and shortwave scattering parameter are related by $A_{\rm B} = (1-\beta_{\rm S})/(1+\beta_{\rm S})$. In Figure \ref{fig:tp3}, we see that increasing the Bond albedo essentially lowers the average temperature and the temperature gradient as well. The effect of varying temperatures is similar to that in Figure \ref{fig:tp1}, except the abundance of \ce{CO} in Figure \ref{fig:tp3}-B is more significantly reduced due to the lower temperatures in the deep levels.


\subsection{The Effects of Scattering in the Infrared (Scattering Greenhouse Effect)}

\begin{figure}
\begin{center}
\includegraphics[width=\columnwidth]{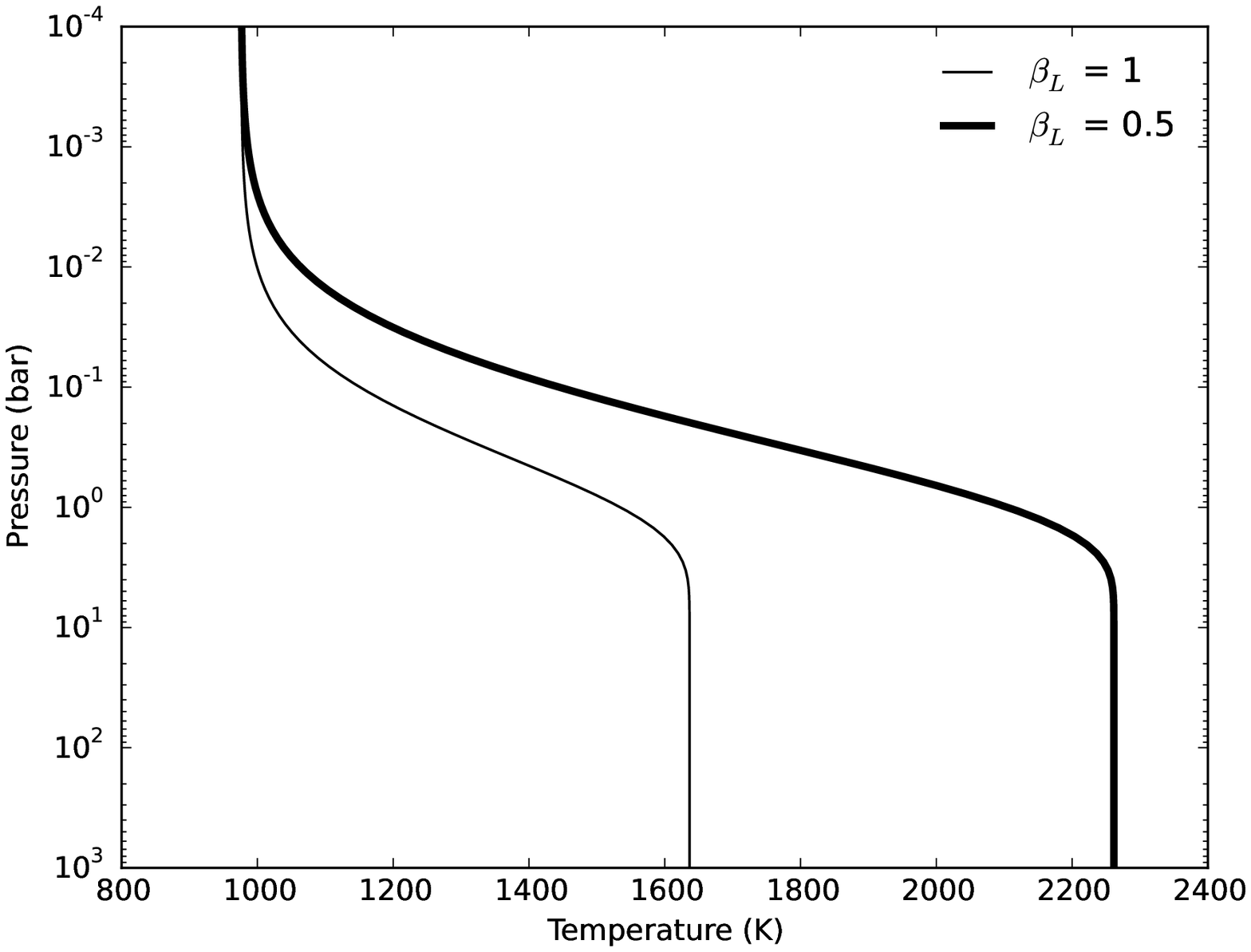}
\includegraphics[width=\columnwidth]{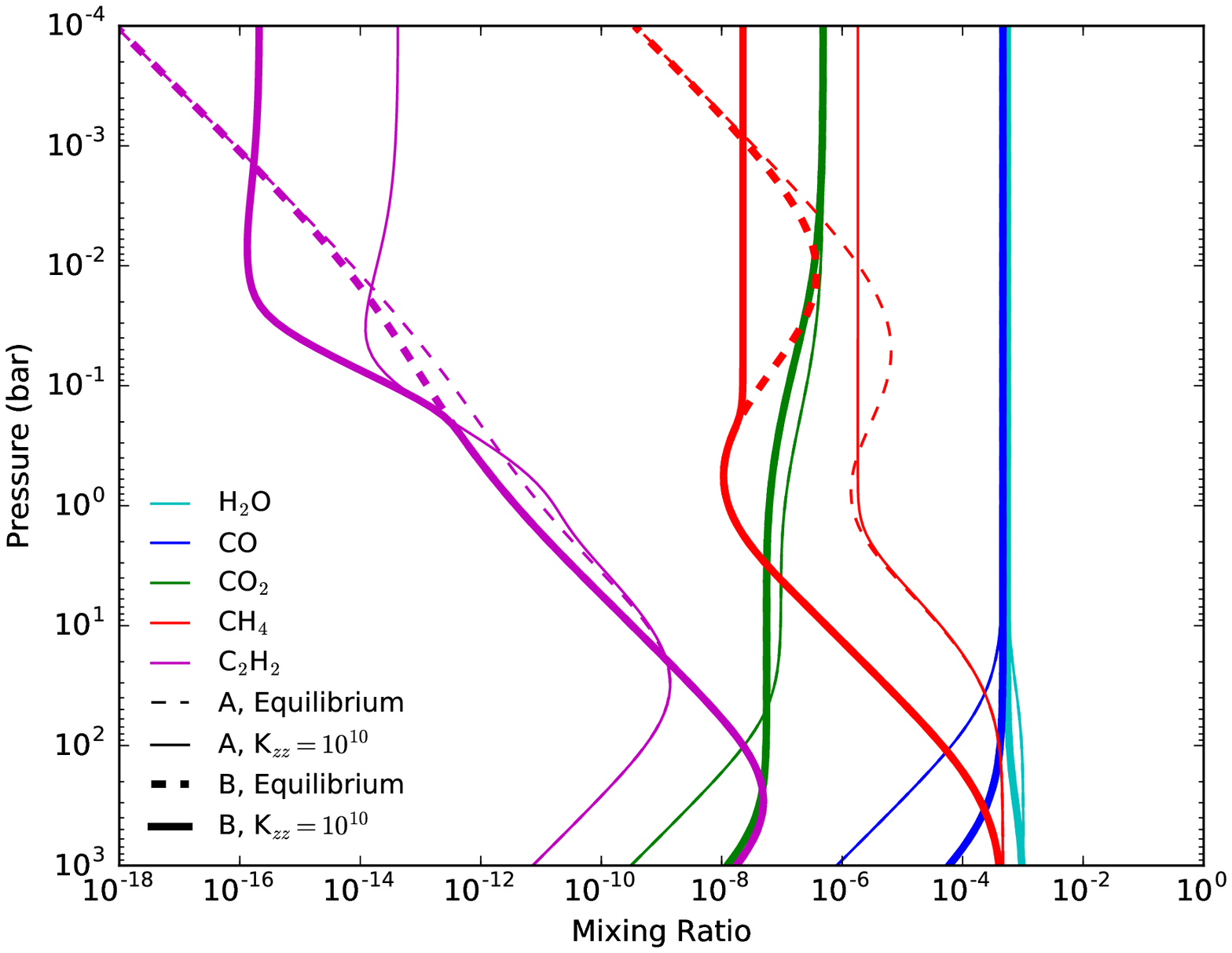}
\end{center}
\caption{Same as Figure \ref{fig:tp1}, but exploring the effects of varying the strength of infrared scattering (which may serve as a proxy for large aerosols).}
\label{fig:tp4}
\end{figure}

The analytical profile of \cite{hml14} includes a generalization to consider scattering in the infrared, which serves as a proxy for the presence of micron-sized (or larger) aerosols, condensates or cloud particles.  This ``scattering greenhouse effect" warms the atmosphere throughout (top panel of Figure \ref{fig:tp4}), which diminishes methane and enhances carbon monoxide (bottom panel of Figure \ref{fig:tp4}). 

\subsection{Revisiting the Quenching Approximation}
\label{subsect:quench}

\begin{figure}
\begin{center}
\includegraphics[width=\columnwidth]{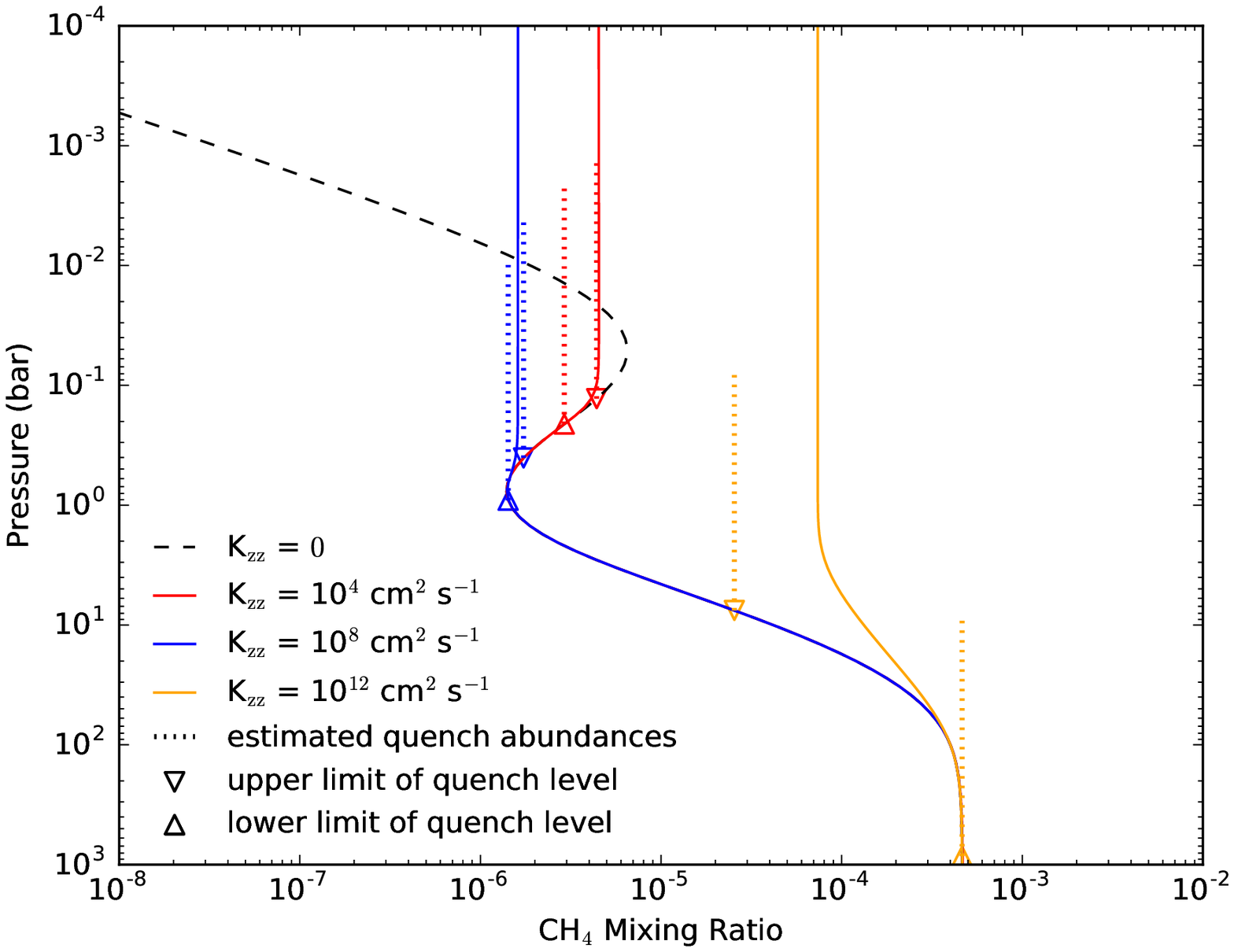}
\includegraphics[width=\columnwidth]{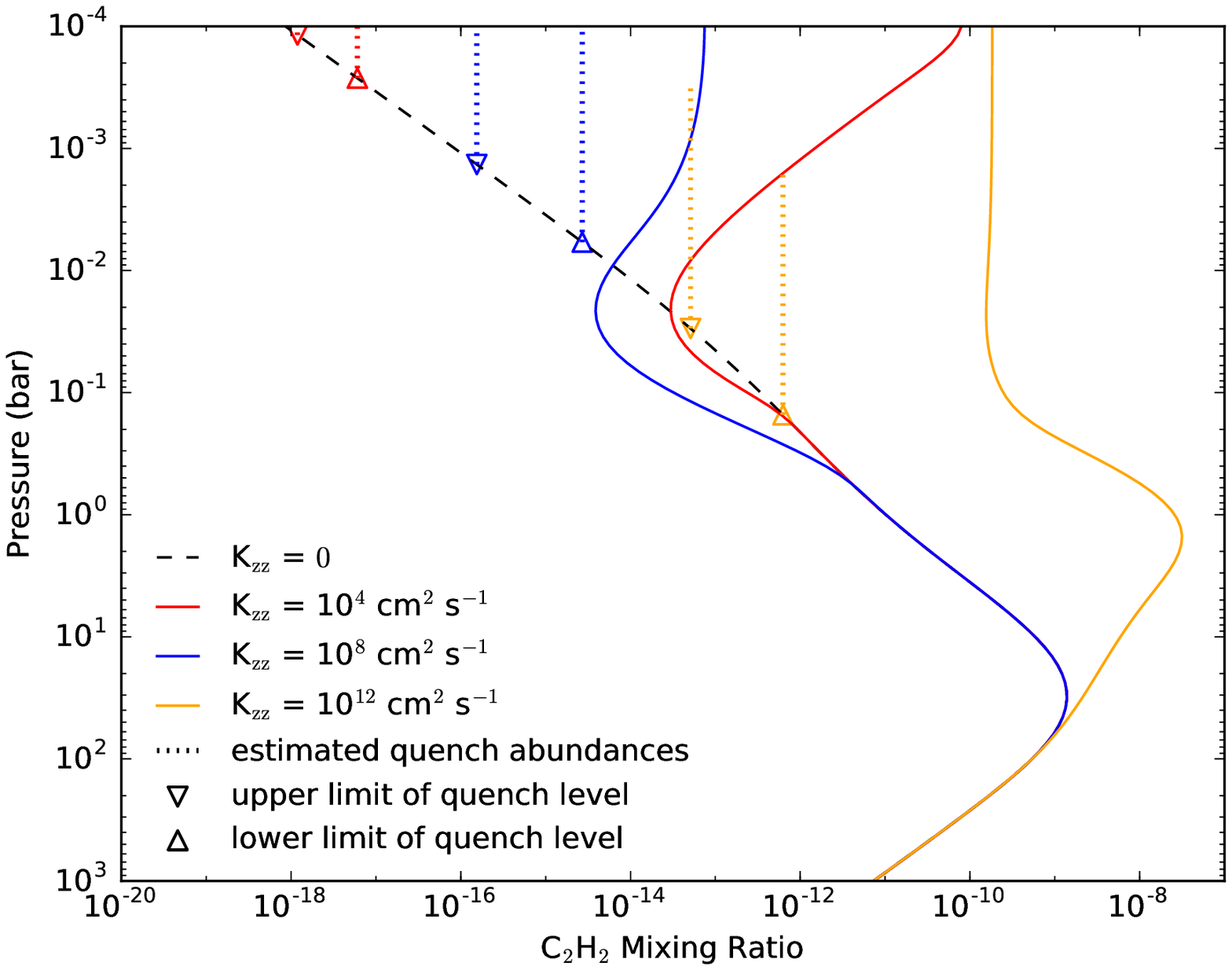}
\includegraphics[width=\columnwidth]{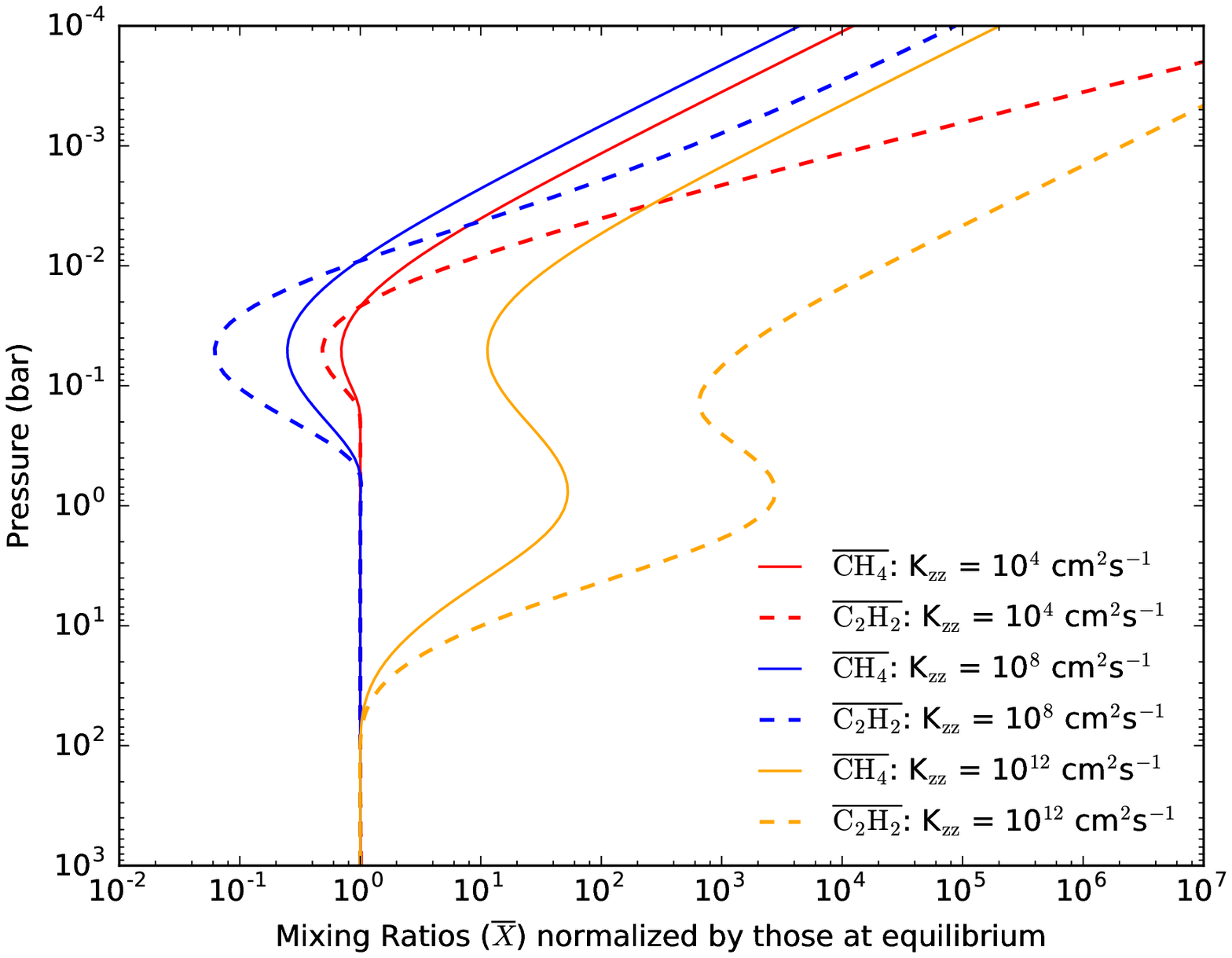}
\end{center}
\caption{Top and middle panels: comparing the full chemical-kinetics calculations (solid curves) with those employing the quenching approximation (dotted lines).  For each solid curve, the pair of triangles marks the range of uncertainties associated with computing the dynamical timescale, namely that $t_{\rm dyn} = L^2/K_{\rm zz}$ and $L=0.1 H$--$H$.  The top and middle panels show the abundances of \ce{CH4} and \ce{C2H2} for the temperature-pressure profile Figure \ref{fig:tp2}-A respectively.  The dashed curves are for calculations with no eddy diffusion (i.e., chemical equilibrium).  Bottom panel: the ratio of abundances of \ce{CH4} or \ce{C2H2} normalized by their chemical equilibrium values and subjected to different degrees of eddy diffusion.}
\label{fig:quench}
\end{figure}

A potential way to describe disequilibrium chemistry, induced by atmospheric motion, without having to perform full calculations of chemical kinetics, is to employ the quenching approximation \citep{pb77}.  It is the notion that there exists a point, within the atmosphere, where the chemical and dynamical timescales are equal, known as the ``quench level".  Below this point, chemistry reacts fast enough that the abundance is simply determined by chemical equilibrium.  Above this point, the abundances of the atoms and molecules are frozen to their equilibrium values at the quench level, because chemistry is slow compared to dynamics.

\cite{madhu11} show a recent example of how the quenching approximation is employed to compute the abundances of major molecules when eddy diffusion is present. In practice, one computes the chemical timescale ($t_{\rm chem}$) given by
\begin{equation}
t_{\rm chem} = n_{i,j} \left(\frac{d n_{i,j}}{d t}\right)^{-1},
\end{equation}
where $d n_{i,j} / d t$ is the control or rate-limiting step.  A caveat is that ambiguities or discrepancies may arise when identifying the rate-limiting reaction and determining the values of the rate coefficients, and one is sometimes forced to choose between competing or even contradictory sources.

The chemical timescale is then compared to the dynamical timescale,
\begin{equation}
t_{\rm dyn} = \frac{L^2}{K_{\rm zz}},
\end{equation}
where $L$ is a characteristic lengthscale, typically taken to be the pressure scale height without justification. \cite{smith98} has discussed how the characteristic lengthscale ranges from being $L=0.1 H$ to $L=H$ (see Table I in \citealt{smith98} and Table 1 in \citealt{vm11}).  This ambiguity translates into an uncertainty of two orders of magnitude in the dynamical timescale.  We note that \cite{madhu11} have assumed $L=H$.  

The uncertainties associated with identifying the control or rate-limiting step and choosing the value of $L$ have been previously discussed in \cite{vm11}.  In this work, we will demonstrate a third uncertainty, which is the interplay between the quenched species and unquenched species.

We focus on two species: methane and acetylene, as they are illustrative and instructive.  We factor in the uncertainty associated with the dynamical timescale.  In Figure \ref{fig:quench}, we compare full calculations of chemical kinetics versus those obtained using the quenching approximation for the temperature-pressure profile in Figure \ref{fig:tp2}-A.  The identification of the key reaction allows us to compute $t_{\rm chem}$.  For \ce{CH4} $\rightarrow$ \ce{CO} conversion, the chemical pathway, at low pressure, consists of
\begin{subequations}
\begin{align}
\begin{split} 
\ce{ CH4 + H} &\rightarrow \ce{ CH3 + H2 } \\
\ce{ CH3 + OH}&\rightarrow \ce{ CH2OH + H  } \\
\ce{ CH2OH + M }&\rightarrow \ce{ H2CO + H + M  } \\
\ce{ H2CO + H }&\rightarrow \ce{ HCO + H2 } \\
\ce{ HCO + M }&\rightarrow \ce{ H + CO + M } \\
\ce{ H + H2O }&\rightarrow \ce{ OH + H2 } \\
\hline \nonumber
\mbox{net} : \ce{ CH4 + H2O} &\rightarrow \ce{CO + 3H2 } 
\end{split} 
\end{align}
\end{subequations}
and at high pressure
\begin{subequations}
\begin{align}
\begin{split} 
\ce{ CH4 + H} &\rightarrow \ce{ CH3 + H2 } \\
\ce{ CH3 + OH + M}&\rightarrow \ce{ CH3OH + M  } \\
\ce{ CH3OH + H }&\rightarrow \ce{ CH3O + H2 } \\
\ce{ CH3O + M }&\rightarrow \ce{ H2CO + H + M } \\
\ce{ H2CO + H }&\rightarrow \ce{ HCO + H2 } \\
\ce{ HCO + M }&\rightarrow \ce{ H + CO + M } \\
\ce{ H + H2O }&\rightarrow \ce{ OH + H2 } \\
\ce{ H2 + M }&\rightarrow \ce{ 2H + M }\\
\hline \nonumber
\mbox{net} : \ce{ CH4 + H2O} &\rightarrow \ce{CO + 3H2 } 
\end{split} 
\end{align}
\end{subequations}
where we found \ce{ CH3 + OH } $\rightarrow$ \ce{ CH2OH + H } and \ce{ CH3 + OH + M } $\rightarrow$ \ce{ CH3OH + M } to be the rate-limiting step at low and high pressures, respectively, in agreement with \cite{moses11}.
 
For \ce{C2H2} $\rightarrow$ \ce{CH4} conversion, the chemical pathway is given by
\begin{subequations}
\begin{align}
\begin{split} 
\ce{C2H2 + H + M } &\rightarrow \ce{ C2H3 + M} \\
\ce{C2H3 + H2 } &\rightarrow \ce{ C2H4 + H} \\
\ce{C2H4 + H + M  } &\rightarrow \ce{ C2H5 + M} \\
\ce{C2H5 + H2 } &\rightarrow \ce{ C2H6 + H} \\
\ce{C2H6 + M  } &\rightarrow \ce{ 2 CH3 + M} \\
2(\ce{CH3 + H2 } &\rightarrow \ce{ CH4 + H}) \\
\ce{ 2H + M } &\rightarrow \ce{ H2 + M } \\
\hline \nonumber
\mbox{net} : \ce{C2H2 + 3H2 } &\rightarrow \ce{ 2 CH4}   
\end{split}
\end{align}
\end{subequations}
where we found the rate-limiting step to be \ce{C2H5 + H2 } $\rightarrow$ \ce{C2H6 + H} or $ \ce{C2H6 + M} \rightarrow$ \ce{ 2 CH3 + M} at low pressure.  

For methane, the approximate mixing ratios lie in the range bounded by the uncertainty associated with $t_{\rm dyn}$.  As the eddy diffusion coefficient becomes larger, the range of methane abundances obtained from the quenching approach spans orders of magnitude, because the gradient of the chemical timescale, across pressure, becomes steep.  The utility of the quenching approximation becomes suspect. 

For acetylene, the quenching approximation does not even produce solutions that are within the range bounded by the uncertainty associated with $t_{\rm dyn}$.  In fact, the accuracy is worse than at the order-of-magnitude level.  The reason is because the unquenched acetylene interacts with the quenched methane, which causes its mixing ratio to deviate from the values predicted by the quench level.  The discrepancy results from manually fixing the quench abundance and negating its interplay with other species.  This reasoning is supported by the bottom panel of Figure \ref{fig:quench}, which demonstrates that the abundances of acetylene and methane, relative to their chemical-equilibrium values, closely track each other.

Our results in Figure \ref{fig:quench} demonstrate that the quenching approximation cannot be applied wholesale to all of the species in a chemical network, because the abundances of some of these species are not determined by their quench levels, but are rather controlled by the abundances of other species.  For this reason, calculations that employ the quenching approximation need to be checked by full calculations of chemical kinetics, at least in the hot exoplanetary atmosphere regime.

For completeness, we include in Tables \ref{tab:methane} and \ref{tab:acetylene} the sets of most relevant reactions for \ce{CH4} and \ce{C2H2}, respectively, for $T=1000$ K and $P=1$ bar.  For the conversion of \ce{CH4} to \ce{CO}, the reactions labeled by R1, R13, R126, R127, R232, R270, and R291 belong to the pathways, and for the conversion of \ce{C2H2} $\rightarrow$ \ce{CH4}, the reactions labeled by R14, R25, R36, R231, R232, R241, R245, and R258 are important, while the rest of the reactions in Table \ref{tab:methane} and \ref{tab:acetylene} do not directly belong to the main \ce{CH4} $\rightarrow$ \ce{CO} and \ce{CH4} $\rightarrow$ \ce{C2H2} pathways.

\subsection{Interplay Between C/O and Atmospheric Mixing}

\begin{figure*}
\begin{center}
\includegraphics[width=\columnwidth]{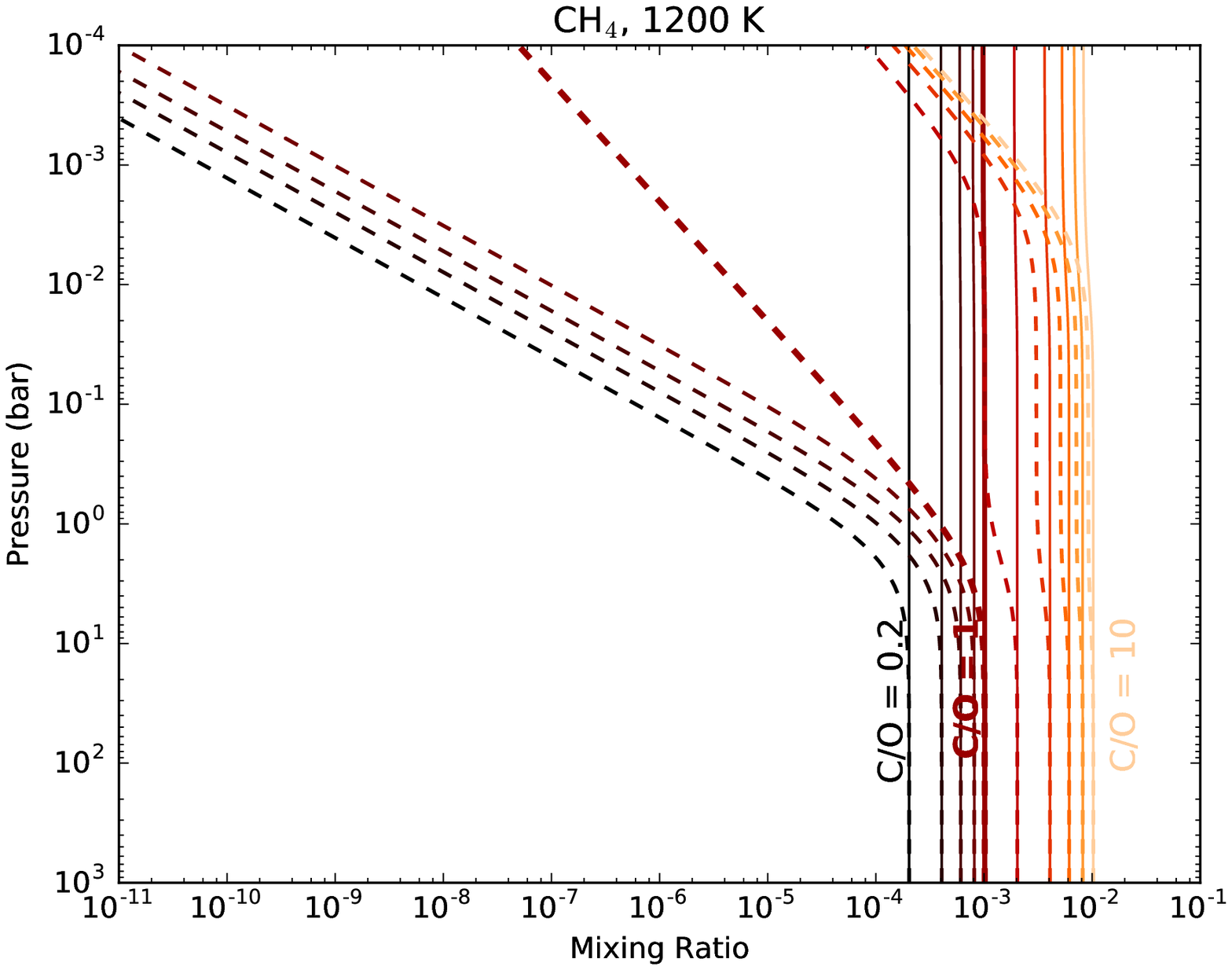}
\includegraphics[width=\columnwidth]{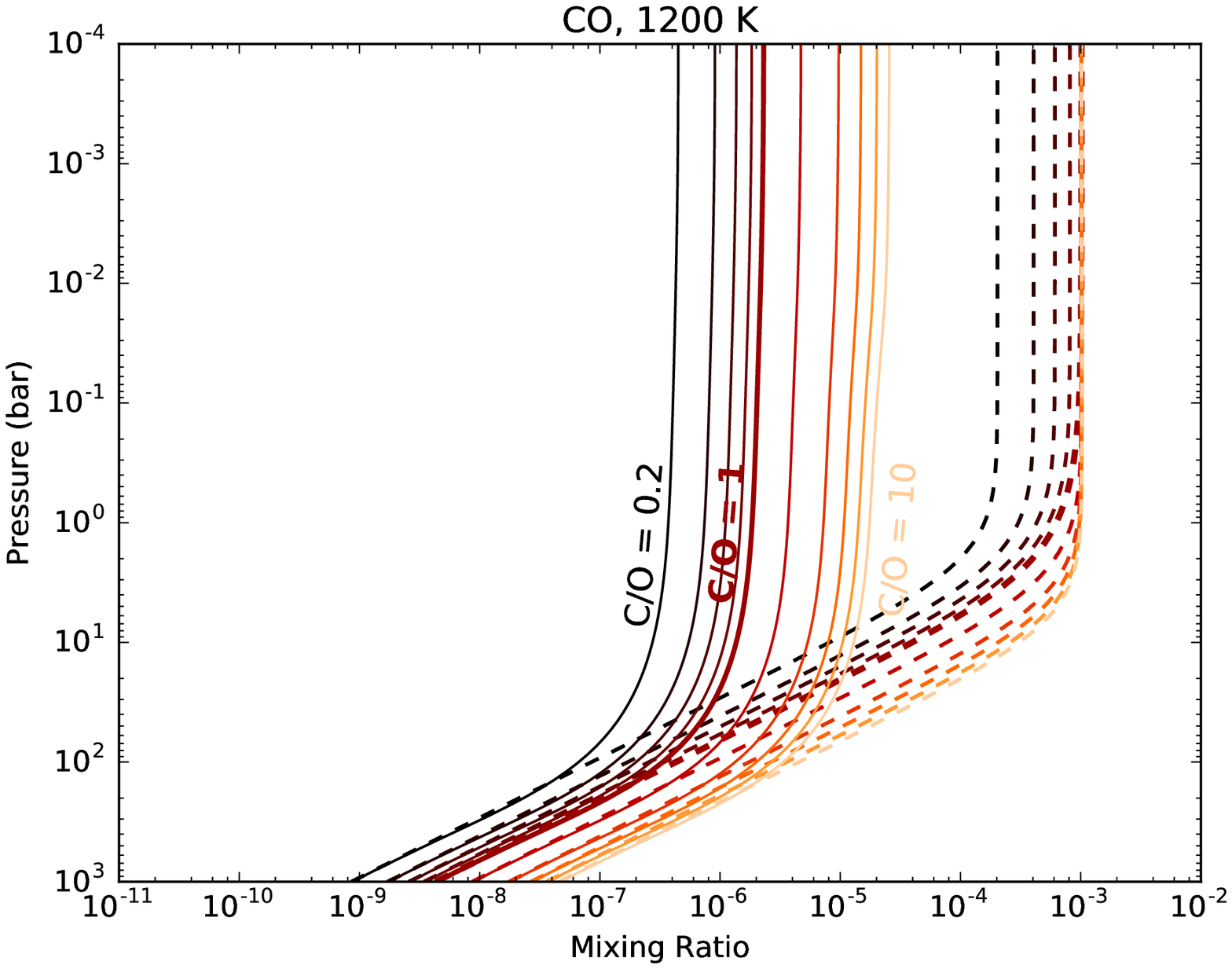}
\includegraphics[width=\columnwidth]{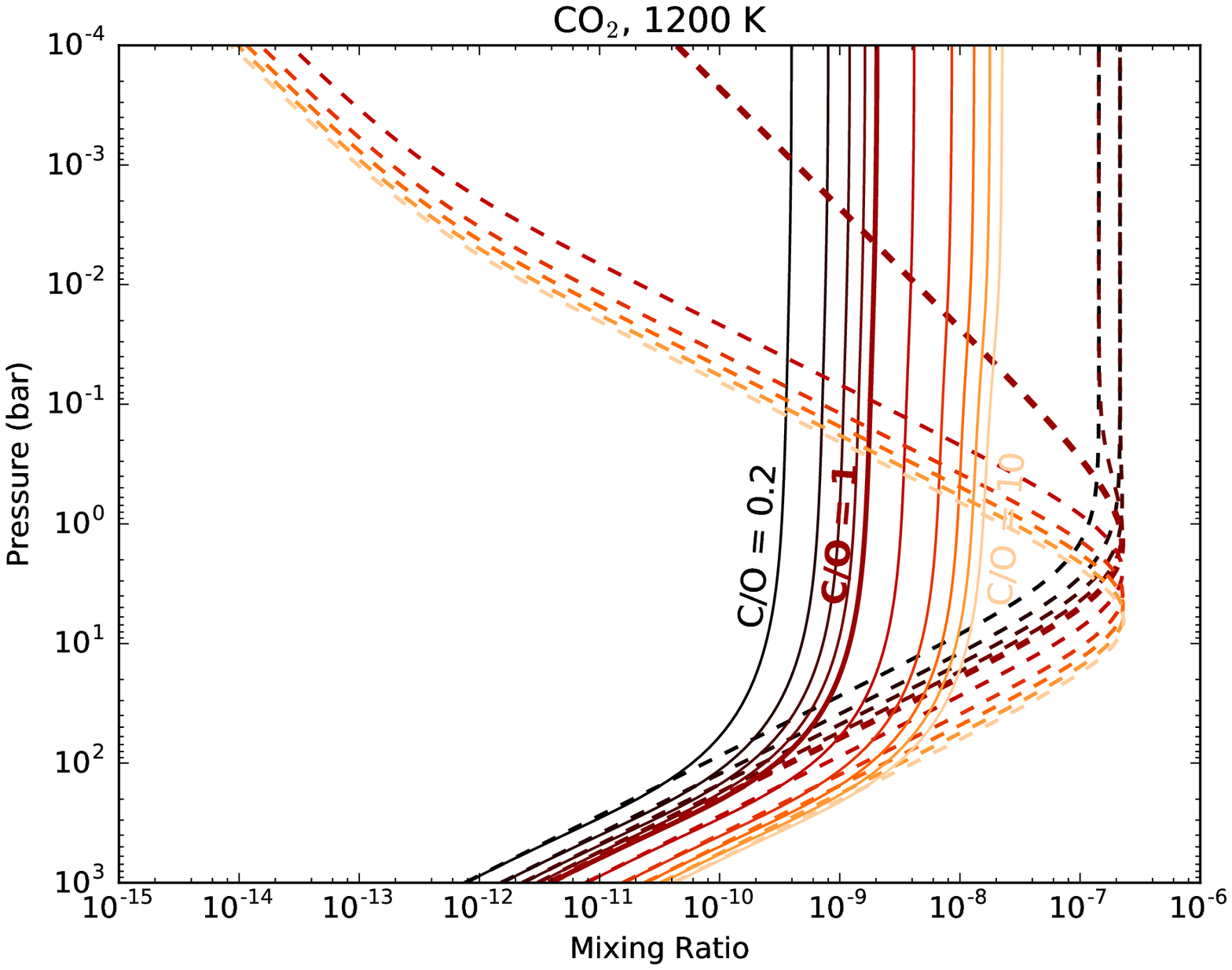}
\includegraphics[width=\columnwidth]{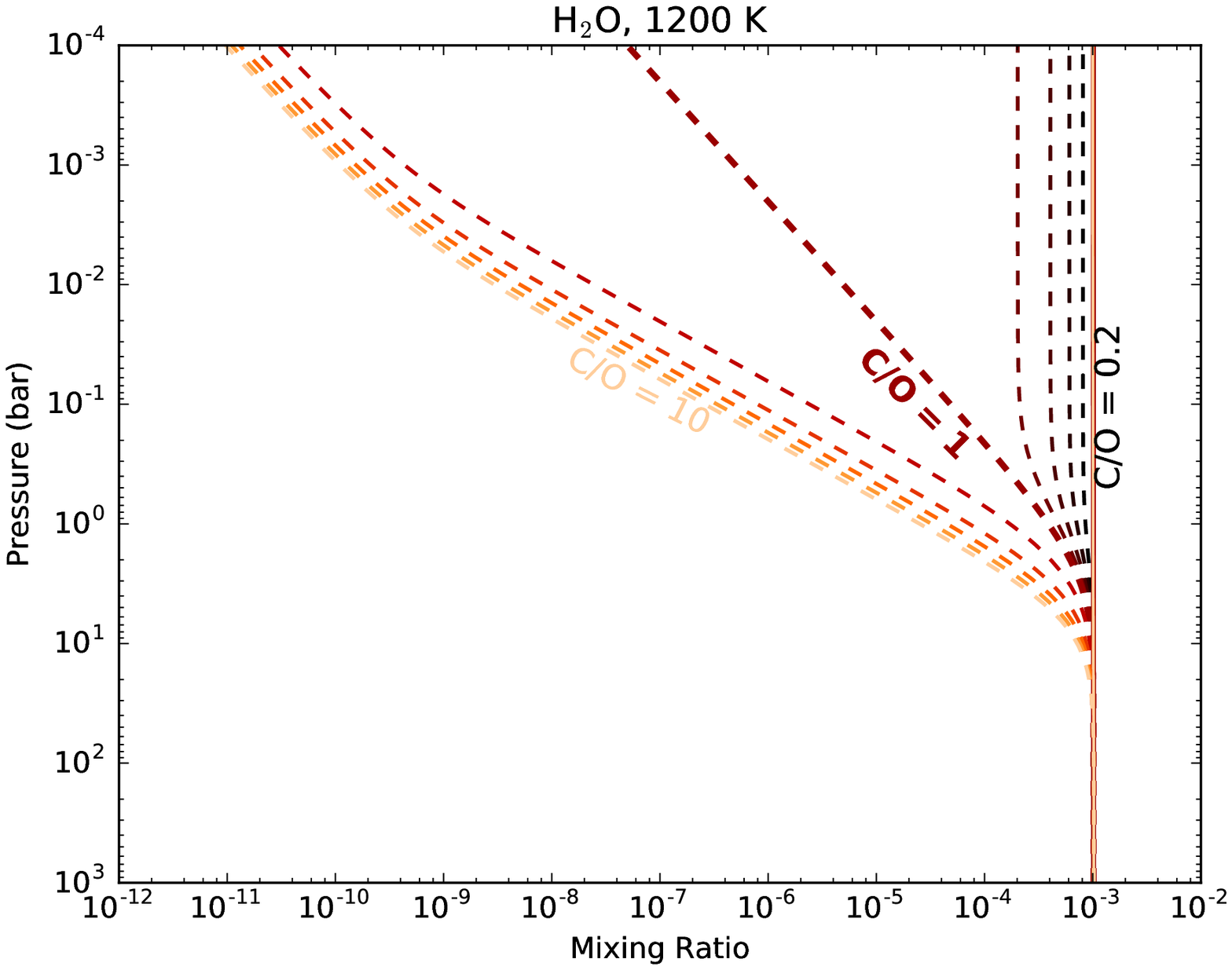}
\includegraphics[width=\columnwidth]{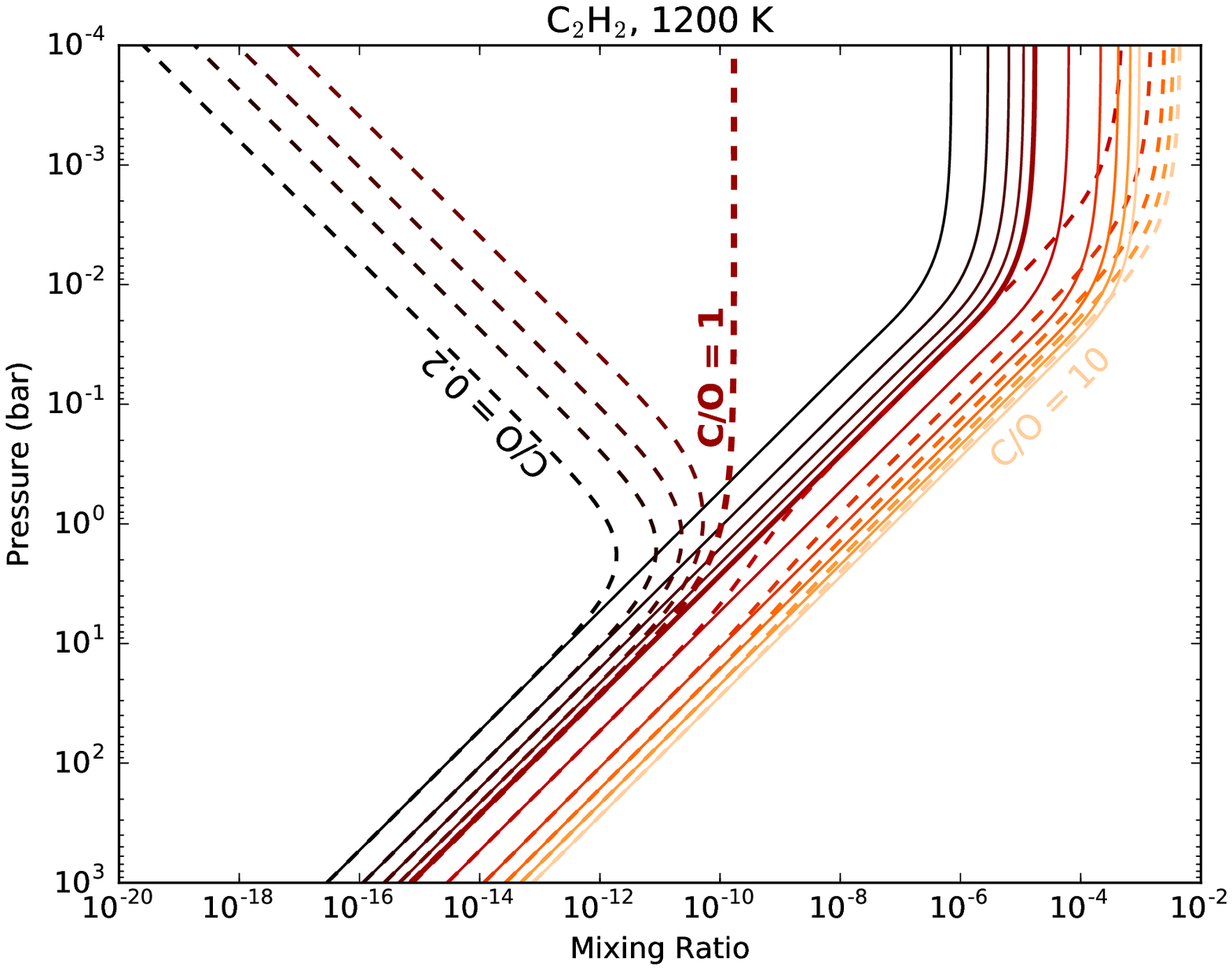}
\end{center}
\caption{The abundances of the major molecules (\ce{CH4}, CO, \ce{CO2}, \ce{H2O} and \ce{C2H2}) for isothermal atmospheres with $T=1200$ K and various C/O values (which are incremented by 0.2 from $\mbox{C/O}=0.2$ to 1, and by 2 from $\mbox{C/O}=2$ to 10).  The dashed and solid curves are for chemical equilibrium and $K_{\rm zz}=10^{10}$ cm$^2$ s$^{-1}$, respectively.}
\label{fig:co1}
\end{figure*}

\begin{figure*}
\begin{center}
\includegraphics[width=\columnwidth]{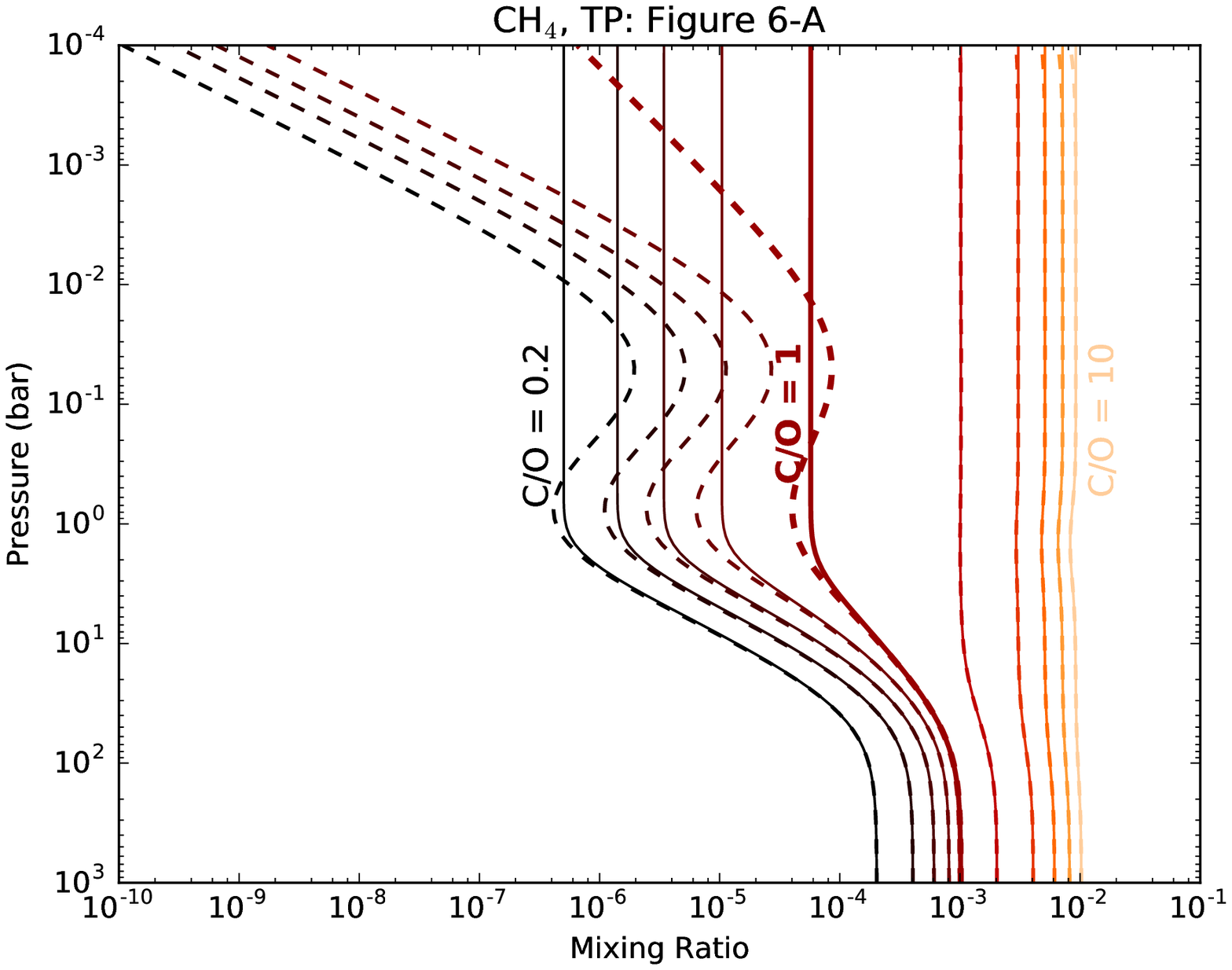}
\includegraphics[width=\columnwidth]{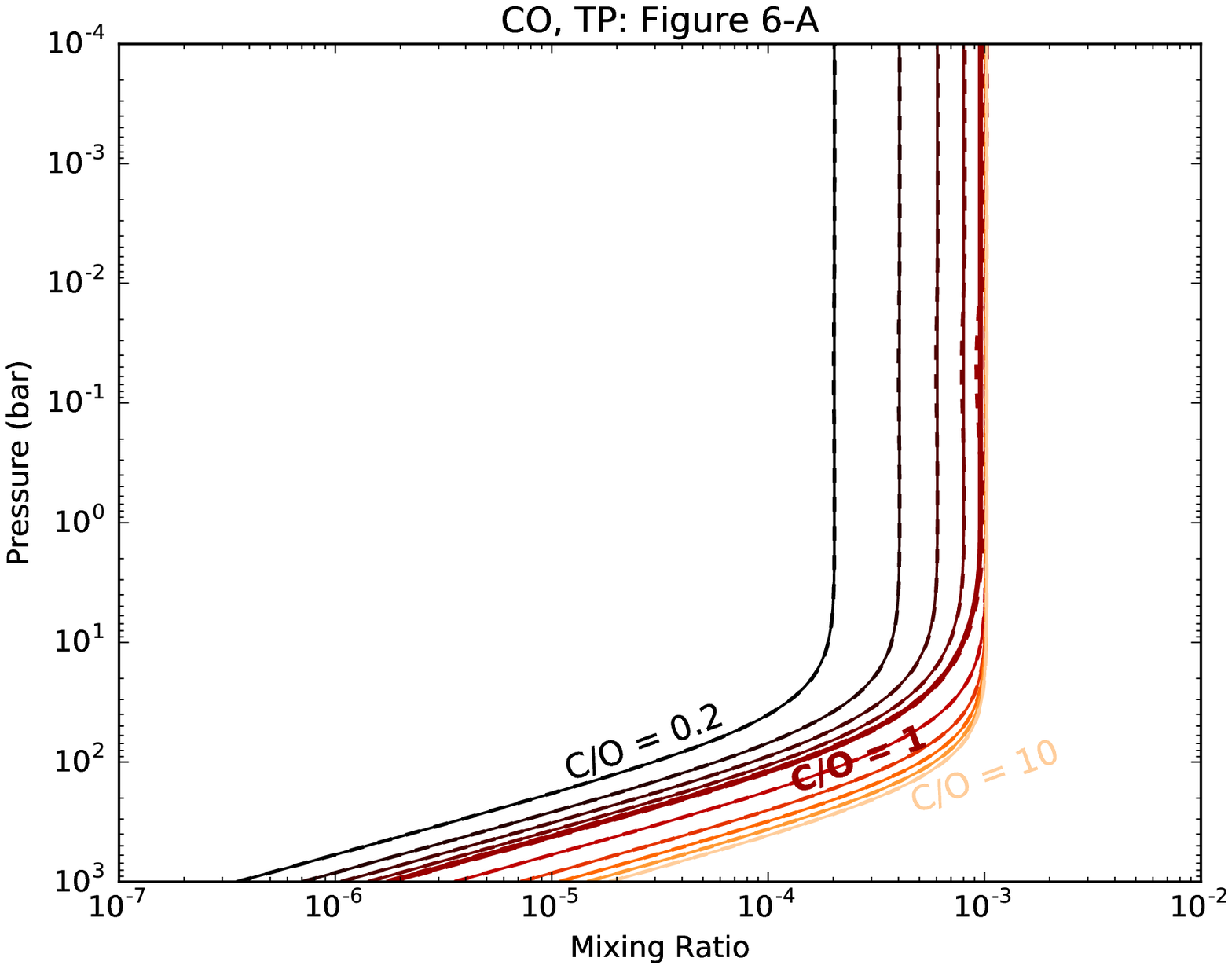}
\includegraphics[width=\columnwidth]{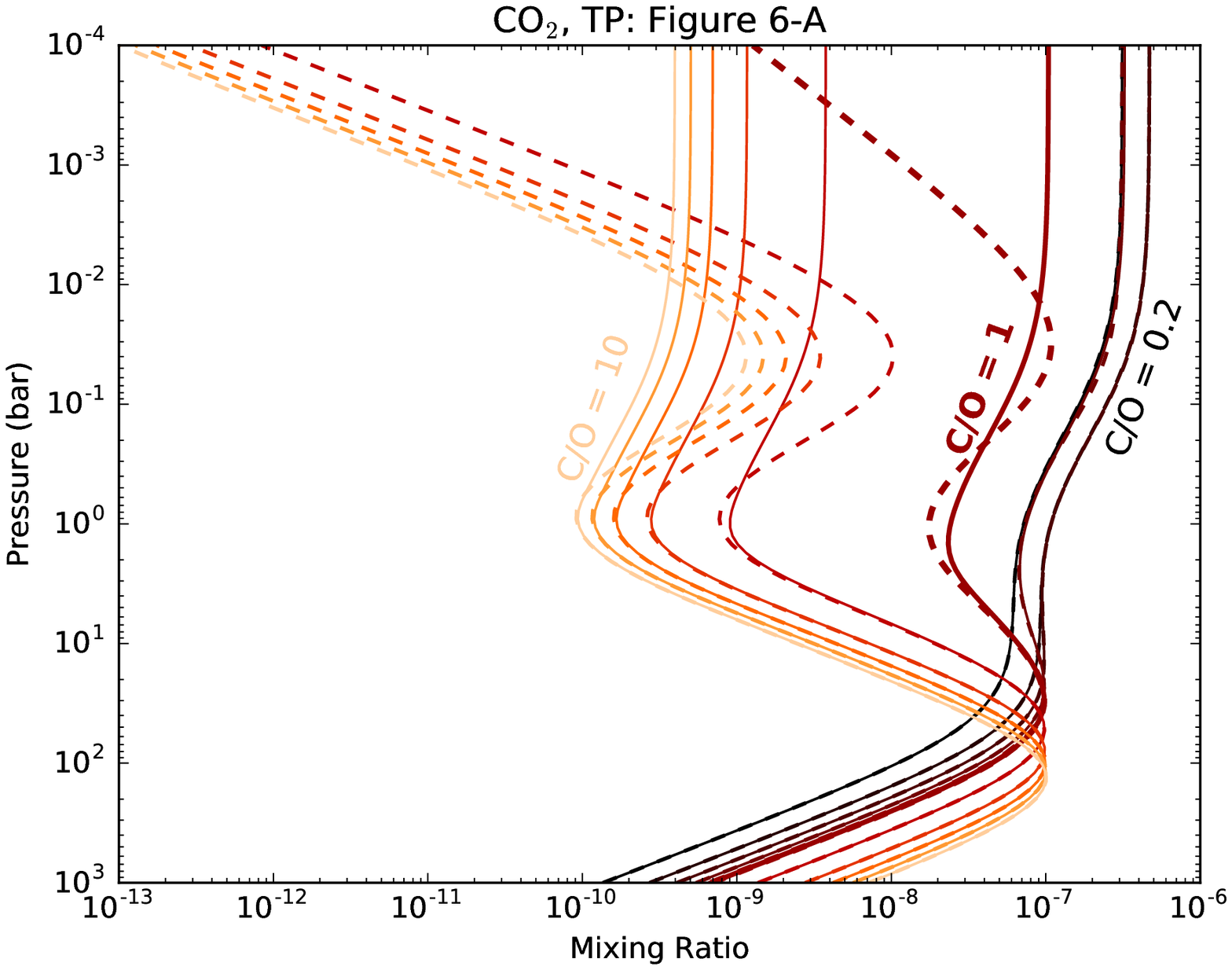}
\includegraphics[width=\columnwidth]{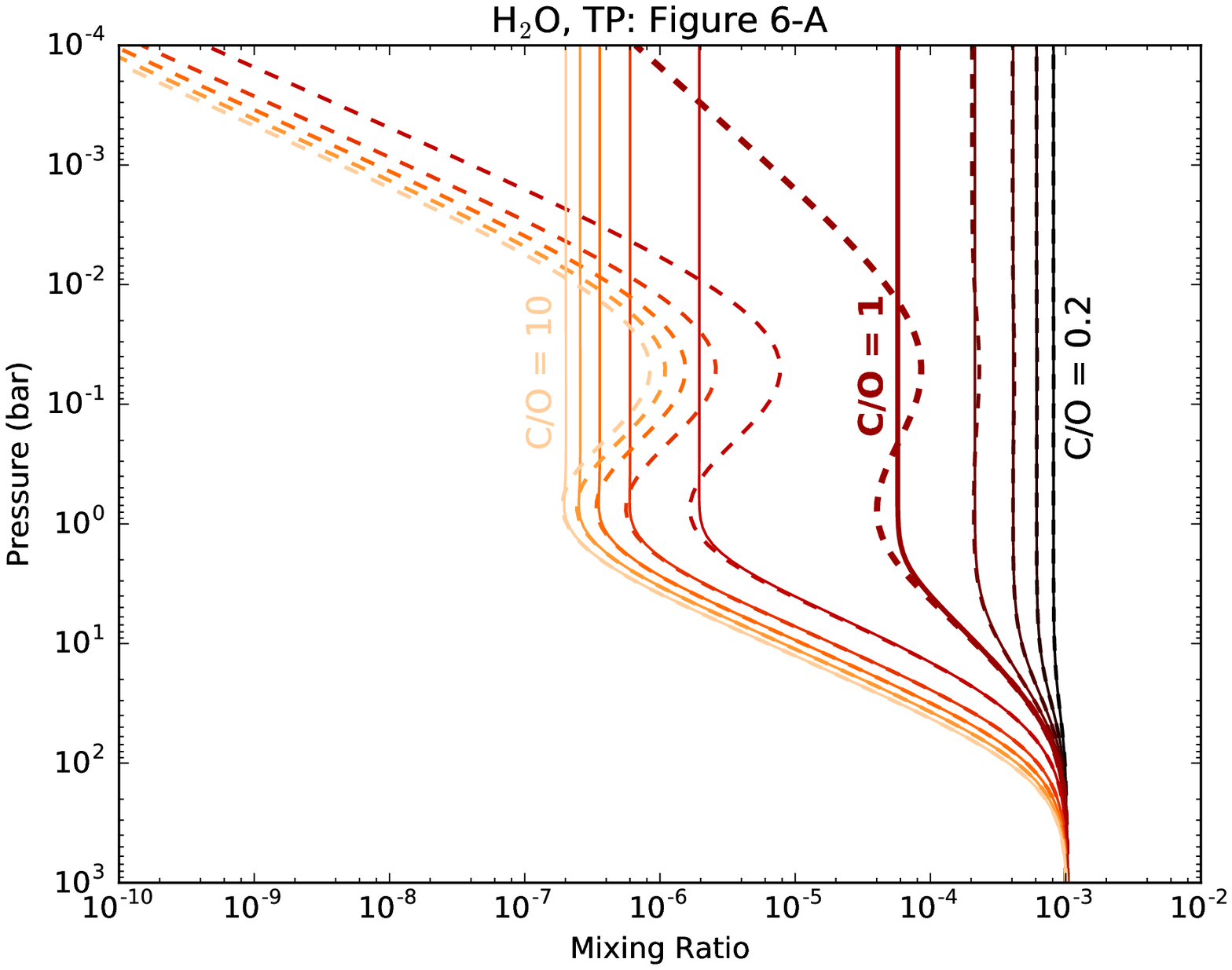}
\includegraphics[width=\columnwidth]{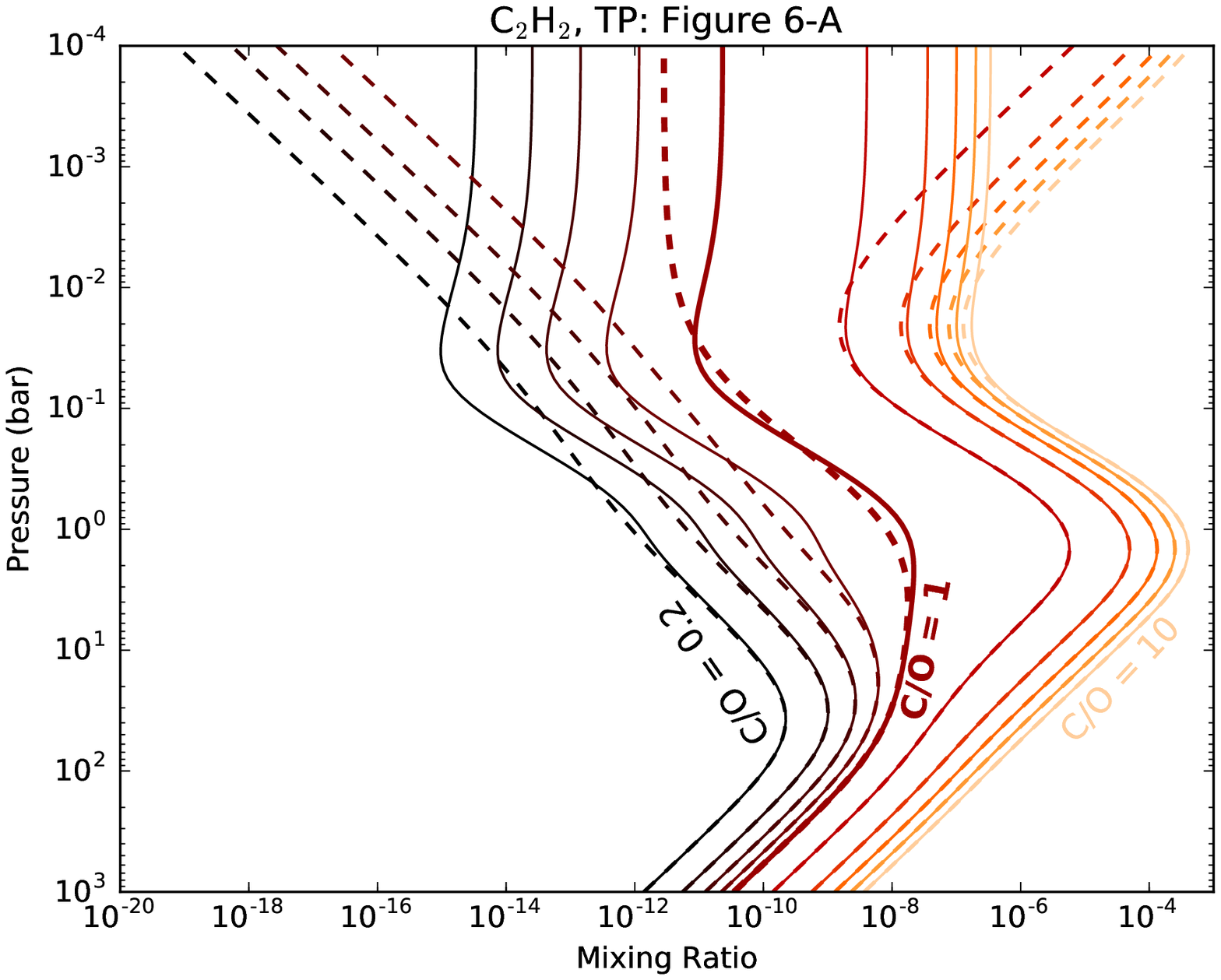}
\end{center}
\caption{Same as Figure \ref{fig:co1}, but for temperature-pressure profile from Figure \ref{fig:tp2}-A.}
\label{fig:co2}
\end{figure*}

\begin{figure*}
\begin{center}
\includegraphics[width=\columnwidth]{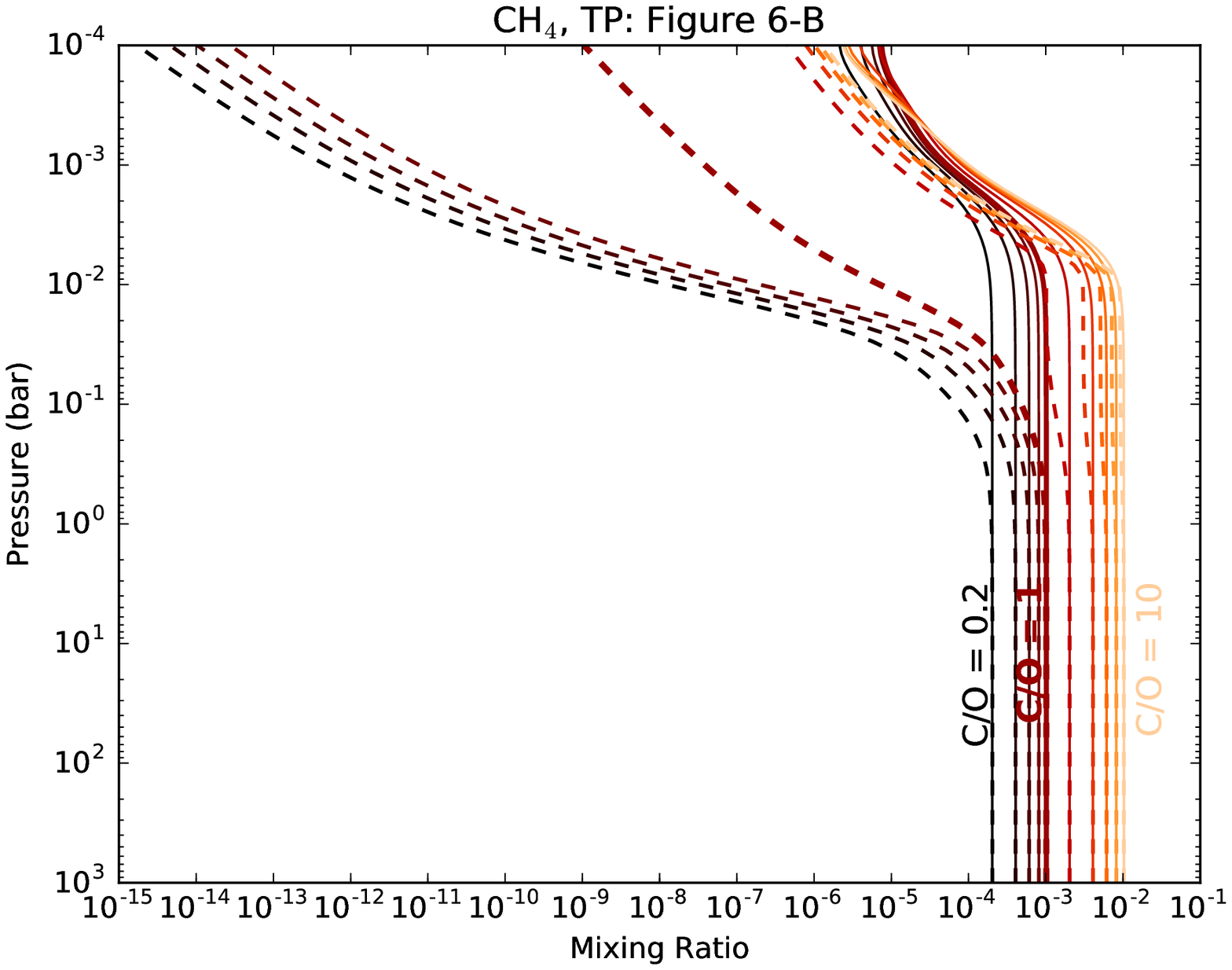}
\includegraphics[width=\columnwidth]{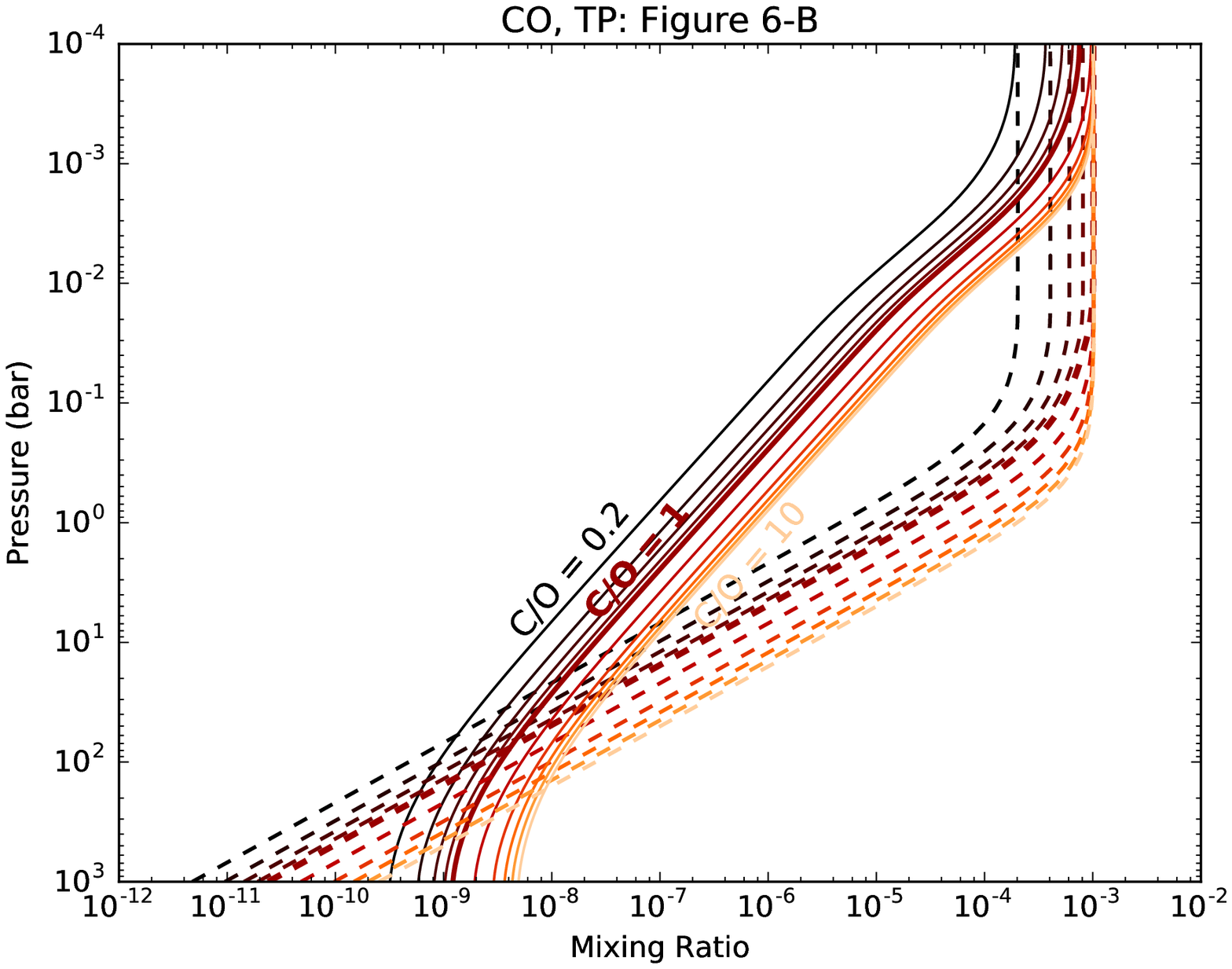}
\includegraphics[width=\columnwidth]{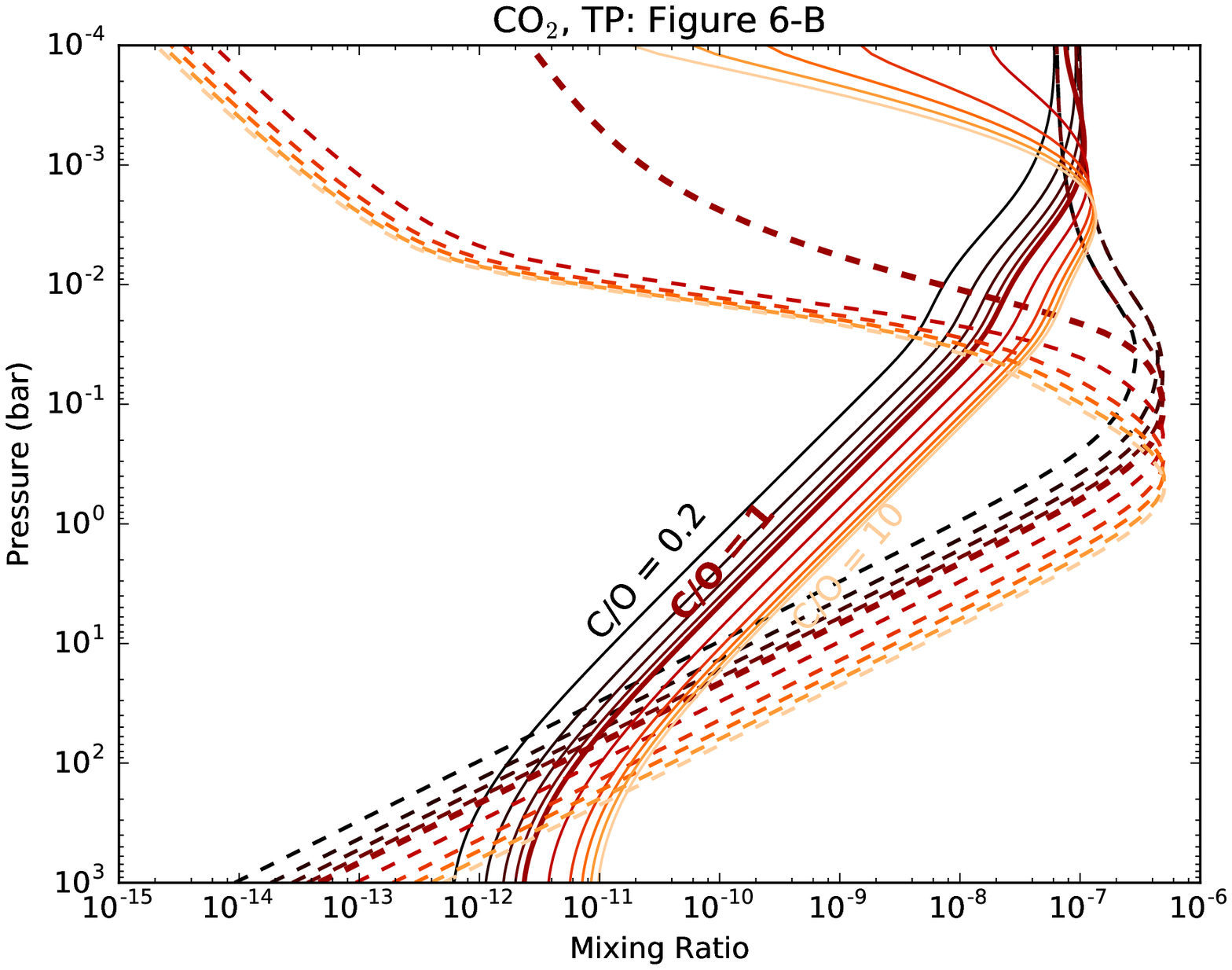}
\includegraphics[width=\columnwidth]{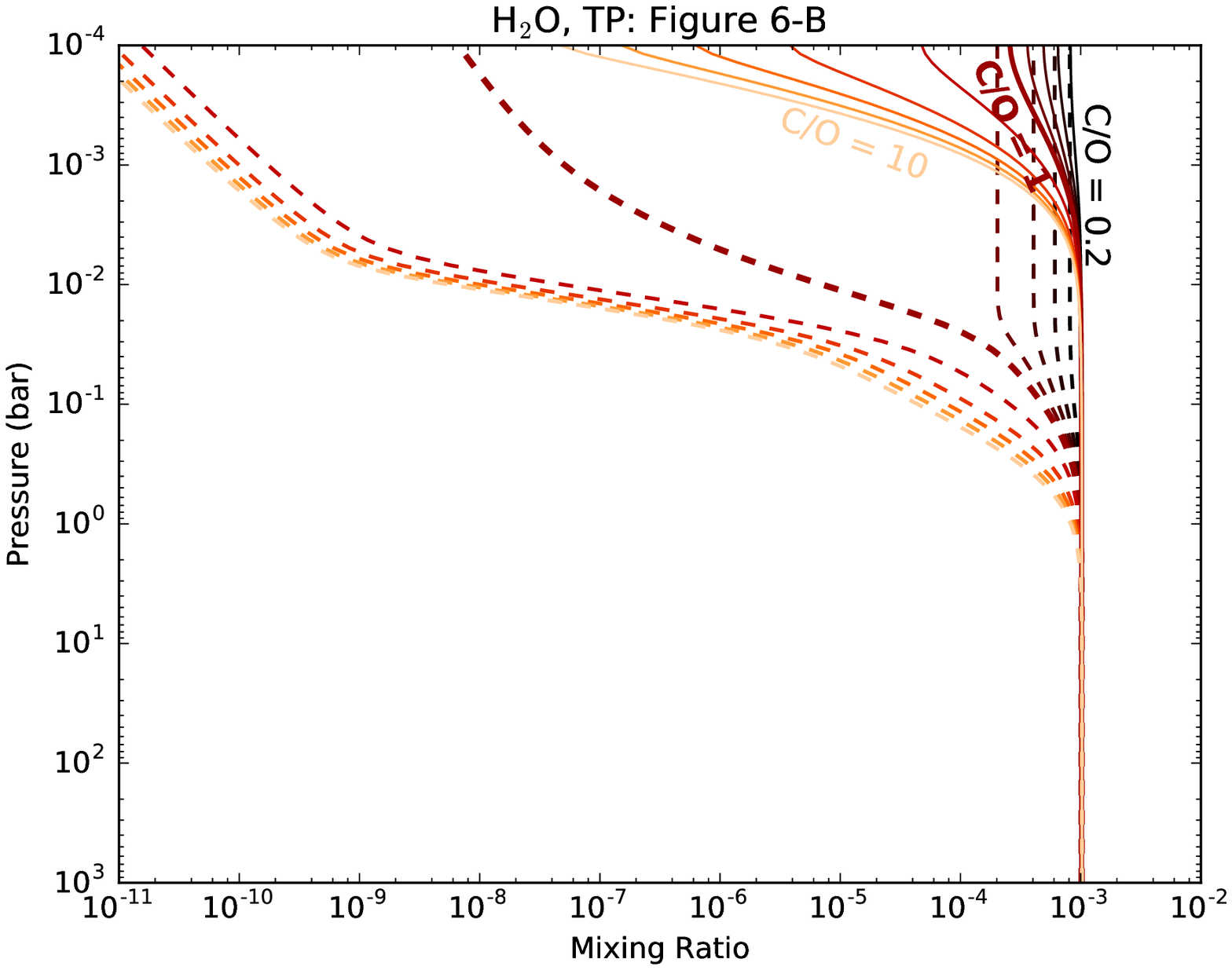}
\includegraphics[width=\columnwidth]{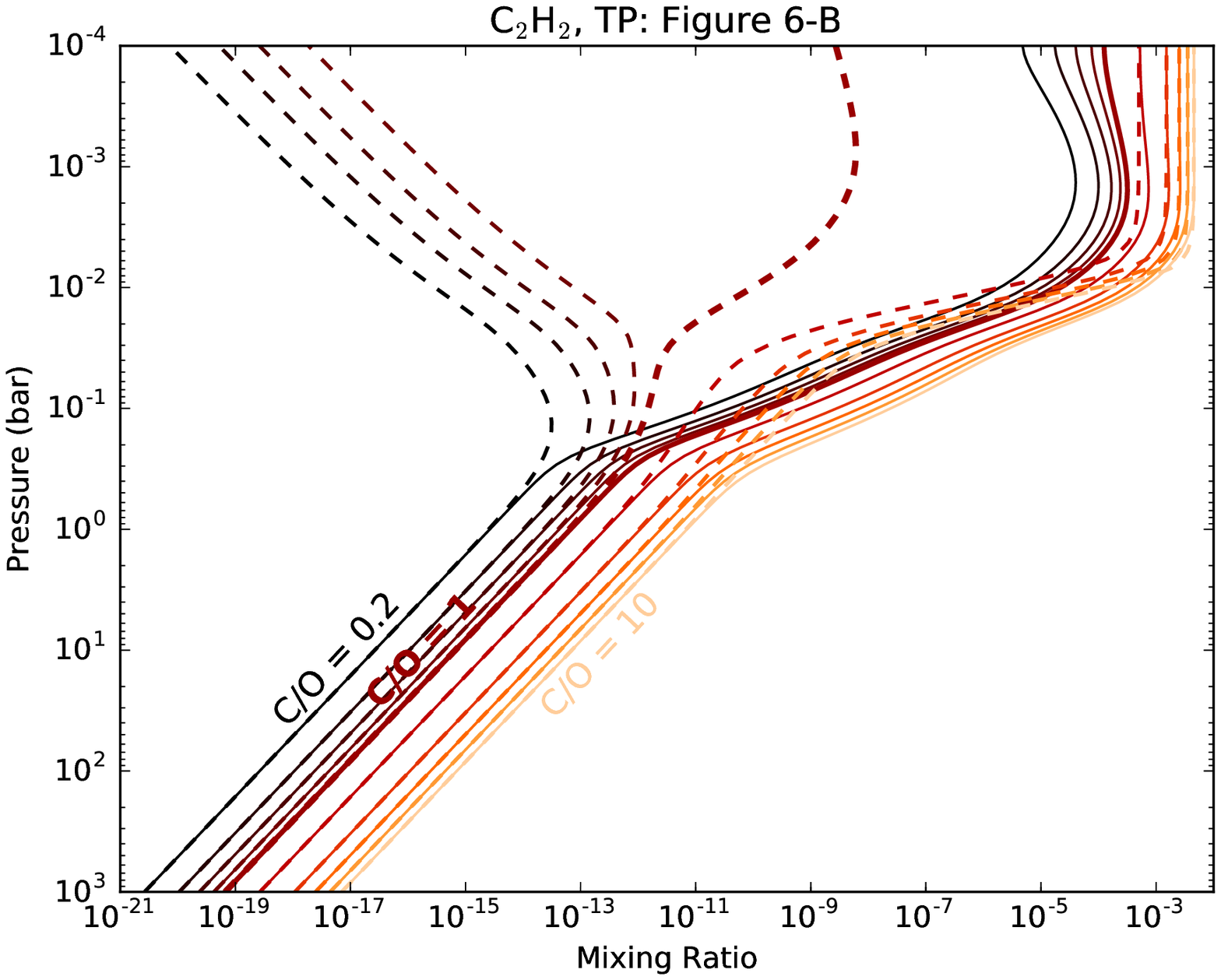}
\end{center}
\caption{Same as Figure \ref{fig:co1}, but for temperature-pressure profile from Figure \ref{fig:tp2}-B.}
\label{fig:co3}
\end{figure*}

\begin{figure}
\begin{center}
\includegraphics[width=\columnwidth]{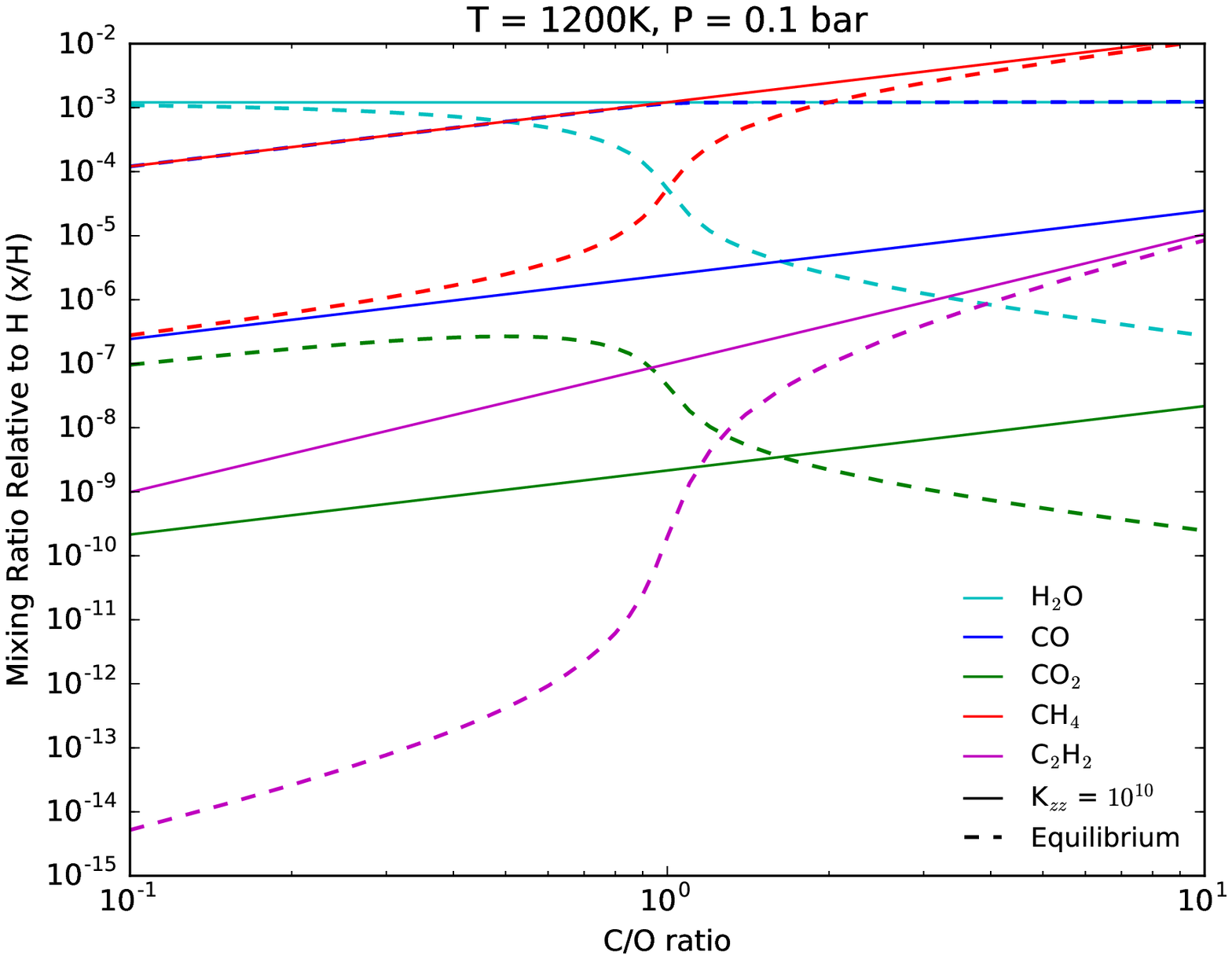}
\end{center}
\caption{Exploring the effects of atmospheric mixing on the relative abundances of various molecules as functions of C/O.  The solid curves are for chemical equilibrium, while the dashed curves are for $K_{\rm zz}=10^{10}$ cm$^2$ s$^{-1}$.  When eddy diffusion is present, a deeper part of the isothermal atmosphere is essentially being sampled.}
\label{fig:co_sum}
\end{figure}

We also explore the interplay between C/O and atmospheric mixing, and the consequences for the photospheric abundances of the major molecules.  In Figures \ref{fig:co1}, \ref{fig:co2} and \ref{fig:co3}, we show the mixing ratios of \ce{CH4}, CO, \ce{CO2}, \ce{H2O} and \ce{C2H2} for isothermal atmospheres at 1200 K, as well as temperature-pressure profiles from Figure \ref{fig:tp2}-A and \ref{fig:tp2}-B, respectively.  We performed calculations from $\mbox{C/O}=0.1$ to 10, both with and without atmospheric mixing\footnote{We note that \cite{moses13b} have previously suggested that graphite formation could limit the C/O value in the gaseous phase to be unity or lower.}.   The chemical-equilibrium molecular abundances are sensitive to the value of C/O and experience a qualitative transition versus pressure at $\mbox{C/O}=1$ \citep{moses13a}.  The carbon and oxygen atoms are bound up in carbon monoxide, because of its relative stability.  Any excess carbon is then bound up in methane if $\mbox{C/O}>1$, and any access oxygen is bound up in  water if $\mbox{C/O}<1$, and these molecules then maintain a uniform mixing ratio across pressure, unaffected by the presence of atmospheric mixing. This trend is broken for methane when a temperature inversion exists, as the higher temperatures favor the deposition of carbon in acetylene (see $\mbox{C/O}>1$ curves in Figure \ref{fig:co3}).

Figure \ref{fig:co_sum} visualizes these trends in a different way by showing the mixing ratios versus C/O.  At 1200 K, the trends of chemical equilibrium displayed by the major molecules are somewhat simple, as already noted by \cite{madhu12} and \cite{hlt16}. When atmospheric mixing is present, one is effectively sampling the molecular abundances originating from a deeper part of the isothermal atmosphere with higher pressures (except for species like \ce{C2H2}). Therefore, the presence of atmospheric mixing may lead to a degeneracy in interpretation if it is focused only on one molecule.  For example, at solar metallicity, a high abundance of methane may be a result of high C/O \textit{or} low C/O in the presence of efficient mixing.  Similarly, a high abundance of water may be the result of low C/O or high C/O in the presence of efficient mixing.  The joint consideration of two or more molecules breaks these degeneracies.  For example, a high abundance of methane combined with a low abundance of water constitutes evidence for a high C/O.

\section{Summary \& Discussion}
\label{sect:discussion}

\subsection{Overall Conclusions}

We have constructed an open-source and validated chemical kinetics code, named \texttt{VULCAN}, for studying the gaseous chemistry of hot (500--2500 K) exoplanetary atmospheres using a reduced C-H-O network of about 300 reactions for 29 species.  We have provided a full description of the rate coefficients and thermodynamic data used.  We have demonstrated that \texttt{VULCAN} is able to reproduce chemical equilibrium as a limiting case and also compared our calculations to the disequilibrium-chemistry models of \cite{moses11} and \cite{rimmer16}.  Specifically, we are able to reproduce the models of HD 189733b and HD 209458b by \cite{moses11}, despite using a reduced chemical network (300 versus nearly 1600 reactions), thus demonstrating that the accuracy of a chemical kinetics calculation is not determined by the sheer size of the network alone.  We further examine trends associated with varying the temperature-pressure profile.  We demonstrate that the quenching approximation cannot always be employed and may result in large errors of several orders of magnitude.  Finally, we show that the abundances of \ce{CH4} and \ce{H2O} depend sensitively on the presence of atmospheric mixing at low and high values of C/O, respectively.

\subsection{Comparison to Previous Work and Opportunities for Future Work}

In terms of its technical setup, \texttt{VULCAN} shares some similarities to the work of \cite{moses11}, \cite{hu12}, \cite{moses13a,moses13b}, \cite{hu15} and \cite{venot12,venot15}.  These codes solve a set of mass continuity equations with chemical source and sink terms, and approximate atmospheric motion by diffusion.  They differ in some details: we have used the Rosenbrock (semi-implicit) method, while \cite{hu12} used the backward Euler method.  Among these studies, we have employed the smallest chemical network, but we have demonstrated that our results are equivalent.  The approach of \cite{rimmer16} is somewhat different from this body of work (see \S\ref{subsect:rimmer}).

An obvious opportunity for future work is to include nitrogen, sulphur and phosphorus in our chemical network.  Another missing ingredient is photochemistry.  It would also be insightful to include the effects of condensation in a setting with disequilibrium chemistry. \cite{bs99} have included condensation in their calculations, but these are restricted to being in chemical equilibrium.

In the long term, it will be necessary to couple radiative transfer, chemistry and atmospheric dynamics, since the temperature-pressure profile of the atmosphere changes with the chemistry, because the relative abundances of the molecules alter the opacities, which in turn change the temperature \cite{drummond16}.  Atmospheric dynamics should be properly represented, instead of being crudely approximated by eddy diffusion, which does not apply to situations where the length scale of atmospheric motion exceeds a pressure scale height.  Initial investigations of the coupling of atmospheric dynamics with chemistry have been performed by \cite{cs06}, \cite{burrows10} and \cite{agundez12}, albeit with (severe) approximations taken.  \cite{cs06} used a single rate coefficient to describe the conversion between carbon monoxide and methane.  \cite{burrows10} post-processed the output of general circulation models to study the relative abundance of methane between the dayside and nightside hemispheres of hot Jupiters.  \cite{agundez12} used a simple dynamical model with solid-body rotation to study the effects of a uniform zonal jet on the chemical kinetics; \cite{agundez14} added photochemistry to the model of \cite{agundez12}.  A fully self-consistent calculation is still missing from the literature.

\section*{Acknowledgements}
ST is grateful to Michael Line for enlightening discussions about constructing chemical networks and useful references in the literature, Olivia Venot for helpful communications on model convergence, and Julianne Moses for sharing the data used for model comparisons.  We acknowledge financial and administrative support from the Center for Space and Habitability (CSH), the PlanetS NCCR framework and the Swiss-based MERAC Foundation. 

\software{\\
\texttt{python} \citep{python}, \\
\texttt{scipy} \citep{scipy}, \\
\texttt{numpy} \citep{numpy}, \\
\texttt{matplotlib} \citep{matplotlib}.
}

\appendix

\section{Description of the Rosenbrock (Semi-implicit) Method}
\label{append:euler}

To describe our method requires that we first concisely review what the explicit, implicit and semi-implicit methods are.  Consider the differential equation,
\begin{equation}
\frac{dn}{dt} = f\left(n\right),
\end{equation}
where $n$ is the dependent variable, $t$ is the independent variable and $f(n)$ is an arbitrary function of $n$.  If we discretize this equation and denote the stepsize by $\Delta t$, then explicit differencing (i.e., the forward Euler method) yields
\begin{equation}
n_{k+1} = n_k + \Delta t ~f\left(n_k\right),
\end{equation}
where $k$ denotes the $k$-th index of the discretized independent variable.  Fully-implicit differencing (i.e., the backward Euler method) gives
\begin{equation}
n_{k+1} = n_k + \Delta t ~f\left(n_{k+1}\right).
\end{equation}
Solving the above non-linear equations for $n_{k+1}$ generally involves using a Newton-Raphson-like iteration method, which is the most computationally expensive part in chemical kinetics. An alternative approach is to perform a Taylor expansion of $f$ for the linear term (i.e., the semi-implicit Euler method) \citep{nr},
\begin{equation}
n_{k+1} = n_k + \Delta t \left[ f(n_k) \left. +  \frac{\partial f}{\partial n} \right\vert_{n_k}  (n_{k+1} - n_k)  \right],
\label{eq:semieuler2}
\end{equation}
which may be expressed more compactly as
\begin{equation}
n_{k+1} = n_k + \Delta t  ( I- \Delta t  J )^{-1} ~f\left(n_k\right).
\end{equation}
where $I$ is the identity matrix and 
\begin{equation}
J \equiv \frac{\partial f}{\partial n} \vert_{n_k}
\end{equation}
represents the Jacobian matrix evaluated at $n_k$.  The semi-implicit Euler method is generally stable when dealing with stiff ODEs.  Since it avoids numerical iteration for solving non-linear equations, it is much more efficient than the backward Euler method. 

The $s$-stage Rosenbrock method generalizes equation (\ref{eq:semieuler2}) \citep{Ros63,nr},
\begin{equation} \label{eq:ros}
\begin{split}
n_{k+1} &= n_k + \Delta t \sum^s_{i=1} b_i g_i, \\
g_i &= ~f\left( n_k + \sum^{i-1}_{j=1} \alpha_{ij} g_j \right) + \Delta t  J \sum^i_{j=1} \gamma_{ij} g_j.
\end{split}
\end{equation}
The quantities $b_i$, $\alpha_{ij}$ and $\gamma_{ij}$ are all scalars (i.e., numbers) chosen to meet the desired stability and accuracy, and $s$ is the number of stages specified. It is common to take $\gamma_{ii}$ = $\gamma$  
 for all stages. For example, if $s=1$, $\gamma=1$ and $b_1$ = 1, we recover the semi-implicit Euler method. To minimize the matrix computation, $g_i$ in (\ref{eq:ros}) can be written as
\begin{equation} 
(I - \Delta t \gamma_{ii} J) g_i = ~f\left( n_k + \sum^{i-1}_{j=1} \alpha_{ij} G_j \right) + \Delta t J ~\sum^{i-1}_{j=1} \gamma_{ij} g_j.
\end{equation} 
\texttt{VULCAN} has been implemented with both a second-order ($s=2$) and a fourth-order ($s=6$) Rosenbrock solver. \cite{ver98} recommended the second-order Rosenbrock method for being stable over large stepsizes, which suits the need of chemical kinetics. With the following coefficients, $b_1=1/2$, $b_2=1/2$, $\alpha_{21}=1$, $\gamma_{11}=\gamma_{22}=1+1/\sqrt{2}$ and $\gamma_{21} = -2$, the second-order Rosenbrock method takes the form,
\begin{equation} 
\begin{split}    
n_{k+1} &= n_k + 3/2 \Delta t g_1 + 1/2 \Delta t g_2, \\
(I - \gamma \Delta t J) g_1 &= f(n_k), \\
(I - \gamma \Delta t J) g_2 &= f(n_k + \Delta t k_1) - 2 k_1.  
\end{split}
\end{equation}
One can easily use the first-order solution $n_{k+1}^*$ to estimate the truncation error by
\begin{equation*}
|n_{k+1} - n_{k+1}^*|,
\end{equation*}
where
\begin{equation*}
n_{k+1}^* \equiv n_k + \Delta t k_1,
\end{equation*}
since
\begin{equation*}
\begin{split}   
\hat{n}_{k+1} &=  n_{k+1}^* + O(\Delta t), \\
\hat{n}_{k+1} &=  n_{k+1} + O(\Delta t^2), \\
{\cal E} &=  | n_{k+1} - n_{k+1}^* | = O(\Delta t^2).
\end{split}
\end{equation*}

The truncation error (${\cal E}$) can be used to adjust the stepsize according to 
\begin{equation*}
\Delta t _{k+1} = 0.9 \Delta t _k ({\cal T}/{\cal E})^{0.5},
\end{equation*}
where ${\cal T}$ is the desired relative error tolerance and 0.9 is simply a safety factor. After each timestep, if ${\cal E}$ is greater than ${\cal T}$, the solution is rejected and the stepsize is reduced. After a successful integration, the above scheme is applied to adjust the stepsize.  Typically, we use values of ${\cal T}$ between 0.01 and 0.1.

\begin{figure*}[!h]
\begin{center}
\includegraphics[width=0.49\columnwidth]{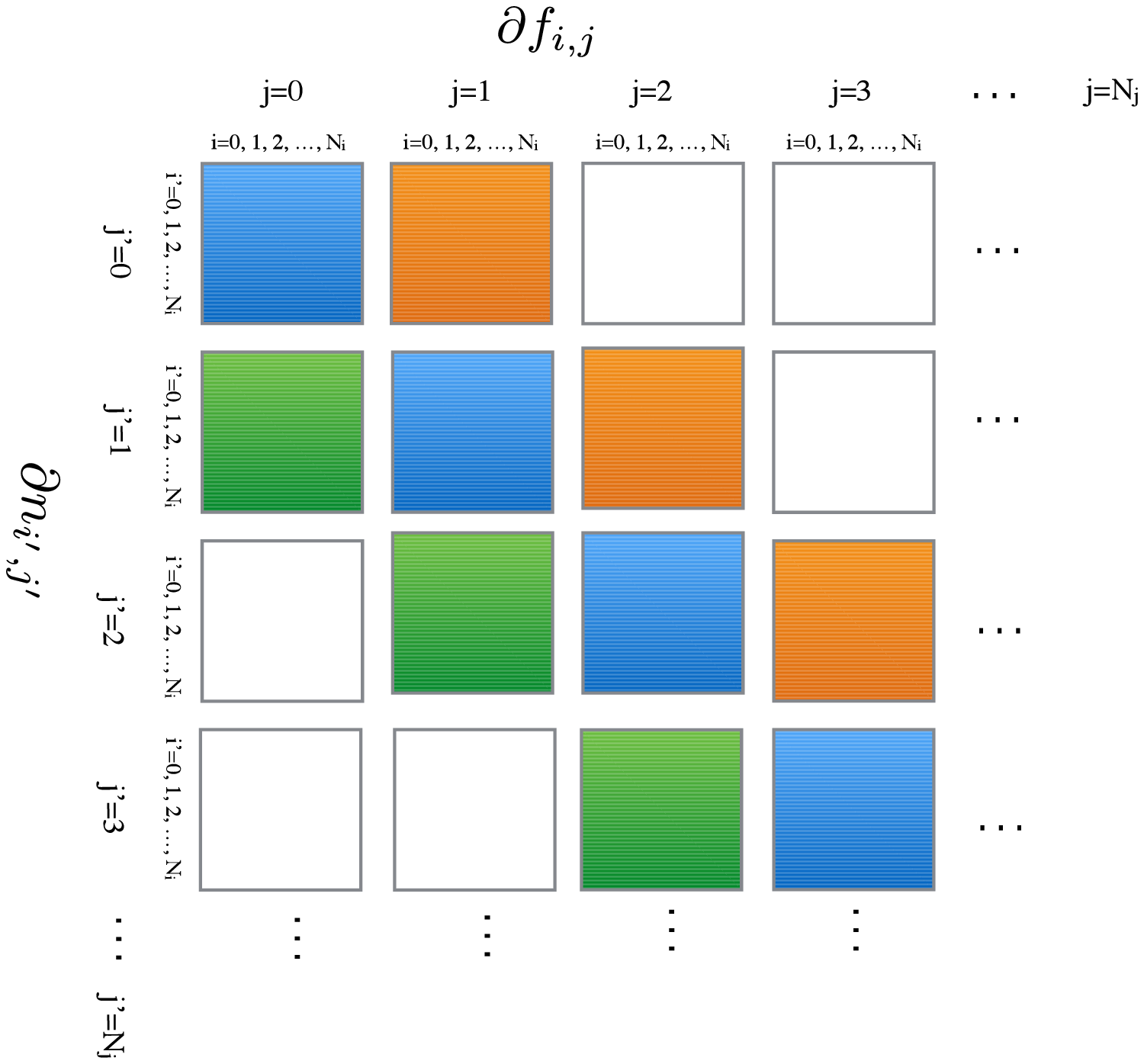}
\end{center}
\caption{Schematic illustration of the Jacobian matrix, which has a ``block tridiagonal" structure.  The diagonal blocks are colored blue and represent the interactions between all of the chemical species in a given layer.  The off-diagonal blocks, which are colored red and green, represent the interactions between the chemical species in a given layer and those located in the layers immediately adjacent to it.  Blank/white squares contain zero values, since there is no coupling between a given atmospheric layer and any layer that is located more than two layers away for our first-order treatment of diffusion.}
\label{fig:jacob}
\end{figure*}

The most computationally expensive part of the above method is evaluating the Jacobian matrix, which has a ``block triadiagonal" structure, as already noted by \cite{hu12}. The entries of the Jacobian matrix are 
\begin{equation} \label{eq:jac}
J_{\alpha \beta} = \frac{\partial f_\alpha}{\partial n_\beta},
\end{equation}
where the indices $\alpha$ and $\beta$ refer to the location of a ``block" or submatrix (Figure \ref{fig:jacob}).  Within each block, $\alpha(i,j)$ and $\beta(i^\prime,j^\prime)$ refer to $f$ for the $i$-th species at the $j$-th level and $n$ for the $i^\prime$-th species at the $j^\prime$-th level, respectively.  Physically, this structure accounts for the interaction of all of the chemical species within a layer and also with those in the other layers.  Since we are employing a first-order discretization of the spatial derivative (the diffusion term), we only need to account for interactions, via diffusion, between a layer and the two layers immediately adjacent to it (i.e., above and below it).  As illustrated in Figure \ref{fig:jacob}, the diagonal blocks account for the interaction between all of the chemical species in a given layer.  If there are $N_i$ species in the chemical network, then each diagonal block has a size of $N_i$ by $N_i$.  The off-diagonal blocks account for interactions between the chemical species located in a layer with those in the adjacent layers and also has a size of $N_i$ by $N_i$.  Other blocks in the Jacobian matrix are empty (contain zeros), because layers that are spaced two layers or more apart do not interact via diffusion in our first-order treatment.  The Jacobian matrix has $N_j$ blocks across each dimension and a size of $N_i N_j$ by $N_i N_j$.  We store the Jacobian matrix as a banded matrix (in order to get rid of most of the zeros; e.g., see Section 2.4 of \citealt{nr}) and handle it using the linear algebra routine \texttt{scipy.linalg.solve\_banded} in \texttt{Python}.

\section{Sensitivity Analysis}
\label{append:sensitivity}

\begin{figure*}[!h]
\begin{center}
\includegraphics[width=0.49\columnwidth]{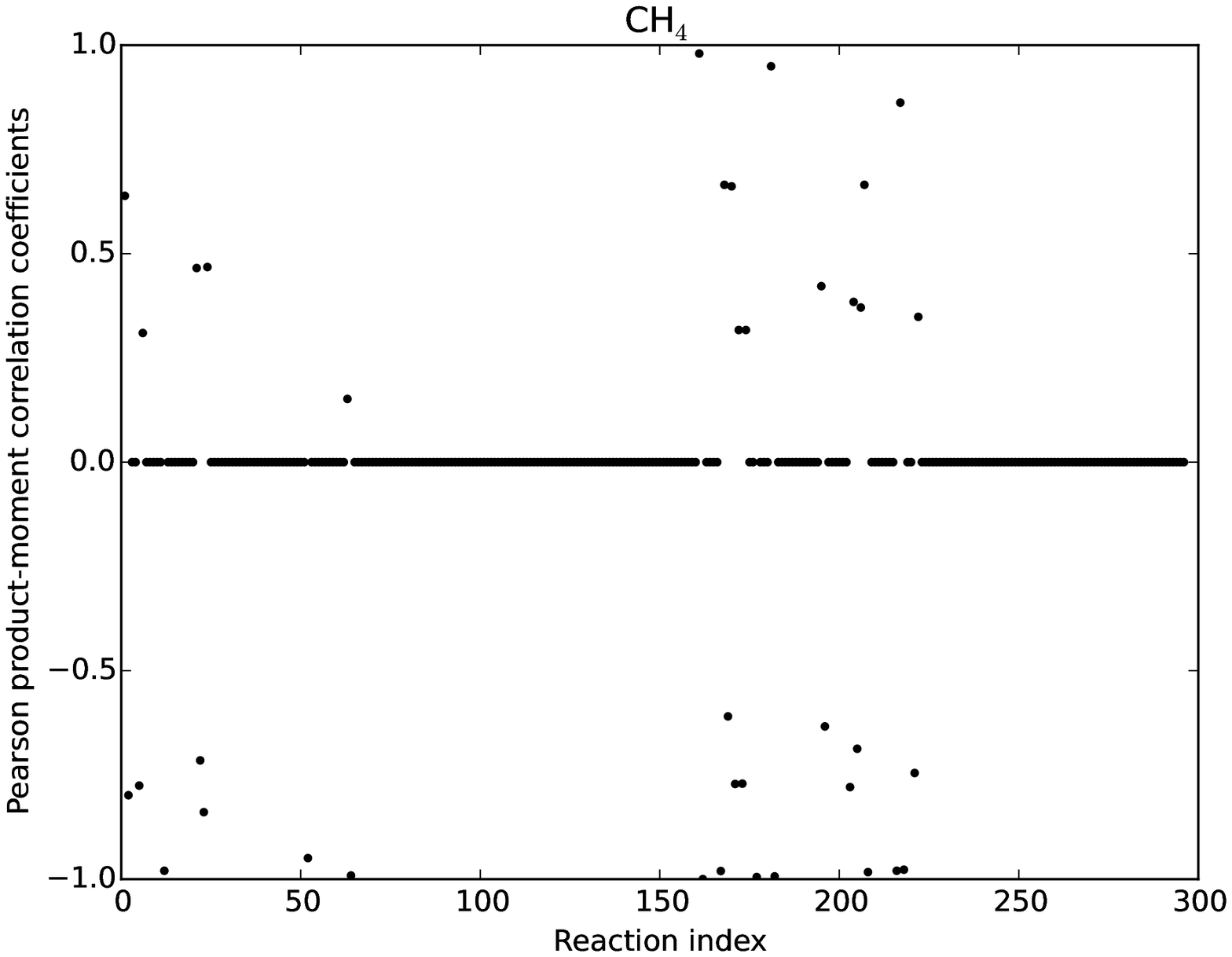}
\includegraphics[width=0.49\columnwidth]{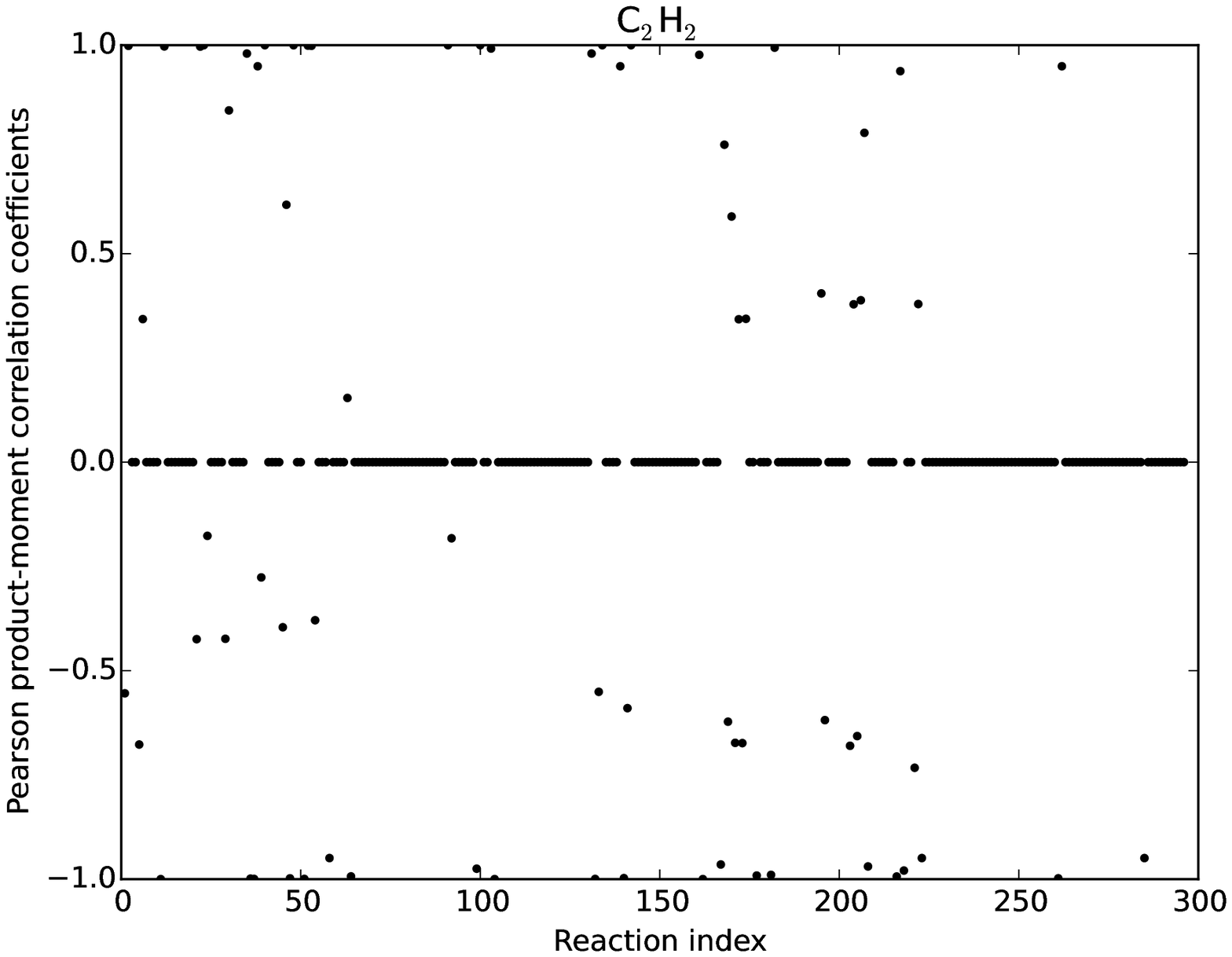}
\includegraphics[width=0.49\columnwidth]{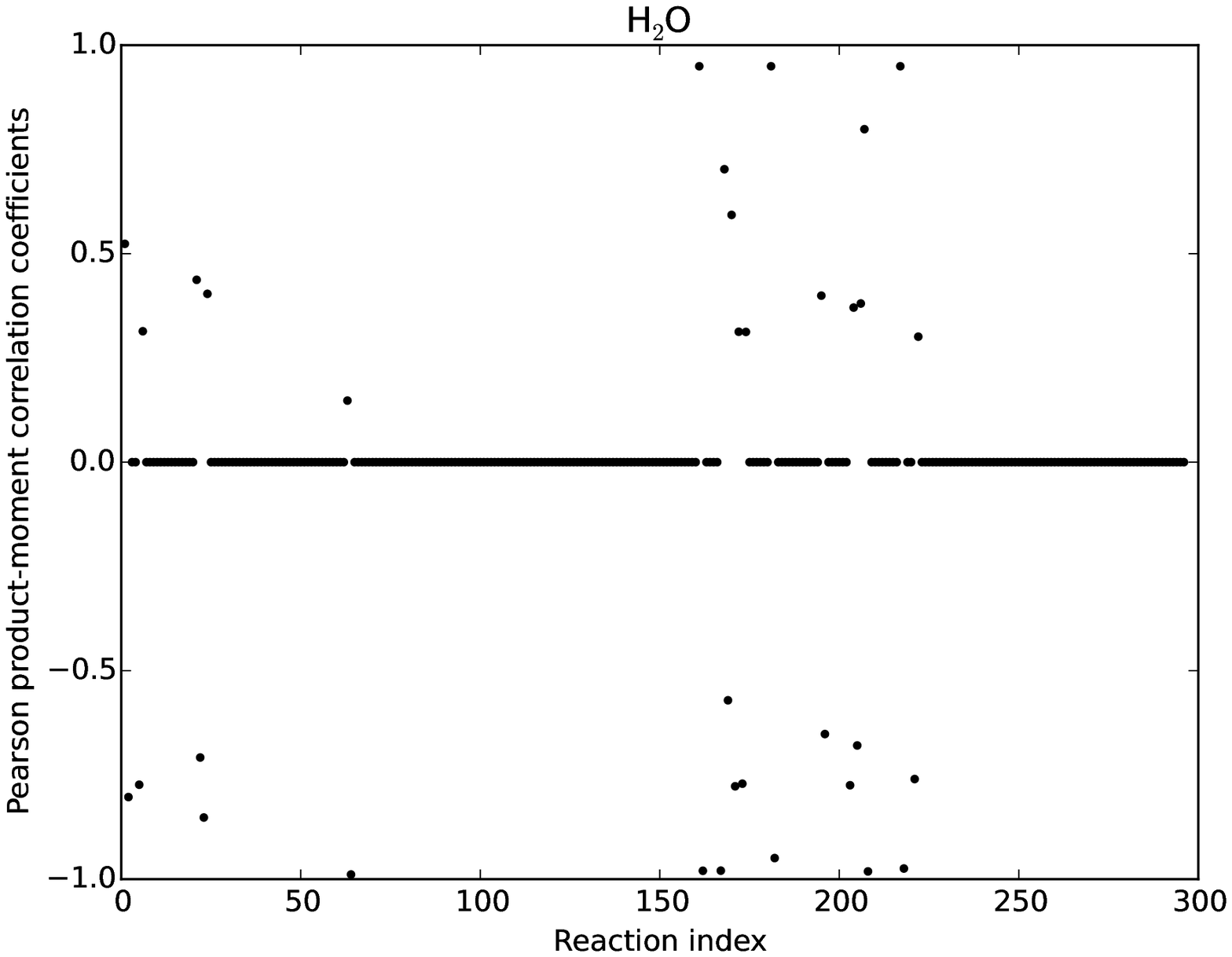}
\includegraphics[width=0.49\columnwidth]{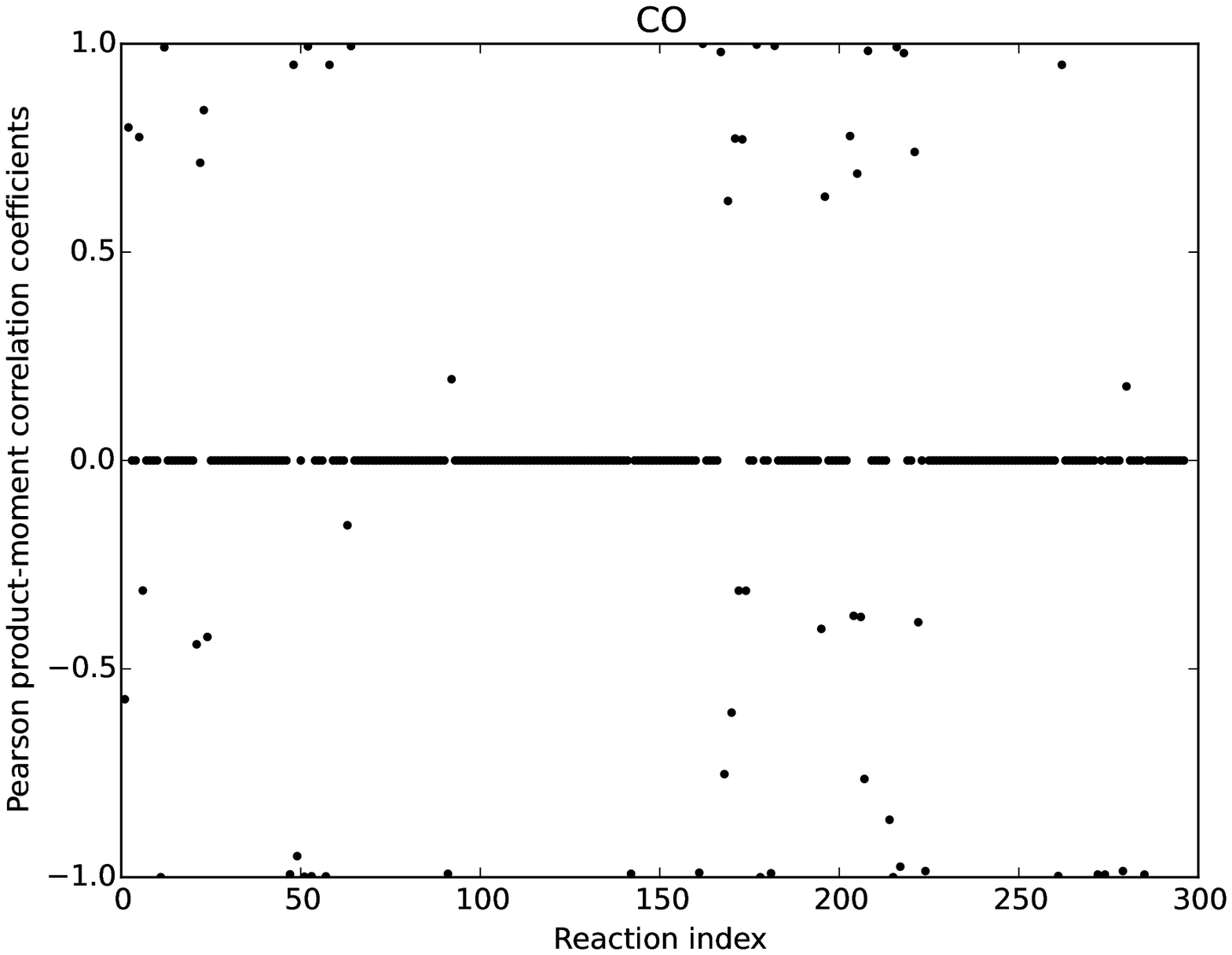}
\end{center}
\caption{Pearson product-moment correlation coefficient between the abundance of a given molecule and methane at its equilibrium value for each reaction in our chemical network.  Values of the coefficient that are markedly non-zero indicate that the abundance of the given molecule is sensitive to a given chemical reaction.}
\label{fig:sensitivity}
\end{figure*}

\begin{table}[!h]
\begin{center}
\caption{Key reactions that determine the abundances of \ce{CH4} at $T=1000$ K and $P=1$ bar (with a Pearson product-moment correlation coefficient greater than 0.75)}
\label{tab:methane}
\begin{tabular}{ll}
\hline
\hline
index & Reaction\\
\hline
R1 & \ce{ H + H2O -> OH + H2 }\\
R13 & \ce{ H + CH4 -> CH3 + H2 }\\
R44 & \ce{ H2O + CO -> H2 + CO2 }\\
R119,120 & \ce{ OH + CH4 <=> H2O + CH3 }\\
R126 & \ce{ OH + CH3 -> CH2OH + H }\\
R127 & \ce{ H2CO + H -> HCO + H2 }\\
R131 & \ce{ H2CO + CH3 -> CH4 + HCO }\\
R135, 136 & \ce{ HCO + H <=> CO + H2 }\\
R150 & \ce{ CH3 + H2O -> CH3OH + H }\\
R157, 158 & \ce{ CH3O + CO <=> CH3 + CO2 }\\
R232 & \ce{ H2 + M -> H + H + M }\\
R236 & \ce{ H2O + M -> OH + H + M }\\
R248 & \ce{ C2H6 + M -> H + C2H5 + M }\\
R270 & \ce{ HCO + M -> H + CO + M }\\
R291 & \ce{ OH + CH3 + M -> CH3OH + M }\\
\hline
\hline
\end{tabular}\\
\end{center}
\end{table}

\begin{table}[!h]
\begin{center}
\caption{Key reactions that determine the abundances of \ce{C2H2} at $T=1000$ K and $P=1$ bar (with a Pearson product-moment correlation coefficient greater than 0.99)}
\label{tab:acetylene}
\begin{tabular}{lll}
\hline
\hline
Index & Reaction\\
\hline
R14 & \ce{ CH3 + H2 -> H + CH4 }\\
R24 & \ce{ C2H2 + H2 -> C2H3 + H }\\
R25 & \ce{ H2  + C2H3 -> H + C2H4 }\\
R31, 32 & \ce{ H + C2H5 <=> 2 CH3 }\\
R36 & \ce{ H + C2H6 -> C2H5 + H2 }\\
R44 & \ce{ H2O + CO -> H2 + CO2 }\\
R70 & \ce{ CH3 + C2H4 -> CH4 + C2H3 }\\
R73, 74 & \ce{ CH3 + C2H6 <=> CH4 + C2H5 }\\
R98 & \ce{ C2H3 + H2O -> C2H4 + OH }\\
R104 & \ce{ C2H5 + H2O -> C2H6 + OH }\\
R120 & \ce{ H2O + CH3 -> OH + CH4 }\\
R131 & \ce{ H2CO + CH3 -> CH4 + HCO }\\
R136 & \ce{ CO + H2 -> HCO + H }\\
R197 & \ce{ HCO + CH3 -> CO + CH4 }\\
R231, 232 & \ce{ H2 + M <=> H + H + M }\\
R235, 236 & \ce{ OH + H + M <=> H2O + M }\\
R240 & \ce{CH4 + M -> H + CH3 + M }\\
R241 & \ce{ C2H2 + H + M -> C2H3 + M }\\
R245 & \ce{ C2H4 + H + M -> C2H5 + M }\\
R247, 248 & \ce{ H + C2H5 + M <=> C2H6 + M }\\
R258 & \ce{ C2H6 + M -> 2 CH3 + M }\\
R261 & \ce{ C2H4 + M -> C2H2 + H2 + M }\\
R263 & \ce{ C2H6 + M -> C2H4 + H2 + M }\\
\hline
\hline
\end{tabular}\\
\end{center}
\end{table}

To apply the quenching approximation, we need a technique for identifying the dominant or key reactions, within our network, that contribute to the production or loss of a molecule.  For each reaction in our network, we vary its rate coefficient and measure the effect on the abundance of a specific molecule.  The influence or importance of each reaction on the molecule in question is quantified via the Pearson product-moment correlation coefficient, also known as ``Pearson's r" \citep{nr}.  

In Figure \ref{fig:sensitivity}, we show our calculations of the Pearson product-moment correlation coefficient for \ce{CH4}, \ce{C2H2}, \ce{H2O} and \ce{CO}.  For each reaction, the Pearson product-moment correlation coefficient is computed from a set of 6 calculations, where each rate coefficient is varied by a factor of $10^{-3}$, $10^{-2}$, $10^{-1}$, 10, $10^2$ and $10^3$.  Positive (negative) values of the Pearson product-moment correlation coefficient mean that the abundance of the molecule in question increases (decreases) when the rate coefficient is increased.

We found this technique to be useful for identifying the key reactions, when we were using the forward and backward rate coefficients and failing to reproduce chemical equilibrium in our initial calculations.  It is also possible to apply this technique when diffusion is present.  In principle, such a technique could also be useful for constructing a reduced chemical network from scratch.  More sophisticated techniques for analyzing and identifying the key reactions may be found in \cite{dob10} and \cite{wakelam10}.

\section{Using a Different Chemical Network for \texttt{VULCAN}}
\label{append:generate}

\texttt{VULCAN} is constructed in a way to allow the user to specify a different chemical network, if desired.  This involves specifying a different set of chemical reactions from the ones we have used in the current study, which also involves writing down their rate coefficients.  The most tedious step is in constructing the Jacobian matrix, described in Appendix \ref{append:euler}.  As part of \texttt{VULCAN}, we have written a complementary code (to the main chemical kinetics solver), also in the \texttt{Python} programming language, that allows a different chemical network to be set up and for its corresponding Jacobian matrix to be constructed symbolically. Hence, as a chemical network is provided, all the relevant numerical functions, including the one for the analytical Jacobian matrix, would be automatically generated to optimize computational efficiency. 

The chemical reactions are read into \texttt{VULCAN} in a user-friendly form,
\begin{equation*}
\begin{split}
&\texttt{Reactions} \qquad\qquad  \texttt{A \quad $b$ \quad E}\\
&\texttt{[ A + B -> C + D ]} \quad  \texttt{X1}	\quad \texttt{X2} \quad \texttt{X3}\\
&\texttt{...}
\end{split}
\end{equation*}
The numbers \texttt{X1}, \texttt{X2}, \texttt{X3} are $A$, $b$, $E$ in the generalized Arrhenius formula in equation (\ref{eq:arr}).  Once the entire list of chemical reactions and rate coefficients are entered, the code allocates the reactions that are relevant for producing and destroying the $i$-th species and constructs the chemical production and loss terms, as well as the diffusion term.  The last step involves using the \texttt{jacobian} routine, from the \texttt{sympy} package, to obtain the algebraical form of the Jacobian matrix related to chemical reactions. Based on this, \texttt{Python} functions of the total Jacobian matrix including diffusion are constructed and can be used for the ODE solver.

\section{Chemical network} 
\label{appendix:rates}

The rate coefficients of bimolecular and termolecular reactions are listed in Table \ref{tab:rates}, where it is understood that $T$ is expressed in units of K. For reactions involving the third body M (termolecular and thermal dissociation reactions), the rate constants depend on the number density of M, or practically the total pressure($k_0$ and $k_{\infty}$ respectively). The rate coefficient at a given pressure is expressed by 
\begin{equation}
k = \frac{k_0 M}{1 + k_0 M/k_{\infty}} F
\end{equation} 
with $F$ being a modification factor to more accurately account for the different collisional efficiency of the different species. A few common choices of $F$ may be found in Appendix C of \cite{venot12}.  Here, we use the Lindemann form, where we simply have $F=1$ \citep{lindemann}.

Following the usual convention, the odd- and even-numbered reactions correspond to the forward and reverse reactions, respectively.  Our reverse rate coefficients are obtained by reversing the forward rate coefficients (see Appendix \ref{append:nasapoly}).

\begin{longtable}[!ht]{|l| p{.26\textwidth} | p{.33\textwidth} | p{.26\textwidth} |}
\caption{Rate Coefficients for the C-H-O Network of \texttt{VULCAN}}
\label{tab:rates}
\endfirsthead
\endhead
\hline
Index  & Reaction & Rate Coefficient & Reference(s) \\
\hline
R1 & \ce{ H + H2O -> OH + H2 } & 7.50 $ \times 10^{-16}$ $ T^{1.600} $  $\exp(-9720.0/T)$ & NIST\footnotemark[8] 1992BAU/COB411-429\\
R3 & \ce{ O + H2 -> OH + H } & 8.52 $ \times 10^{-20}$ $ T^{2.670} $  $\exp(-3160.0/T)$ & NIST 1992BAU/COB411-429\\
R5 & \ce{ O + H2O -> 2OH } & 8.20 $ \times 10^{-14}$ $ T^{0.950} $  $\exp(-8570.0/T)$ & NIST 1991LIF/MIC59-67\\
R7 & \ce{ H + CH -> H2 + C } & 1.31 $ \times 10^{-10}$ $\exp(-80.0/T)$ &NIST 1993HAR/GUA5472-5481\\
R9 & \ce{ H + CH2 -> CH + H2 } & 1.00 $ \times 10^{-11}$ $\exp( 900.0/T)$ & NIST 1992BAU/COB411-429\\
R11 & \ce{ CH2 + H2 -> H + CH3 } & 7.32 $ \times 10^{-19}$ $ T^{2.300} $  $\exp(-3699.0/T)$ & NIST 2010LU/MAT5493-550\\
R13 & \ce{ H + CH4 -> CH3 + H2 } & 2.20 $ \times 10^{-20}$ $ T^{3.000} $  $\exp(-4040.0/T)$ & NIST 1992BAU/COB411-429\\
R15 & \ce{ C + CH -> C2 + H } & 1.05 $ \times 10^{-12}$ $ T^{0.500} $ & NSRDS 67\footnotemark[9]\\
R17 & \ce{ CH3 + C -> H + C2H2 } & 1.00 $ \times 10^{-10}$ & NSRDS 67\\
R19 & \ce{ H2 + C2H -> H + C2H2 } & 9.20 $ \times 10^{-18}$ $ T^{2.170} $  $\exp(-478.0/T)$ & \cite{os96}\\
R21 & \ce{ CH + CH2  -> H + C2H2 } & 6.64 $ \times 10^{-11}$ & \cite{braun81}\\
R23 & \ce{ H + C2H3 -> C2H2 + H2 } & 2.01 $ \times 10^{-11}$ & NIST 1992BAU/COB411-429\\
R25 & \ce{ H2 + C2H3 -> H + C2H4 } & 5.00 $ \times 10^{-20}$ $ T^{2.630} $  $\exp(-4300.0/T)$ & NIST 1986TSA/HAM1087\\
R27 & \ce{ CH + CH4 -> H + C2H4 } & 5.00 $ \times 10^{-11}$ $\exp( 200.0/T)$ & \cite{bau94}\\
R29 & \ce{ CH2 + CH3 -> H + C2H4 } & 7.01 $ \times 10^{-11}$ & NIST 1992BAU/COB411-429\\
R31 & \ce{ H + C2H5 -> 2CH3 } & 5.99 $ \times 10^{-11}$ & NIST 1992BAU/COB411-429\\
R33 & \ce{ H + C2H5 -> C2H4 + H2 } & 3.01 $ \times 10^{-12}$ & NIST 1986TSA/HAM1087\\
R35 & \ce{ H + C2H6 -> C2H5 + H2 } & 9.19 $ \times 10^{-22}$ $ T^{3.500} $  $\exp(-2600.0/T)$ & NIST 1986TSA/HAM1087\\
R37 & \ce{ C2H + C2H2 -> H + C4H2 } & 1.30 $ \times 10^{-10}$ & \cite{vak01}\\
R39 & \ce{ OH + CO -> H + CO2 } & 1.05 $ \times 10^{-17}$ $ T^{1.500} $  $\exp(-259.0/T)$ & NIST 1992BAU/COB411-429\\
R41 & \ce{ CH + CH3 -> H2 + C2H2 } & 1.00 $ \times 10^{-11}$ & NSRDS 67\\
R43 & \ce{ H2 + CO2 -> H2O + CO } & 1.66 $ \times 10^{-15}$ $ T^{0.500} $  $\exp(-7550.0/T)$ & NSRDS 67\\
R45 & \ce{ C2 + O -> C + CO } & 1.05 $ \times 10^{-12}$ & NSRDS 67\\
R47 & \ce{ 2CH2 -> C2H2 + 2H } & 1.80 $ \times 10^{-10}$ $\exp(-400.0/T)$ & \cite{bau92}\\
R49 & \ce{ CH2 + CH2 -> CH + CH3 } & 3.98 $ \times 10^{-10}$ $\exp(-5000.0/T)$ &NSRDS 67\\
R51 & \ce{ CH2 + CH2 -> H + C2H3 } & 3.32 $ \times 10^{-11}$ & NSRDS 67\\
R53 & \ce{ CH2 + CH4 -> 2CH3  } & 4.09 $ \times 10^{-18}$ $ T^{2.000} $  $\exp(-4162.0/T)$ & \cite{bo89}\\
R55 & \ce{ CH2 + C2H5 -> CH3 + C2H4 } & 3.01 $ \times 10^{-11}$ & NIST 1986TSA/HAM1087\\
R57 & \ce{ CH3 + OH -> CH2 + H2O } & 1.85 $ \times 10^{-21}$ $ T^{3.000} $  $\exp(-1400.0/T)$ & NIST 1998WIL/BAL1625-1631\\
R59 & \ce{ C2H + OH -> CH2 + CO  } & 3.01 $ \times 10^{-11}$ & NIST 1986TSA/HAM1087\\
R61 & \ce{ C2H2 + O -> CH2 + CO  } & 6.78 $ \times 10^{-16}$ $ T^{1.500} $  $\exp(-854.0/T)$ & NIST 1987CVE261\\
R63 & \ce{ CH3 + C2H -> C2H2 + CH2 } & 1.00 $ \times 10^{-11}$ & NSRDS 67\\
R65 & \ce{ CH4 + C2H -> CH3 + C2H2 } & 3.01 $ \times 10^{-12}$ $\exp(-250.0/T)$ &NIST 1986TSA/HAM1087\\
R67 & \ce{ CH3 + C2H3 -> CH4 + C2H2 } & 6.51 $ \times 10^{-13}$ & NIST 1986TSA/HAM1087\\
R69 & \ce{ CH4 + C2H3 -> CH3 + C2H4 } & 2.13 $ \times 10^{-24}$ $ T^{4.020} $  $\exp(-2754.0/T)$ & NIST 1986TSA/HAM1087\\
R71 & \ce{ CH3 + C2H5 -> CH4 + C2H4 } & 3.25 $ \times 10^{-11}$ $ T^{-0.500} $ & NIST 1986TSA/HAM1087\\
R73 & \ce{ CH3 + C2H6 -> CH4 + C2H5 } & 9.12 $ \times 10^{-25}$ $ T^{4.000} $  $\exp(-4170.0/T)$ & NIST 1986TSA/HAM1087\\
R75 & \ce{ C2H2 + OH -> CH3 + CO } & 8.04 $ \times 10^{-28}$ $ T^{4.000} $  $\exp(-1010.0/T)$ & NIST 1989MIL/MEL1031-1039\\
R77 & \ce{ C2 + H2 -> H + C2H } & 1.10 $ \times 10^{-10}$ $\exp(-4000.0/T)$ &NIST 1997KRU/ROT2138-2146\\
R79 & \ce{ C2 + CH4 -> CH3 + C2H } & 5.05 $ \times 10^{-11}$ $\exp(-297.0/T)$ & \cite{pitts82}\\
R81 & \ce{ C2H + CH2 -> CH + C2H2 } & 3.01 $ \times 10^{-11}$ & NIST 1986TSA/HAM1087\\
R83 & \ce{ C2H + C2H6 -> C2H2 + C2H5 } & 5.99 $ \times 10^{-12}$ & NIST 1986TSA/HAM1087\\
R85 & \ce{ C2H + O -> CH + CO } & 1.69 $ \times 10^{-11}$ & NIST 1992BAU/COB411-429\\
R87 & \ce{ C2H + OH -> C2H2 + O } & 3.01 $ \times 10^{-11}$ & NIST 1986TSA/HAM1087\\
R89 & \ce{ C2H + H2O -> C2H2 + OH } & 2.20 $ \times 10^{-21}$ $ T^{3.050} $  $\exp(-376.0/T)$ & NIST 2005CAR/NGU114307\\
R91 & \ce{ 2C2H3 -> C2H4 + C2H2 } & 1.60 $ \times 10^{-12}$ & NIST 1986TSA/HAM1087\\
R93 & \ce{ C2H3 + C2H5 -> 2C2H4 } & 8.00 $ \times 10^{-13}$ & NIST 1986TSA/HAM1087\\
R95 & \ce{ C2H3 + C2H5 -> C2H6 + C2H2 } & 8.00 $ \times 10^{-13}$ & NIST 1986TSA/HAM1087\\
R97 & \ce{ C2H4 + OH -> C2H3 + H2O } & 2.60 $ \times 10^{-20}$ $ T^{2.750} $  $\exp(-2100.0/T)$ & NIST 1986TSA/HAM1087\\
R99 & \ce{ 2C2H5 -> C2H4 + C2H6 } & 2.31 $ \times 10^{-12}$ & NIST 1986TSA/HAM1087\\
R101 & \ce{ C2H5 + C2H4 -> C2H3 + C2H6 } & 1.05 $ \times 10^{-21}$ $ T^{3.130} $  $\exp(-9060.0/T)$ & NIST 1986TSA/HAM1087\\
R103 & \ce{ C2H6 + OH -> C2H5 + H2O } & 1.47 $ \times 10^{-14}$ $ T^{1.040} $  $\exp(-913.0/T)$ & NIST 1986TSA/HAM1087\\
R105 & \ce{ CH + O -> OH + C } & 2.52 $ \times 10^{-11}$ $\exp(-2381.0/T)$ & NIST 1986MUR/ROD267\\
R107 & \ce{ O + CH -> H + CO } & 6.59 $ \times 10^{-11}$ & NIST 1992BAU/COB411-429\\
R109 & \ce{ O + CH3 -> CH2 + OH } & 1.00 $ \times 10^{-11}$ $\exp(-3970.0/T)$ &NSRDS 67\\
R111 & \ce{ CH3 + OH -> O + CH4 } & 1.16 $ \times 10^{-19}$ $ T^{2.200} $  $\exp(-2240.0/T)$ & NIST 1991COH/WES1211-1311\\
R113 & \ce{ O + C2H6 -> OH + C2H5 } & 2.00 $ \times 10^{-12}$ $ T^{0.600} $  $\exp(-3680.0/T)$ & NIST 1986TSA/HAM1087\\
R115 & \ce{ OH + C -> CO + H } & 1.05 $ \times 10^{-12}$ $ T^{0.500} $ & NSRDS 67\\
R117 & \ce{ OH + CH2 -> H2O + CH } & 1.43 $ \times 10^{-18}$ $ T^{2.020} $  $\exp(-3410.0/T)$ & NIST  2007JAS/KLI3932-3950\\
R119 & \ce{ OH + CH4 -> H2O + CH3 } & 3.19 $ \times 10^{-19}$ $ T^{2.400} $  $\exp(-1060.0/T)$ & NIST 1986TSA/HAM1087\\
R121 & \ce{ OH + C2H3 -> H2O + C2H2 } & 5.00 $ \times 10^{-11}$ & NIST 1986TSA/HAM1087\\
R123 & \ce{ OH + C2H5 -> H2O + C2H4 } & 4.00 $ \times 10^{-11}$ & NIST 1986TSA/HAM1087\\
R125 & \ce{ CH2OH + H -> OH + CH3 } & 1.60 $ \times 10^{-10}$ & NIST 1987TSA471\\
R127 & \ce{ H2CO + H -> HCO + H2 } & 3.64 $ \times 10^{-16}$ $ T^{1.770} $  $\exp(-1510.0/T)$ & NIST 1986TSA/HAM1087\\
R129 & \ce{ O + C2H4 -> HCO + CH3 } & 2.19 $ \times 10^{-16}$ $ T^{1.550} $  $\exp(-215.0/T)$ & NIST 1986TSA/HAM1087\\
R131 & \ce{ H2CO + CH3 -> CH4 + HCO } & 9.20 $ \times 10^{-21}$ $ T^{2.810} $  $\exp(-2950.0/T)$ & NIST 1986TSA/HAM1087\\
R133 & \ce{ CH3 + CH2OH -> H2CO + CH4 } & 4.00 $ \times 10^{-12}$ & NIST 1987TSA471\\
R135 & \ce{ HCO + H -> CO + H2 } & 1.50 $ \times 10^{-10}$ & NIST 1992BAU/COB411-429\\
R137 & \ce{ HCO + OH -> CO + H2O } & 1.69 $ \times 10^{-10}$ & NIST 1992BAU/COB411-429\\
R139 & \ce{ CO2 + CH -> HCO + CO } & 5.71 $ \times 10^{-12}$ $\exp(-345.1/T)$ & \cite{bau92}\\
R141 & \ce{ CH3 + O -> H2CO + H } & 1.40 $ \times 10^{-10}$ & NIST 1992BAU/COB411-429\\
R143 & \ce{ CH3O + O -> H2CO + OH } & 1.00 $ \times 10^{-11}$ & NIST 1992BAU/COB411-429\\
R145 & \ce{ CH3O + OH -> H2CO + H2O } & 3.01 $ \times 10^{-11}$ & NIST 1992BAU/COB411-429\\
R147 & \ce{ CH3OH + H -> CH3O + H2 } & 6.82 $ \times 10^{-20}$ $ T^{2.685}$ $\exp(-4643.0/T)$ &\cite{moses11}\\
R149 & \ce{ CH3OH + H -> CH3 + H2O } & 4.91 $ \times 10^{-19}$ $ T^{2.485} $  $\exp(-10380.0/T)$ & \cite{moses11}\\
R151 & \ce{ CH2 + O -> CO + H + H } & 1.33 $ \times 10^{-10}$ & NIST 1986FRA422\\
R153 & \ce{ CH2 + OH -> H2CO + H } & 3.01 $ \times 10^{-11}$ & NIST 1986TSA/HAM1087\\
R155 & \ce{ CO2 + CH2 -> H2CO + CO } & 3.90 $ \times 10^{-14}$ & NIST 1986TSA/HAM1087\\
R157 & \ce{ CH3O + CO -> CH3 + CO2 } & 2.61 $ \times 10^{-11}$ $\exp(-5940.0/T)$ &NIST 1986TSA/HAM1087\\
R159 & \ce{ CH3OH + H -> CH2OH + H2 } & 1.09 $ \times 10^{-19}$ $ T^{2.728} $  $\exp(-2240.0/T)$ & \cite{moses11}\\
R161 & \ce{ HCO + C2H -> C2H2 + CO } & 1.00 $ \times 10^{-10}$ & NIST 1986TSA/HAM1087\\
R163 & \ce{ CH2OH + C2H -> H2CO + C2H2 } & 5.99 $ \times 10^{-11}$ & NIST 1987TSA471\\
R165 & \ce{ CH3O + C2H -> H2CO + C2H2 } & 4.00 $ \times 10^{-11}$ & NIST 1986TSA/HAM1087\\
R167 & \ce{ CH3OH + C2H -> CH2OH + C2H2 } & 1.00 $ \times 10^{-11}$ & NIST 1987TSA471\\
R169 & \ce{ CH3OH + C2H -> CH3O + C2H2 } & 2.01 $ \times 10^{-12}$ & NIST 1987TSA471\\
R171 & \ce{ O + C2H3 -> C2H2 + OH } & 1.76 $ \times 10^{-12}$ $ T^{0.200} $  $\exp(-215.2/T)$ & NIST 2005HAR/KLI985-993\\
R173 & \ce{ CH + C2H4 -> C2H2 + CH3 } & 2.23 $ \times 10^{-10}$ $\exp( 173.0/T)$ & NIST 1992BAU/COB411-429\\
R175 & \ce{ CH2 + C2H3 -> C2H2 + CH3 } & 3.00 $ \times 10^{-11}$ & NIST 1986TSA/HAM1087\\
R177 & \ce{ CH3 + C -> C2H2 + H } & 8.31 $ \times 10^{-11}$ & \cite{miller89}\\
R179 & \ce{ O + CH2 -> CO + H2 } & 9.96 $ \times 10^{-11}$ & NIST 1988FRA/BHA885-893\\
R181 & \ce{ O + C2H3 -> HCO + CH2 } & 2.00 $ \times 10^{-11}$ & NIST 1992BAU/COB411-429\\
R183 & \ce{ HCO + CH2 -> CO + CH3 } & 3.01 $ \times 10^{-11}$ & NIST 1986TSA/HAM1087\\
R185 & \ce{ O + C2H4 -> H2CO + CH2 } & 1.35 $ \times 10^{-17}$ $ T^{1.800} $  $\exp(-90.0/T)$ & NIST 1994BAU/COB847-1033\\
R187 & \ce{ CH2OH + CH2 -> OH + C2H4 } & 4.00 $ \times 10^{-11}$ & NIST 1986TSA/HAM1087\\
R189 & \ce{ CH2OH + CH2 -> H2CO + CH3 } & 2.01 $ \times 10^{-12}$ & NIST 1986TSA/HAM1087\\
R191 & \ce{ CH3O + CH2 -> H2CO + CH3 } & 3.00 $ \times 10^{-11}$ & NIST 1986TSA/HAM1087\\
R193 & \ce{ CH3OH + CH2 -> CH3O + CH3 } & 2.39 $ \times 10^{-23}$ $ T^{3.100} $  $\exp(-3490.0/T)$ & NIST 1987TSA471\\
R195 & \ce{ CH3OH + CH2 -> CH2OH + CH3 } & 5.29 $ \times 10^{-23}$ $ T^{3.200} $  $\exp(-3609.0/T)$ & NIST 1986TSA/HAM1087\\
R197 & \ce{ HCO + CH3 -> CO + CH4 } & 2.01 $ \times 10^{-10}$ & NIST 1986TSA/HAM1087\\
R199 & \ce{ CH3O + CH3 -> H2CO + CH4 } & 4.00 $ \times 10^{-11}$ & NIST 1986TSA/HAM1087\\
R201 & \ce{ H2CO + CH -> CO + CH3 } & 8.00 $ \times 10^{-11}$ $\exp( 260.0/T)$ & \cite{bau92}\\
R203 & \ce{ CH3OH + CH3 -> CH3O + CH4 } & 2.39 $ \times 10^{-23}$ $ T^{3.100} $  $\exp(-3490.0/T)$ & NIST 1987TSA471\\
R205 & \ce{ CH3CO + H -> HCO + CH3 } & 3.32 $ \times 10^{-11}$ & NIST 1984WAR197C\\
R207 & \ce{ CH3CO + CH3 -> CO + C2H6 } & 4.90 $ \times 10^{-11}$ & \cite{ada81}\\
R209 & \ce{ O + OH -> O2 + H } & 7.47 $ \times 10^{-10}$ $ T^{-0.500} $  $\exp(-30.0/T)$ & NIST 1986TSA/HAM1087\\
R211 & \ce{ H + CH3O -> H2CO + H2  } & 3.01 $ \times 10^{-11}$ & NIST 1986TSA/HAM1087\\
R213 & \ce{ H + H2CCO -> CO + CH3  } & 1.29 $ \times 10^{-15}$ $ T^{1.450} $  $\exp(-1399.0/T)$ & NIST 2006SEN/KLI5772-5781\\
R215 & \ce{ O + C2H3 -> H2CCO + H  } & 1.60 $ \times 10^{-10}$ & NIST 1986TSA/HAM1087\\
R217 & \ce{ C2H2 + O -> HCCO + H  } & 1.50 $ \times 10^{-11}$ $\exp(-2280.0/T)$ &NIST 1986TSA/HAM1087\\
R219 & \ce{ HCCO + H -> CO + CH2  } & 2.49 $ \times 10^{-10}$ & NIST 1988FRA/BHA885-893\\
R221 & \ce{ O + H2CO -> HCO + OH } & 6.85 $ \times 10^{-13}$ $ T^{0.570} $  $\exp(-1390.0/T)$ & \cite{bau92}\\
R223 & \ce{ HCO + HCO -> H2CO + CO } & 3.01 $ \times 10^{-11}$ & NIST 1987TSA471\\
R225 & \ce{ OH + CH3O -> H2O + H2CO } & 3.00 $ \times 10^{-11}$ & NIST 1986TSA/HAM1087\\
R227 & \ce{ CH2OH + CH3O -> H2CO + CH3OH } & 4.00 $ \times 10^{-11}$ & NIST 1987TSA471\\
R229 & \ce{ CH3O + CH3O -> H2CO + CH3OH } & 1.00 $ \times 10^{-10}$ & NIST 1986TSA/HAM1087\\
R231 & \ce{ H + H  + M -> H2 + M } & $k_0$  = 2.70 $ \times 10^{-31}$ $ T^{-0.600} $ \newline $k_{\infty}$ = 3.31 $ \times 10^{-06}$ $T^{-0.600} $ & NIST 1992BAU/COB411-429 \newline NIST \\
R233 & \ce{ H + O + M -> OH + M } &  $k_0$  = 1.30 $ \times 10^{-29}$ $ T^{-1.000} $ \newline $k_{\infty}$ = 1.00 $ \times 10^{-11}$  &  NIST 1986TSA/HAM1087\newline \cite{moses11}\\
R235 & \ce{ OH + H + M -> H2O + M } &  $k_0$  = 3.89 $ \times 10^{-25}$ $ T^{-2.000} $ \newline $k_{\infty}$ = 4.26 $ \times 10^{-11}$ $ T^{-0.230} $ &   NIST 1992BAU/COB411-429 \newline NIST 2008SEL/GEO5085-5095\\
R237 & \ce{ H + CH + M -> CH2 + M } &  $k_0$  = 2.76 $ \times 10^{-29}$ $ T^{-1.000} $ \newline $k_{\infty}$ = 1.00 $ \times 10^{-12}$ &   NSRDS 67 \newline Estimated from R253\\
R239 & \ce{ H + CH3 + M -> CH4 + M } &  $k_0$  = 1.76 $ \times 10^{-24}$ $ T^{-1.800} $ \newline $k_{\infty}$ = 2.06 $ \times 10^{-10}$ $ T^{-0.400} $ &   NIST 1992BAU/COB411-429 \newline NIST 1986TSA/HAM1087\\
R241 & \ce{ H + C2H2 + M -> C2H3 + M } &  $k_0$  = 1.05 $ \times 10^{-07}$ $ T^{-7.270} $  $\exp(-3630.0/T)$ \newline $k_{\infty}$ = 9.13 $ \times 10^{-12}$ $\exp(-1220.0/T)$&  NIST 1986TSA/HAM1087 \newline NIST 1984WAR197C\\
R243 & \ce{ H + C2H3 + M -> C2H4 + M } &  $k_0$  = 1.50 $ \times 10^{-11}$ \newline $k_{\infty}$ = 6.40 $ \times 10^{-11}$ $ T^{0.200} $& \cite{far91}\newline \cite{harding05}\\
R245 & \ce{ H + C2H4 + M -> C2H5 + M } &  $k_0$  = 7.69 $ \times 10^{-30}$ $\exp(-380.0/T)$ \newline $k_{\infty}$ = 1.27 $ \times 10^{-15}$ $ T^{1.490}$ $\exp(-499.0/T)$  &   NIST 1994BAU/COB847-1033 \newline NIST 1986TSA/HAM1087\\
R247 & \ce{ H + C2H5 + M -> C2H6 + M } &  $k_0$  = 4.00 $\times 10^{-19}$ $T^{-3.000}$ $\exp(-600.0/T)$ \newline $k_{\infty}$ = 9.04 $\times 10^{-11}$ $T^{0.160}$ & \cite{moses05} \newline NIST 2005HAR/GEO4646-4656\\
R249 & \ce{ H2 + C + M -> CH2 + M } &  $k_0$  = 6.89 $ \times 10^{-11}$ \newline $k_{\infty}$ = 2.06 $ \times 10^{-11}$ $\exp(-57.0/T)$ &  NIST 1989FUJ/SAG5474-5478 \newline 1993HAR/GUA5472-5481 \\
R251 & \ce{ CH + M -> C + H + M } &  $k_0$  = 3.16 $ \times 10^{-10}$ $\exp(-33700.0/T)$ \newline $k_{\infty}$ = 1.00 $ \times 10^{-12}$ &   NIST 1992DEA/HAN517-532 \newline Estimated from R253\\
R253 & \ce{ CH2 + H + M ->  CH3 + M } &  $k_0$  = 9.00 $ \times 10^{-32}$ $\exp( 550.0/T)$ \newline $k_{\infty}$ = 8.55 $ \times 10^{-12}$ $ T^{0.150} $&  Estimate from R255 \newline  Estimate from R255\\
R255 & \ce{ CH + H2 + M -> CH3 + M } &  $k_0$  = 3.40 $ \times 10^{-31}$ $\exp( 736.0/T)$ \newline $k_{\infty}$ = 7.3 $ \times 10^{-11}$ &   \cite{moses00a}\newline \cite{moses00a}\\
R257 & \ce{ 2CH3 + M -> C2H6 + M } &  $k_0$  = 3.50 $ \times 10^{-07}$ $ T^{-7.000} $  $\exp(-1390.0/T)$ \newline $k_{\infty}$ = 1.58 $ \times 10^{-09}$ $ T^{-0.540} $  $\exp(-68.0/T)$&  NIST 1994BAU/COB847-1033\newline NIST 2006KLI/GEO1133-1147 \\
R259 & \ce{ C2H + H + M -> C2H2 + M } &  $k_0$  = 1.26 $ \times 10^{-18}$ $ T^{-3.100} $  $\exp(-721.0/T)$ \newline $k_{\infty}$ = 3.00 $ \times 10^{-10}$ &  NIST 1986TSA/HAM1087 \newline NIST 1986TSA/HAM1087\\
R261 & \ce{ C2H4 + M -> C2H2 + H2 + M } &  $k_0$  = 5.8 $ \times 10^{-08}$ $\exp(-36000/T)$ \newline $k_{\infty}$ = 7.95 $ \times 10^{12}$ $ T^{0.440} $  $\exp(-44700.0/T)$&   NIST 1986TSA/HAM1087 \newline NIST 1986TSA/HAM1087\\
R263 & \ce{ C2H6 + M -> C2H4 + H2 + M } &  $k_0$  = 3.80 $ \times 10^{-07}$ $\exp(-34000.0/T)$ \newline $k_{\infty}$ = 1.32 $ \times 10^{+15}$ $\exp(-36800.0/T)$&   NIST 1985SCH/KLO88\newline NIST 1931MAR/MCC878-881\\
R265 & \ce{ CO + O + M -> CO2 + M } &  $k_0$  = 1.70 $ \times 10^{-33}$ $\exp(-1510.0/T)$ \newline $k_{\infty}$ = 2.66 $ \times 10^{-14}$ $\exp(-1459.0/T)$&   NIST 1986TSA/HAM1087\newline \cite{simon72}\\
R267 & \ce{ CH2OH + M -> H + H2CO + M  } &  $k_0$  = 1.66 $ \times 10^{-10}$ $\exp(-12630.0/T)$ \newline $k_{\infty}$ = 3.00 $ \times 10^{+09}$ $\exp(-14600.0/T)$ &  NIST 1992CRI/DOV169-185\newline NIST 1975BOW343\\
R269 & \ce{ H + CO + M -> HCO + M } &  $k_0$  = 5.29 $ \times 10^{-34}$ $\exp(-370.0/T)$ \newline $k_{\infty}$ = 1.96 $ \times 10^{-13}$ $\exp(-1360.0/T)$&   NIST 1994BAU/COB847-1033\newline \cite{arai81}\\
R271 & \ce{ H2O + CH + M -> CH2OH + M } &  $k_0$  = 1.00 $ \times 10^{-31}$ \newline $k_{\infty}$ = 9.48 $ \times 10^{-12}$ $\exp(380.0/T)$&  \cite{moses00b}\newline \cite{zab88}\\
R273 & \ce{ CH3O + M -> H + H2CO + M } &  $k_0$  = 9.00 $ \times 10^{-11}$ $\exp(-6790.0/T)$ \newline $k_{\infty}$ = 1.56$\times 10^{15}$ $T^{-0.390}$ $\exp(-13300.0/T)$&   NIST 1986TSA/HAM1087 \newline NIST 2006CUR250-275\\
R275 & \ce{ CH2OH + H + M -> CH3OH + M } &  $k_0$  = 1.20 $\times 10^{-16}$ $T^{-4.650}$  $\exp(-2557.1/T)$ \newline $k_{\infty}$ = 2.30 $\times 10^{-10}$ $T^{0.040}$ &  \cite{combustion} \newline NIST  2007JAS/KLI3932-3950\\
R277 & \ce{ OH + C2H2 + M -> CH3CO + M } &  $k_0$  = 4.99 $ \times 10^{-25}$ $ T^{-2.000} $ \newline $k_{\infty}$ = 1.06 $ \times 10^{-07}$ $ T^{-1.900} $&   \cite{bau92}\newline \cite{bau92}\\
R279 & \ce{ CO + CH3 + M -> CH3CO + M } &  $k_0$  = 3.95 $ \times 10^{-10}$ $ T^{-7.500} $  $\exp(-5490.0/T)$ \newline $k_{\infty}$ = 5.14 $ \times 10^{-19}$ $ T^{2.200} $  $\exp(-3030.0/T)$&  NIST 1986TSA/HAM1087\newline \cite{huy08}\\
R281 & \ce{ HCO + H + M -> H2CO + M  } &  $k_0$  = 7.33 $ \times 10^{-24}$ $ T^{-2.570} $  $\exp(-215.0/T)$ \newline $k_{\infty}$ = 7.77 $ \times 10^{-14}$ $\exp(2280.0/T)$& NIST 1998EIT/YU5196-5205 \newline NIST 1981TSU/KAT985\\
R283 & \ce{ CO + H2 + M -> H2CO + M} &  $k_0$ = 2.80$\times 10^{-20}$ $ T^{-3.420} $ $\exp(-42445.0/T)$ \newline $k_{\infty}$ = 7.14 $\times 10^{-17}$ $T^{1.500}$ $\exp(-40055.0/T)$ & \cite{wf97}\newline \cite{wf97}\\
R285 & \ce{ C + C + M -> C2 + M } & $k_0$ = 4.97 $ \times 10^{-27}$ $ T^{-1.600} $ & Estimated \\
R287 & \ce{ C2H + M -> C2 + H + M } &  $k_0$  = 2.92 $ \times 10^{+11}$ $ T^{-5.160} $  $\exp(-57400.0/T)$ \newline $k_{\infty}$ = 1.00 $ \times 10^{-12}$ &  NIST 1997KRU/ROT2138-2146 \newline Estimated \\
R289 & \ce{ O + C + M -> CO + M } & $k_0$ = 9.10 $ \times 10^{-22}$ $ T^{-3.100} $  $\exp(-2114.0/T)$ & NSRDS 67\\
R291 & \ce{ OH + CH3 + M -> CH3OH + M } & $k_0$  = 1.932 $\times 10^{3}$ $T^{-9.880}$ $\exp(-7544.0/T)$ + \newline 5.109$\times 10^{-11}$ $T^{-6.25}$ $\exp(-1433.0/T)$ \newline  $k_{\infty}$ = 1.031 $\times 10^{-10}$ $T^{-0.018}$  $\exp(16.74/T)$ &\cite{moses11}\newline \cite{moses11}\\

\hline
\end{longtable}
\footnotetext[8]{Shown by Squib in the NIST database}
\footnotetext[9]{http://www.nist.gov/data/nsrds/NSRDS-NBS67.pdf}

\section{Thermodynamics Data from the NASA Polynomials}
\label{append:nasapoly}

The ``dimensional" equilibrium constant is used to reverse the forward reaction rate (e.g., \citealt{hlt16})
\begin{equation} 
\frac{k_f}{k_r} = K_{\rm eq} \left(\frac{k_B T}{P_0}\right)^{\Delta \mu}
\label{eq:rev1}
\end{equation}
with $k_f$ and $k_r$ being the forward and reverse rate constant, $k_B$ being the Boltzmann constant, $\Delta \mu$ being the difference between the forward and reverse stoichiometric coefficients, $P_0$ being the standard-state pressure (1 bar for ideal gas). The equilibrium constant can be expressed by the standard Gibbs free as
\begin{equation}
K_{\rm eq} = \mbox{exp} \left(- \frac{\Delta G^0}{{\cal R}T} \right) =  \exp \left(- \frac{\Delta H^0 - T \Delta s^0}{{\cal R}T} \right),
\label{eq:rev2}
\end{equation}
where $\Delta G^0$ is the change of standard Gibbs free energy from reactants to products, 
\begin{equation} 
\Delta G^0 = \Delta G[\mbox{products}] - \Delta G[\mbox{reactants}],
\end{equation}
with $\Delta H^0$ and $\Delta s^0$ being the changes in standard enthalpy and standard entropy, respectively, defined in the same way.  Equations (\ref{eq:rev1}) and (\ref{eq:rev2}) allow us to calculate the reverse rate coefficient,
\begin{equation} \label{eq:rev3}
k_r = \frac{k_f}{\exp \left[- \left( \Delta H^0 - T \Delta s^0 \right)/{\cal R}T \right]} \left( \frac{k_B T}{P_0} \right)^{-\Delta \mu}.
\end{equation}
The standard enthalpy and standard entropy of each species are expressed in terms of the NASA polynomials, which we archive in Table \ref{tab:nasa9}.  For each species, the first row is for $T=200$--1000 K.  The second row is for $T=1000$--6000 K.  The enthalpy and entropy are given by
\begin{equation}
\frac{H^0}{{\cal R} T} = -a_1 T^{-2} + a_2 \frac{\ln T}{T}  + a_3 +
\frac{a_4}{2} T + \frac{a_5}{3} T^2 + \frac{a_6}{4} T^3 +
\frac{a_7}{5} T^4 + \frac{a_8}{T},
\end{equation}
\begin{equation}
\frac{s^0}{{\cal R}} = - \frac{a_1}{2} T^{-2} - a_2 T^{-1}  + a_3 \ln T +
a_4 T + \frac{a_5}{2} T^2 + \frac{a_6}{3} T^3 + \frac{a_7}{4} T^4 +
a_9.
\end{equation}
For completeness, we note that
\begin{equation}
\frac{c_p^0}{{\cal R}} = a_1 T^{-2} + a_2 T^{-1} + a_3 + a_4 T + a_5 T^2
+ a_6 T^3 + a_7 T^4.
\end{equation}

\begin{table}[!ht]
\caption{Thermodynamics data from the NASA polynomials}
\label{tab:nasa9}
\begin{tiny}
\begin{tabular}{|c|c|c|c|c|c|c|c|c|c|}
\hline
 & $a_1$ & $a_2$ & $a_3$ & $a_4$ & $a_5$ & $a_6$ & $a_7$ & $a_8$ & $a_9$ \\
\hline
\ce{H} & 0.00000000E+00 & 0.00000000E+00 & 2.50000000E+00 &
0.00000000E+00 & 0.00000000E+00 & 0.00000000E+00 & 0.00000000E+00 &
2.54737080E+04 & -4.46682853E-01 \\
\hline
 & 6.07877425E+01 & -1.81935442E-01 & 2.50021182E+00 & -1.22651286E-07
& 3.73287633E-11 & -5.68774456E-15 & 3.41021020E-19 & 2.54748640E+04 &
-4.48191777E-01 \\
\hline
\ce{H2} & 4.07832321E+04 & -8.00918604E+02 & 8.21470201E+00 &
-1.26971446E-02 & 1.75360508E-05 & -1.20286027E-08 & 3.36809349E-12 &
2.68248466E+03 & -3.04378884E+01 \\
\hline
 & 5.60812801E+05 & -8.37150474E+02 & 2.97536453E+00 & 1.25224912E-03
& -3.74071619E-07 & 5.93662520E-11 & -3.60699410E-15 & 5.33982441E+03
& -2.20277477E+00 \\
\hline
\ce{O} & -7.95361130E+03 & 1.60717779E+02 & 1.96622644E+00 &
1.01367031E-03 & -1.11041542E-06 & 6.51750750E-10 & -1.58477925E-13 &
2.84036244E+04 & 8.40424182E+00 \\
\hline
 & 2.61902026E+05 & -7.29872203E+02 & 3.31717727E+00 & -4.28133436E-04
& 1.03610459E-07 & -9.43830433E-12 & 2.72503830E-16 & 3.39242806E+04 &
-6.67958535E-01 \\
\hline
\ce{OH} & -1.99885899E+03 & 9.30013616E+01 & 3.05085423E+00 &
1.52952929E-03 & -3.15789100E-06 & 3.31544618E-09 & -1.13876268E-12 &
2.99121423E+03 & 4.67411079E+00 \\
\hline
 & 1.01739338E+06 & -2.50995728E+03 & 5.11654786E+00 & 1.30529993E-04
& -8.28432226E-08 & 2.00647594E-11 & -1.55699366E-15 & 2.01964021E+04
& -1.10128234E+01 \\
\hline
\ce{H2O} & -3.94796083E+04 & 5.75573102E+02 & 9.31782653E-01 &
7.22271286E-03 & -7.34255737E-06 & 4.95504349E-09 & -1.33693325E-12 &
-3.30397431E+04 & 1.72420578E+01 \\
\hline
 & 1.03497210E+06 & -2.41269856E+03 & 4.64611078E+00 & 2.29199831E-03
& -6.83683048E-07 & 9.42646893E-11 & -4.82238053E-15 & -1.38428651E+04
& -7.97814851E+00 \\
\hline
\ce{CH} & 2.22059013E+04 & -3.40541153E+02 & 5.53145229E+00 &
-5.79496426E-03 & 7.96955488E-06 & -4.46591159E-09 & 9.59633832E-13 &
7.24078327E+04 & -9.10767305E+00 \\
\hline
 & 2.06076344E+06 & -5.39620666E+03 & 7.85629385E+00 & -7.96590745E-04
& 1.76430830E-07 & -1.97638627E-11 & 5.03042951E-16 & 1.06223659E+05 &
-3.15475744E+01 \\
\hline
\ce{C} & 6.49503147E+02 & -9.64901086E-01 & 2.50467548E+00 &
-1.28144803E-05 & 1.98013365E-08 & -1.60614403E-11 & 5.31448341E-15 &
8.54576311E+04 & 4.74792429E+00 \\
\hline
 & -1.28913647E+05 & 1.71952857E+02 & 2.64604439E+00 & -3.35306895E-04
& 1.74209274E-07 & -2.90281783E-11 & 1.64218238E-15 & 8.41059785E+04 &
4.13004742E+00 \\
\hline
\ce{CH2} & 3.21892173E+04 & -2.87760181E+02 & 4.20358382E+00 &
3.45540596E-03 & -6.74619334E-06 & 7.65457164E-09 & -2.87032842E-12 &
4.73362471E+04 & -2.14362860E+00 \\
\hline
 & 2.55041803E+06 & -7.97162539E+03 & 1.22892449E+01 & -1.69912292E-03
& 2.99172860E-07 & -2.76700749E-11 & 1.05134174E-15 & 9.64221689E+04 &
-6.09473991E+01 \\
\hline
\ce{CH3} & -2.87618881E+04 & 5.09326866E+02 & 2.00214395E-01 &
1.36360583E-02 & -1.43398935E-05 & 1.01355673E-08 & -3.02733194E-12 &
1.40827182E+04 & 2.02277279E+01 \\
\hline
 & 2.76080266E+06 & -9.33653117E+03 & 1.48772961E+01 & -1.43942977E-03
& 2.44447795E-07 & -2.22455578E-11 & 8.39506576E-16 & 7.48180948E+04 &
-7.91968240E+01 \\
\hline
\ce{CH4} & -1.76685100E+05 & 2.78618102E+03 & -1.20257785E+01 &
3.91761929E-02 & -3.61905443E-05 & 2.02685304E-08 & -4.97670549E-12 &
-2.33131436E+04 & 8.90432275E+01 \\
\hline
 & 3.73004276E+06 & -1.38350148E+04 & 2.04910709E+01 & -1.96197476E-03
& 4.72731304E-07 & -3.72881469E-11 & 1.62373721E-15 & 7.53206691E+04 &
-1.21912489E+02 \\
\hline
\ce{C2} & 5.55963451E+05 & -9.98012644E+03 & 6.68162037E+01 &
-1.74343272E-01 & 2.44852305E-04 & -1.70346758E-07 & 4.68452773E-11 &
1.44586963E+05 & -3.44822970E+02 \\
\hline
 & -9.68926793E+05 & 3.56109299E+03 & -5.06413893E-01 & 2.94515488E-03
& -7.13944119E-07 & 8.67065725E-11 & -4.07690681E-15 & 7.68179683E+04
& 3.33998524E+01 \\
\hline
\ce{C2H2} & 1.59811209E+05 & -2.21664412E+03 & 1.26570781E+01 &
-7.97965108E-03 & 8.05499275E-06 & -2.43330767E-09 & -7.52923318E-14 &
3.71261906E+04 & -5.24433890E+01 \\
\hline
 & 1.71384741E+06 & -5.92910666E+03 & 1.23612794E+01 & 1.31418699E-04
& -1.36276443E-07 & 2.71265579E-11 & -1.30206620E-15 & 6.26657897E+04
& -5.81896059E+01 \\
\hline
\ce{C2H3} & -3.34789687E+04 & 1.06410410E+03 & -6.40385706E+00 &
3.93451548E-02 & -4.76004609E-05 & 3.17007135E-08 & -8.63340643E-12 &
3.03912265E+04 & 5.80922618E+01 \\
\hline
 & 2.71808009E+06 & -1.03095683E+04 & 1.83657981E+01 & -1.58013115E-03
& 2.68059494E-07 & -2.43900400E-11 & 9.20909639E-16 & 9.76505559E+04 &
-9.76008686E+01 \\
\hline
\ce{C2H} & 1.34366949E+04 & -5.06797072E+02 & 7.77210741E+00 &
-6.51233982E-03 & 1.03011785E-05 & -5.88014767E-09 & 1.22690186E-12 &
6.89226999E+04 & -1.87188163E+01 \\
\hline
 & 3.92233457E+06 & -1.20475170E+04 & 1.75617292E+01 & -3.65544294E-03
& 6.98768543E-07 & -6.82516201E-11 & 2.71926279E-15 & 1.43326663E+05 &
-9.56163438E+01 \\
\hline
\ce{C2H4} & -1.16360584E+05 & 2.55485151E+03 & -1.60974643E+01 &
6.62577932E-02 & -7.88508186E-05 & 5.12522482E-08 & -1.37034003E-11 &
-6.17619107E+03 & 1.09333834E+02 \\
\hline
 & 3.40876367E+06 & -1.37484790E+04 & 2.36589807E+01 & -2.42380442E-03
& 4.43139566E-07 & -4.35268339E-11 & 1.77541063E-15 & 8.82042938E+04 &
-1.37127811E+02 \\
\hline
\ce{C2H5} & -1.41131255E+05 & 2.71428509E+03 & -1.53497773E+01 &
6.45167258E-02 & -7.25914396E-05 & 4.59911601E-08 & -1.21836754E-11 &
5.98141884E+02 & 1.09096652E+02 \\
\hline
 & 4.16922040E+06 & -1.66298214E+04 & 2.79544213E+01 & -3.05171576E-03
& 5.68516004E-07 & -5.68286360E-11 & 2.35564856E-15 & 1.13701009E+05 &
-1.63935800E+02 \\
\hline
\ce{C2H6} & -1.86204416E+05 & 3.40619186E+03 & -1.95170509E+01 &
7.56583559E-02 & -8.20417322E-05 & 5.06113580E-08 & -1.31928199E-11 &
-2.70293289E+04 & 1.29814050E+02 \\
\hline
 & 5.02578213E+06 & -2.03302240E+04 & 3.32255293E+01 & -3.83670341E-03
& 7.23840586E-07 & -7.31918250E-11 & 3.06546870E-15 & 1.11596395E+05 &
-2.03941058E+02 \\
\hline
\ce{C4H2} & 2.46754257E+05 & -3.89785564E+03 & 2.36608046E+01 &
-2.20807780E-02 & 2.78110114E-05 & -1.57734001E-08 & 3.42316546E-12 &
7.08690782E+04 & -1.10917356E+02 \\
\hline
 & 2.32817991E+06 & -8.92518609E+03 & 2.11432688E+01 & -1.36887128E-03
& 2.32750316E-07 & -2.12451762E-11 & 8.05331302E-16 & 1.05778842E+05 &
-1.08831357E+02 \\
\hline
\ce{CO} & 1.48904533E+04 & -2.92228594E+02 & 5.72452717E+00 &
-8.17623503E-03 & 1.45690347E-05 & -1.08774630E-08 & 3.02794183E-12 &
-1.30313188E+04 & -7.85924135E+00 \\
\hline
 & 4.61919725E+05 & -1.94470486E+03 & 5.91671418E+00 & -5.66428283E-04
& 1.39881454E-07 & -1.78768036E-11 & 9.62093557E-16 & -2.46626108E+03
& -1.38741311E+01 \\
\hline
\ce{CO2} & 4.94365054E+04 & -6.26411601E+02 & 5.30172524E+00 &
2.50381382E-03 & -2.12730873E-07 & -7.68998878E-10 & 2.84967780E-13 &
-4.52819846E+04 & -7.04827944E+00 \\
\hline
 & 1.17696242E+05 & -1.78879148E+03 & 8.29152319E+00 & -9.22315678E-05
& 4.86367688E-09 & -1.89105331E-12 & 6.33003659E-16 & -3.90835059E+04
& -2.65266928E+01 \\
\hline
\ce{CH2OH} & -1.56007624E+05 & 2.68544628E+03 & -1.34202242E+01 &
5.75713947E-02 & -7.28444999E-05 & 4.83664886E-08 & -1.29349260E-11 &
-1.59682041E+04 & 9.96303370E+01 \\
\hline
 & 2.25034951E+06 & -8.17318606E+03 & 1.59963918E+01 & -8.70413372E-04
& 6.06918395E-08 & 4.40834946E-12 & -5.70230950E-16 & 4.64531343E+04 &
-7.83515845E+01 \\
\hline
\ce{H2CO} & -1.17391634E+05 & 1.87362885E+03 & -6.89028857E+00 &
2.64156167E-02 & -2.18638930E-05 & 1.00569301E-08 & -2.02347695E-12 &
-2.30735177E+04 & 6.42042055E+01 \\
\hline
 & 1.70082541E+06 & -7.62085384E+03 & 1.47244755E+01 & -1.64911175E-03
& 3.29214472E-07 & -3.49504977E-11 & 1.52613500E-15 & 3.14681295E+04 &
-7.38647850E+01 \\
\hline
\ce{HCO} & -1.18985189E+04 & 2.15153611E+02 & 2.73022403E+00 &
1.80651611E-03 & 4.98430057E-06 & -5.81456792E-09 & 1.86968989E-12 &
2.90575564E+03 & 1.13677254E+01 \\
\hline
 & 6.94960612E+05 & -3.65622338E+03 & 9.60473117E+00 & -1.11712928E-03
& 2.87532802E-07 & -3.62624774E-11 & 1.80832960E-15 & 2.54370444E+04 &
-3.58247372E+01 \\
\hline
\ce{CH3O} & 8.65711766E+04 & -6.63168525E+02 & 2.25745567E+00 &
2.26628379E-02 & -2.97056640E-05 & 2.19934135E-08 & -6.58804338E-12 &
4.17410213E+03 & 8.17477790E+00 \\
\hline
 & 2.10118824E+06 & -8.84196880E+03 & 1.82264573E+01 & -1.74348503E-03
& 3.34043427E-07 & -3.43067316E-11 & 1.47389777E-15 & 5.30958206E+04 &
-9.42250059E+01 \\
\hline
\ce{CH3OH} & -2.41664289E+05 & 4.03214719E+03 & -2.04641544E+01 &
6.90369807E-02 & -7.59893269E-05 & 4.59820836E-08 & -1.15870674E-11 &
-4.43326117E+04 & 1.40014219E+02 \\
\hline
 & 3.41157076E+06 & -1.34550020E+04 & 2.26140762E+01 & -2.14102918E-03
& 3.73005054E-07 & -3.49884639E-11 & 1.36607344E-15 & 5.63608156E+04 &
-1.27781428E+02 \\
\hline
\ce{CH3CO} & -7.19389413E+04 & 1.46446517E+03 & -6.63227613E+00 &
4.10846838E-02 & -4.22625664E-05 & 2.48576682E-08 & -6.29255848E-12 &
-9.30937081E+03 & 6.42289762E+01 \\
\hline
 & 2.48538815E+06 & -1.12071420E+04 & 2.27752544E+01 & -2.31426055E-03
& 4.53618917E-07 & -4.74263555E-11 & 2.04466390E-15 & 6.38008841E+04 &
-1.21535093E+02 \\
\hline
\ce{O2} & -3.42556342E+04 & 4.84700097E+02 & 1.11901096E+00 &
4.29388924E-03 & -6.83630052E-07 & -2.02337270E-09 & 1.03904002E-12 &
-3.39145487E+03 & 1.84969947E+01 \\
\hline
 & -1.03793902E+06 & 2.34483028E+03 & 1.81973204E+00 & 1.26784758E-03
& -2.18806799E-07 & 2.05371957E-11 & -8.19346705E-16 & -1.68901093E+04
& 1.73871651E+01 \\
\hline
\ce{H2CCO} & 3.54959809E+04 & -4.06306283E+02 & 3.71892192E+00 &
1.58350182E-02 & -1.72619569E-05 & 1.15737696E-08 & -3.30584263E-12 &
-5.20999258E+03 & 3.83960422E+00 \\
\hline
 & 2.01356492E+06 & -8.20088746E+03 & 1.75969407E+01 & -1.46454452E-03
& 2.69588697E-07 & -2.66567484E-11 & 1.09420452E-15 & 4.17777688E+04 &
-8.72580358E+01 \\
\hline
\ce{HCCO} & 6.95961270E+04 & -1.16459440E+03 & 9.45661626E+00 &
-2.33124063E-03 & 5.16187360E-06 & -3.52616997E-09 & 8.59914323E-13 &
2.53500399E+04 & -2.72635535E+01 \\
\hline
 & 1.09392200E+06 & -4.49822821E+03 & 1.24644643E+01 & -6.34331740E-04
& 1.10854902E-07 & -1.12548868E-11 & 5.68915194E-16 & 4.65228030E+04 &
-5.09907043E+01 \\
\hline
\end{tabular}
\end{tiny}
\end{table}

\section{Licensing and Permission to use the \texttt{TEA} Code}

We thank the developers of the Thermochemical Equilibrium Abundances (\texttt{TEA}) code \citep{blecic16}, initially developed at the University of Central Florida, Orlando, Florida, USA.  The Reproducible Research Compendium (RRC) is available at \texttt{http://github.com/exoclime/VULCAN}.


\label{lastpage}

\end{document}